\documentclass{article}

\usepackage[final,nonatbib]{neurips_2021}
\usepackage[utf8]{inputenc} 
\usepackage[T1]{fontenc}    
\usepackage{mathrsfs}
\usepackage{multirow}
\usepackage{afterpage}
\usepackage{url}            
\usepackage{booktabs}       
\usepackage{amsfonts}       
\usepackage{nicefrac}       
\usepackage{microtype}      
\usepackage{verbatim}
\usepackage{subfig}
\usepackage{makecell}
\usepackage{caption}
\usepackage{amsfonts}
\usepackage{amsmath}
\numberwithin{equation}{section}
\usepackage{color}
\usepackage[symbol]{footmisc}

\usepackage{amssymb}
\usepackage{bbm}
\usepackage{bm}
\usepackage{autobreak}
\usepackage{graphics}
\usepackage{float}
\usepackage[export]{adjustbox}
\usepackage{tablefootnote}
\usepackage{relsize}
\usepackage[usenames,dvipsnames]{xcolor}
\usepackage{graphicx,amsthm}
\usepackage[backend=biber,citestyle=numeric-comp,style=phys,eprint=true,sorting=none,natbib=true]{biblatex}
\usepackage[hidelinks,colorlinks=true,citecolor=MidnightBlue,linkcolor=BrickRed,urlcolor=blue,breaklinks=true]{hyperref}
\usepackage{pifont}
\usepackage{setspace}

\def\*#1{\mathbf{#1}}
\newcommand*\diff{\mathop{}\!\mathrm{d}}

\theoremstyle{definition}

\theoremstyle{remark}

\usepackage[capitalise]{cleveref}

\newcommand{\Mod}[1]{\ (\mathrm{mod}\ #1)}

\def\thetab{\boldsymbol{\theta}}

\def\varphib{\boldsymbol{\varphi}}

\def\etab{\boldsymbol{\eta}}

\def\varop{\mathop{\mathrm{Var}}}
\def\avgE{\mathbb{E}}

\def\<{\langle} \def\>{\rangle}

\DeclareRobustCommand{\argmin}{\operatorname*{arg \ min}}

\newcommand{\cmark}{\ding{51}}
\newcommand{\xmark}{\ding{55}}

\usepackage[linesnumbered,ruled,vlined]{algorithm2e}
\DontPrintSemicolon

\SetKwComment{Comment}{\color{green!50!black}// }{}

\newcommand{\assign}{\leftarrow}

\SetKwProg{Function}{function}{}{}

\usepackage{lipsum}

\title{Supervised Learning and the Finite-Temperature String Method for Computing Committor Functions and Reaction Rates}

\author{
  Muhammad R. Hasyim$^{1,}\hspace{-0.05ex}\thanks{These two authors contributed equally to this work.} \ \ ^{,}$\thanks{\texttt{muhammad\_hasyim@berkeley.edu}}\ \ ,   Clay H. Batton$^{1,*,}$\thanks{\texttt{chbatton@berkeley.edu}}\ \ , Kranthi K. Mandadapu$^{1,2,}$\thanks{\texttt{kranthi@berkeley.edu}} \\
  $^{1}$Department of Chemical \& Biomolecular Engineering, University of California at Berkeley\\
  $^{2}$Chemical Sciences Division, Lawrence Berkeley National Laboratory\\
}

\begin{document}

\maketitle

\begin{abstract}
A central object in the computational studies of rare events is the committor function.
Though costly to compute, the committor function encodes complete mechanistic information of the processes involving rare events, including reaction rates and transition-state ensembles.
Under the framework of transition path theory (TPT), recent work \cite{rotskoff2020learning} proposes an algorithm where a feedback loop couples a neural network that models the committor function with importance sampling, mainly umbrella sampling, which collects data needed for adaptive training. 
In this work, we show additional modifications are needed to improve the accuracy of the algorithm. 
The first modification adds elements of supervised learning, which allows the neural network to improve its prediction by fitting to sample-mean estimates of committor values obtained from short molecular dynamics trajectories. 
The second modification replaces the committor-based umbrella sampling with the finite-temperature string (FTS) method, which enables homogeneous sampling in regions where transition pathways are located. 
We test our modifications on low-dimensional systems with non-convex potential energy where reference solutions can be found via analytical or the finite element methods, and show how combining supervised learning and the FTS method yields accurate computation of committor functions and reaction rates. We also provide an error analysis for algorithms that use the FTS method, using which reaction rates can be accurately estimated during training with a small number of samples.
The methods are then applied to a molecular system in which no reference solution is known, where accurate computations of committor functions and reaction rates can still be obtained. 
\end{abstract}

\section{Introduction}
A fundamental problem in chemistry is to discover the mechanistic pathways governing kinetic processes at the microscopic level. 
These processes include phase transitions in colloidal systems \citep{lu2013colloidal}, chemical reactions at aqueous interfaces \citep{jungwirth2008ions}, and protein folding \citep{dill2008protein}. 
While diverse in context, they exhibit a common bottleneck in the form of high-energy barriers, which separate the reactant and product states of the pathway. 
Despite remarkable progress in high-performance molecular simulations \citep{plimpton1995fast,pronk2013gromacs,anderson2020hoomd}, finding these pathways is difficult due to the rarity of barrier-crossing events at timescales achievable by current computational resources. 
Studying these rare events constitute identifying the transition pathways, and sampling them is an important part of obtaining a mechanistic understanding of the problem. 

Several strategies exist for capturing rare barrier-crossing events, one of which is transition path sampling (TPS) \cite{dellago1998transition,bolhuis2002transition}; an importance sampling technique for generating an ensemble of transition pathways. An alternative strategy is to rely on transition path theory (TPT) \cite{weinan2006towards,vanden2010transition}, which can outline
various computational methods to obtain an average characteristic pathway, e.g., the finite-temperature
string (FTS) method \cite{weinan2005finite,vanden2009revisiting}.
Both strategies involve the calculation of the committor function $q(\* x)$; the probability that a trajectory starting from some initial configuration $\* x$ enters the product state before the reactant state. 
The committor function can be further used to obtain reaction rates and transition-state ensembles. 
Its standard computation entails generating many trajectories for every initial configuration 
$\* x$, which may become prohibitively expensive \cite{peters2006}. 

In the framework of TPT, the committor function can be computed by solving a high-dimensional partial differential equation (PDE) in configuration space, called the backward Kolmogorov equation (BKE) \cite{weinan2006towards,vanden2010transition,onsager1938ions}. 
The complexity in solving the high-dimensional BKE may be reduced by constructing a low-dimensional set of collective variables (CVs) \cite{maragliano2006}, but they are not known \textit{a priori} and require exhaustive trial-and-error to obtain ones that best describe a reaction pathway \cite{peters2016}. 
On the other hand, one does not need to solve the BKE over the entire configuration space to obtain reaction rates and transition-state ensembles but focuses on important regions across the transition path.
One way to target these regions is importance sampling \cite{frenkel2001understanding} where molecular simulations are biased to generate configurations according to target values of the committor function in regions across the transition path.
However, since the committor function has no closed-form expression as a function of configuration $ \* x $ and intrinsically involves 
averages over finite-time trajectories, it is impractical to use it in conjunction with existing importance sampling techniques. 
Modern machine learning (ML) approaches can alleviate this issue by representing committor functions via artificial neural networks.
This is the strategy used in recent work \cite{rotskoff2020learning} to create an ML algorithm that adopts a feedback loop between importance sampling and neural network training, which involves minimizing a loss function derived from the BKE. 
The feedback loop uses the neural network to acquire high-quality data from short molecular dynamics (MD) or Monte Carlo (MC) simulations via umbrella sampling \cite{torrie1977nonphysical} where a bias potential built from the neural network enhances sampling of the transition state. 
However, as will be shown in this work, umbrella sampling poorly explores regions across the transition path, which may result in an inaccurate computation of committor functions and thereby inaccurate, high-variance estimates of the reaction rates. 
This issue may be mitigated by a careful fine-tuning of the parameters used in umbrella sampling, which is a non-trivial task, or increasing the number of samples used during training, which may require long molecular simulations to reach the desired accuracy. 
Furthermore, the bias potential built from the neural network can lead to prohibitively expensive simulation due to the non-local many-body nature and size of the neural network.

In this work, we improve the algorithm in Ref.~\cite{rotskoff2020learning} to increase its accuracy.
The accuracy is evaluated by computing the error in the committor function and reaction rate, 
with both errors evaluated between the neural network and a solution of the BKE computed either using analytical methods or the finite element method with fine resolution for low-dimensional problems.
We show that accuracy in committor functions can be improved by
adding elements of supervised learning, where the neural network is trained on estimates of committor values generated via short trajectories. Accuracy in reaction rates can be improved by replacing the committor-based umbrella sampling with the FTS method \cite{vanden2009revisiting}, which samples configurations homogeneously across the transition path, and enables accurate low-variance on-the-fly estimation of reaction rates. 
The resulting algorithm with the FTS method is also amenable to error analysis, enabling accurate estimation of reaction rates with a lower number of samples. 
We also demonstrate the applicability of this method to a molecular system with a high-dimensional configuration space and demonstrate that accurate computations of the committor function and reaction rate can be obtained.

Our paper is organized as follows: in \cref{sec:abridgedtpt}, we review the framework of TPT to introduce the BKE and construct an optimization problem from the BKE that is feasible to solve using ML. 
In \cref{sec:base}, we review the ML algorithm proposed in Ref.~\cite{rotskoff2020learning}, and describe how it uses umbrella sampling with feedback loops. 
We propose modifications to this algorithm starting with the addition of supervised learning elements in \cref{sec:basesl} and ending
with the review and use of the FTS method for importance sampling in \cref{sec:basefts}. 
In \cref{sec:1dbrownian,sec:2dmullerbrown}, we test all algorithms to problems corresponding to a particle diffusing in non-convex potential energies, showcasing how our modifications lead to a more accurate low-variance computation of the committor function and reaction rates. In \cref{sec:lognorm}, we provide an error analysis for algorithms that use the FTS method, demonstrating that the sampling distribution of the estimated reaction rates obeys a log-normal distribution, which can be used to remove the sampling error in these estimates.
In \cref{sec:dimer_solvent}, we apply the algorithms to a molecular system, i.e., a solvated dimer undergoing a transition between a compact to an extended state, and find the previously seen trends in low-dimensional systems to be applicable to such a high-dimensional system. 

\section{Theory and Algorithms} \label{sec:algs}

\subsection{From Transition Path Theory to Machine Learning} \label{sec:abridgedtpt}

\begin{figure}[t]
  \centering
  \includegraphics[width=0.75\linewidth]{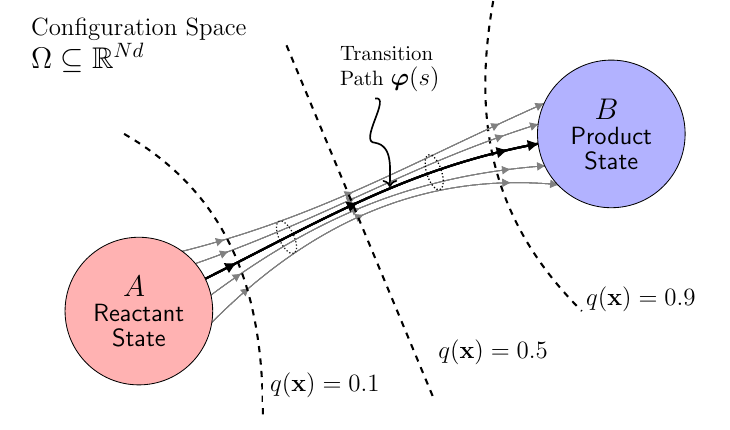}
  \caption{A schematic of transition path theory (TPT). Gray lines are flow lines of the probability flux $\* J(\* x)$, and the transition tube, i.e., the region of high flux, is localized around the transition path $\varphib(s)$. Dashed lines are isocommittor surfaces, with the middle dashed line defining the transition-state ensemble where $q(\* x)=0.5$.} \label{fig:tptillus2}
\end{figure}

To review TPT, consider a $d$-dimensional system with $N$-many particles at equilibrium that interact with a potential energy function $V(\* x)$, where $\* x \in \Omega $ is a configuration of the system and $\Omega \subset \mathbb{R}^{Nd}$ is the configuration space. 
Equilibrium properties can be computed via ensemble averages $\langle \ldots \rangle =  \int_\Omega \diff \* x \* \rho(\* x) \ldots$ over the Boltzmann distribution $\rho(\* x) = e^{-\beta V( \* x)}/Z$ where $\beta = 1/k_\mathrm{B} T$ with $k_\mathrm{B}$ being the Boltzmann constant, $T$ the temperature, and $Z = \int_\Omega \diff \* x \ e^{-\beta V(\* x)}$ the partition function. 
Given this model system, TPT can be used to analyze the system's transition from a reactant state $A \subset \Omega$ to a product state $B \subset \Omega$ \cite{weinan2006towards,vanden2010transition,berezhkovskii2013diffusion}; see \cref{fig:tptillus2} for a schematic of the problem. 
Central to TPT is the calculation of the  committor function $q(\* x)$, which is defined as the probability to first reach $B$ before $A$ given that the system initially starts at $\* x_0=\* x$. 
The formula for $q(\* x)$ is given by
\begin{equation}
    q(\* x) = \avgE \left[h_{B}\left(\* x_{\tau}\right) \mid \* x_0 = \* x \right]; \quad \tau = \argmin_{t \in [0,+\infty)} \{\* x_t \in A \cup B : \* x_0 = \* x \} \,, \label{eq:forwardqdef}
\end{equation}
where $ \avgE [ \ldots \mid   \* x_0 = \* x]$ is an average over all trajectories starting from $\* x$, $\tau$ is the first-passage time, and $h_C(\* x)=1$ if $\* x \in C$ and zero otherwise. 
Using stochastic calculus \cite{durrett1996stochastic}, one may compute the committor as a solution to the steady-state backward Kolmogorov equation (BKE)
\begin{equation}
    \nabla_{\* x} \cdot \* J(\* x) = 0 \,,  \label{eq:bke}  
\end{equation}
with $\* J(\* x) = \rho (\* x) \* D ( \* x ) \nabla_{\* x} q(\* x)$ being the probability flux and $\* D (\* x) $ being the position-dependent diffusivity matrix, subjected to the boundary conditions
\begin{equation}
    q(\* x) = 0, \ \* x \in \partial A; \quad q(\* x) = 1, \ \* x \in  \partial B \label{eq:bkebc} \,,
\end{equation}
where $\partial A$ and $\partial B$ are the boundaries of $A$ and $B$ respectively. 

Solving the BKE for the committor function allows us to evaluate many quantities including transition paths, transition-state ensembles, and reaction rates. 
The transition path is a curve $\varphib(s)$ that encodes how the system, on average, moves from $A$ to $B$ in the configuration space. 
For every value of $s$, one can compute $\varphib(s)$ self-consistently as the average configuration weighted by the flux $|\*  J(\* x)| = \rho(\* x)\frac{k_\mathrm{B} T}{\gamma}|\nabla_{\* x} q(\* x)|$ at a chosen level set of the committor function $q(\* x)$, i.e.,
\begin{equation}
    \varphib(s) = \frac{\int_{P} \diff S |\* J(\* x)| \* x}{\int_{P} \diff S |\* J(\* x)| } = \frac{\int_{P} \diff S \rho(\* x) |\nabla_{\* x}q(\* x)|  \* x}{\int_{P} \diff S \rho(\* x) |\nabla_{\* x}q(\* x)| } \,, \label{eq:pathdef} 
\end{equation}
where $\int_{P} \diff S $ is a surface integral over the level set $P= \{ \* x \in \Omega : q(\* x)=q(\varphib(s)) \}$ \cite{weinan2006towards, vanden2010transition}. 
Note that for processes involving high-energy barriers the region of high flux typically forms a tubular region called the transition tube, which is localized around $\varphib(s)$; see \cref{fig:tptillus2}. 
The level sets of $q(\* x)$ are also referred to as the isocommittor surfaces, where the isocommittor surface corresponding to the level set $\{ \* x \in \Omega : q(\* x)=\frac{1}{2} \}$ defines the transition-state ensemble. The reaction rate $\nu_R$, defined as the frequency with which a system transitions from $A$ to $B$, can be evaluated as \cite{weinan2006towards}
\begin{equation}
    \nu_R = \frac{k_\mathrm{B} T}{\gamma}\int_\Omega \diff \* x \ \rho(\* x) |\nabla_{\* x} q(\* x)|^2  = \frac{k_\mathrm{B} T}{\gamma} \left\langle |\nabla_{\* x} q(\* x)|^2  \right\rangle \label{eq:rxnrate} \,.
\end{equation}

The BKE, which is a high-dimensional PDE, is infeasible to solve via standard finite difference/elements for large molecular systems, as the number of grid points/elements grows exponentially with system size $N$.
However, it is in these situations that methods inspired by ML may hold a feasible alternative, where the committor function can be approximated by a neural network whose model parameters can be solved by transforming the BKE into an optimization problem \cite{khoo2019solving,li2019computing,rotskoff2020learning,li2021semigroup}.
To this end, we begin by constructing a variational form of the BKE. 
Following the standard procedure for elliptic PDEs \cite{panosfemnotes}, we consider a variation of the committor function $ \delta q (\* x) $, which obeys the constraints $ \delta q (\* x ) = 0 $ for $ \* x \in \partial A $ and $ \* x \in \partial B $ to satisfy the boundary conditions in \cref{eq:bkebc}.
Multiplying \cref{eq:bke} by $ \delta q (\* x) $, integrating over $ \Omega \setminus A \cup B$, and then integrating by parts yields
\begin{equation}
    \int_{\Omega \setminus A \cup B} \diff \* x \ \delta q(\* x) \nabla_{\* x} \left[ \rho(\* x) \nabla_{\* x} q(\* x)\right] = - \int_{\Omega  \setminus A \cup B }\diff \* x \ \rho(\* x) \ \nabla_{\* x}\delta q(\* x)  \cdot \nabla_{\* x} q(\* x) = 0 \,. \label{eq:integratebkewithq}
\end{equation}
Applying Vainberg's theorem \cite{panosfemnotes} to \cref{eq:integratebkewithq} leads to the following
functional: 
\begin{equation}
    L \left[ \tilde{q} \right] = \frac{1}{2} \int_{\Omega \setminus A \cup B} \diff \* x \rho(\* x) |\nabla_{\* x} \tilde{q}(\* x)|^2  = \frac{1}{2} \left\langle |\nabla_{\* x} \tilde{q}(\* x)|^2 \right\rangle_{\Omega  \setminus A \cup B } \label{eq:variational_estimate}
\end{equation}
whose extremization over the space of admissible functions $\tilde{q}(\* x)$ subject to boundary conditions \cref{eq:bkebc} leads to the solution of the BKE. 
The variational form in \cref{eq:variational_estimate} therefore transforms the strong form of BKE into a problem of functional optimization, where the committor function satisfies
\begin{equation}
    q (\* x) = \argmin_{\tilde{q}} L \left[ \tilde{q} \right] \quad \text{s.t.} \quad \tilde{q}(\* x) = 0, \ \* x \in  \partial A; \quad \tilde{q}(\* x) = 1, \ \* x \in \partial B \,. \label{eq:variational_problem}
\end{equation}

\Cref{eq:variational_problem} guides a new ML-based optimization problem, where
we may approximate the committor function with a neural network model $ q(\* x) \approx \hat{q}(\* x; \thetab)$ with the model parameters $\thetab$. 
Introducing the BKE loss function as
\begin{equation}
    \ell(\* x; \thetab) = \frac{1}{2} |\nabla_{\* x} \hat{q}(\* x; \thetab)|^2 \label{eq:bkeloss}
\end{equation}
and imposing boundary conditions in \cref{eq:bkebc} by the penalty method \cite{nocedal2006numerical} with the loss functions
\begin{align}
    \ell_\mathrm{A}(\* x_A; \thetab) &= \frac{1}{2}(\hat{q}(\* x_A; \thetab))^2 \,, \label{eq:reactantloss} \\ 
    \ell_\mathrm{B}(\* x_B; \thetab) &= \frac{1}{2}(\hat{q}(\* x_B; \thetab)-1)^2 \,, \label{eq:productloss}
\end{align}
where $ \* x_A \in A $ and $ \* x_B \in B $, the model parameters $\thetab$ can be obtained by extremizing the following objective function:
\begin{gather}
L(\thetab) = \left\langle  \ell(\* x; \thetab) \right\rangle + \lambda_\mathrm{A} \left\langle \ell_\mathrm{A}(\* x; \thetab) \right\rangle_A+ \lambda_\mathrm{B} \left\langle \ell_\mathrm{B}(\* x; \thetab) \right\rangle_B \,. \label{eq:variationalbke}
\end{gather}
Here, $\left\langle \ldots \right\rangle_{C}$ denotes ensemble averaging constrained in a region $C \subset \Omega$, and $\lambda_\mathrm{A}$ and $\lambda_\mathrm{B}$ control the penalty strengths that enforce boundary conditions at $A$ and $B$, respectively. Note that the ensemble average of the BKE loss function $ \left\langle  \ell(\* x; \thetab) \right\rangle$ is proportional to the reaction rate in \cref{eq:rxnrate} up to a constant factor $2 k_\mathrm{B} T/ \gamma$, and thus it is crucial for any ML approach that solves the BKE to be able to compute $ \left\langle  \ell(\* x; \thetab) \right\rangle$ accurately.

The task of minimizing \cref{eq:variationalbke} may not yet be feasible in large system sizes, since the ensemble averages involve high-dimensional integrals, which may be evaluated via standard quadrature but their computational cost grows exponentially with system size. 
To resolve this issue, one may approximate the ensemble averages in \cref{eq:variationalbke} with averages over samples obtained via molecular dynamics (MD) or Monte Carlo (MC) simulations. 
In this case, \cref{eq:variationalbke} can be evaluated as 
\begin{align}
    \hat{L}(\thetab; \mathcal{S}, \mathcal{A}, \mathcal{B}) & =  \frac{1}{|\mathcal{S}|} \sum_{\* x \in \mathcal{S}} \ell(\* x; \thetab) +\frac{\lambda_\mathrm{A}}{|\mathcal{A}|} \sum_{\* x \in \mathcal{A}}  \ell_\mathrm{A}(\* x; \thetab) + \frac{\lambda_\mathrm{B}}{|\mathcal{B}|} \sum_{\* x \in \mathcal{B}}  \ell_\mathrm{B}(\* x; \thetab) \,, \label{eq:erm_bke}
 \end{align}
where $\mathcal{A}$, $\mathcal{B}$ and $\mathcal{S}$ are batches of samples obtained in the reactant state $A$, product state $B$ and configuration space $\Omega $, respectively, and the operator $| \cdot |$  denotes the size of each batch. 
The outlined strategy is the basis behind some of the recent ML approaches for solving the BKE \cite{khoo2019solving,li2019computing,rotskoff2020learning,li2021semigroup} though earlier works can be found that utilize a different objective function to train a neural network that takes collective variables as input and is trained on data obtained from transition path sampling \cite{ma2005rc,peters2006likelihood}.
The main challenge inherent in these approaches is sampling; since the first term in \cref{eq:variationalbke} is proportional to the magnitude of the flux $|\*J(\* x; \thetab)| = \rho(\* x) \frac{k_\mathrm{B} T}{\gamma} |\nabla_{\* x} \hat{q}(\* x; \thetab)|$, the optimization problem is dominated by the rare configurations found in regions of high flux, e.g. the transition-state ensemble. 
An inadequate sampling of the transition-state ensemble may lead to poor estimates of the average BKE loss function in \cref{eq:bkeloss}, resulting in an inaccurate computation of committor functions and reaction rates in \cref{eq:rxnrate}. 
Inadequate sampling may also lead to poor estimates of the gradient $ \nabla_{\thetab} L $, which may negatively impact the performance of the neural network training.
In Ref.~\cite{rotskoff2020learning}, this sampling problem is partially resolved via an importance sampling technique, namely umbrella sampling, that is coupled with the neural network model in a feedback loop. 

\subsection{Solving the BKE with Umbrella Sampling and Feedback Loops} \label{sec:base}

In this section, we review the algorithm in Ref.~\cite{rotskoff2020learning} that utilizes umbrella sampling for obtaining the committor functions. 
To this end, consider a system that evolves via discrete overdamped Langevin dynamics with noise $ \* w_t $ that has zero mean and unit variance.
Umbrella sampling biases the system's dynamics by adding a potential of the form $ W (\* x; \thetab) = \frac{1}{2} \kappa ( \hat{q}(\* x; \thetab)-q_0)^2 $ to the potential energy function $ V ( \* x ) $, where $ q_0 $ is the target committor value and $ \kappa $ is the bias strength.
This bias leads to modified equations of motion
\begin{equation}
\* x_{t+1} = \* x_t-\gamma^{-1} \nabla_{\* x}\left[ V(\* x_t) + W (\* x_t; \thetab) \right]\Delta t + \sqrt{2 k_\mathrm{B} T \Delta t \gamma^{-1}} \* w_t \,, \label{eq:ussamplinglangevin}
\end{equation}
which sample a target distribution given by $\rho (\* x; \thetab) \propto e^{-\beta[V(\* x)+W (\* x; \thetab)]}$ as $\Delta t  \to 0$. With a suitable choice of $q_0$ and $\kappa$, the system may explore configurations $ \* x $ and values of $ \hat{q}( \* x ; \thetab ) $ that are rare according to the unbiased equilibrium distribution $\rho(\* x) \sim e^{-\beta V(\* x)}$.
In Ref.~\cite{rotskoff2020learning}, this strategy is expanded to target a range of $ \hat{q}(\* x; \thetab) $ values between zero and one by introducing $M$-many simulation systems, each of which uses a biasing potential with a unique target value and biasing strength.
Referring to these simulation systems as replicas and enumerating them via an indexing variable $ \alpha \in \{ 1,\ldots, M\}$,
the bias potential for each replica can be written as $ W_\alpha (\* x; \thetab)  = \frac{1}{2}\kappa_\alpha ( \hat{q}(\* x; \thetab)-q_\alpha)^2 $, which induces a biased distribution  $\rho_\alpha(\* x; \thetab) \propto e^{-\beta[V(\* x)+W_\alpha (\* x; \thetab)]}$. 
The set of target committor values and biasing strengths is denoted as $\{(\kappa_\alpha, q_\alpha )\}_{\alpha=1}^M$. 
Note that the configurations corresponding to the target distributions can also be generated via MC or other MD methods instead of \cref{eq:ussamplinglangevin}. 

The algorithm for solving the BKE is a closed feedback loop between the replica dynamics and any chosen optimizer, such as stochastic gradient descent (SGD) \cite{robbins1951stochastic}, Heavy-Ball \cite{polyak1964some}, or Adam \cite{kingma2014adam}, to obtain model parameters $\thetab$ that extremize \cref{eq:erm_bke}.
At the $k$-th iteration, replicas generate samples that are stored into a collection of batches $ \{ \mathcal{M}^\alpha_k \}_{\alpha=1}^M$, where the $\alpha$-th batch $\mathcal{M}^\alpha_k$ consists of samples obtained from a short MD/MC trajectory run of the $\alpha$-th replica. 
This data is then used to compute the gradient $\nabla_{\thetab} \hat{L}$ in order to update the model parameters $\thetab_{k} \to \thetab_{k+1}$. 
At the $(k+1)$-th iteration, the process repeats by using $ \hat{q}( \* x ; \thetab_{k+1} ) $ to obtain new samples for further optimization.

\begin{figure}[t]
\centering
\begin{minipage}{0.5\linewidth}
\begin{algorithm}[H]
\caption{The BKE--US Method \citep{rotskoff2020learning} \label{alg:base}}
\KwData{Initial conditions $\thetab_0$. Reactant and product batches $\mathcal{A}$ and  $\mathcal{B}$. Hyperparameters for optimizer $\etab$. Bias potential parameters $\{(\kappa_\alpha, q_\alpha)\}_{\alpha=1}^M$. Penalty strengths $\lambda_\mathrm{A}$ and $\lambda_\mathrm{B}$.}
{
\For {$k=0,\dots,K$}{
    \For {$\alpha=1,\dots,M$ in parallel}{
        \For {$m=1,\dots,|\mathcal{M}_k^\alpha|$}{
            Sample $\* x^{\alpha}_{m} \sim \rho_\alpha(\* x; \thetab_k)$ with MD/MC simulation, e.g., \cref{eq:ussamplinglangevin}.
        
            Store $\* x^{\alpha}_{m}$ in batch $\mathcal{M}_k^\alpha$.
            
            }
        }
    
    Sample mini-batch $\mathcal{A}_k \subset \mathcal{A}$ and $\mathcal{B}_k \subset \mathcal{B}$.
    
    Compute $z_\alpha$ with a free-energy method, e.g., FEP \cref{eq:expestimator}.
    
    Compute $\nabla_{\thetab} \hat{L}(\thetab_k; \{ (\mathcal{M}^\alpha_k, z_\alpha) \}, \mathcal{A}_k, \mathcal{B}_k)$ with \cref{eq:basemethod_grad}.
    
    Update $\thetab_k \to \thetab_{k+1}$ with optimizer.
    
}
}
\end{algorithm}
\end{minipage}
\quad
\begin{minipage}{0.45\linewidth}
\includegraphics[width=\linewidth]{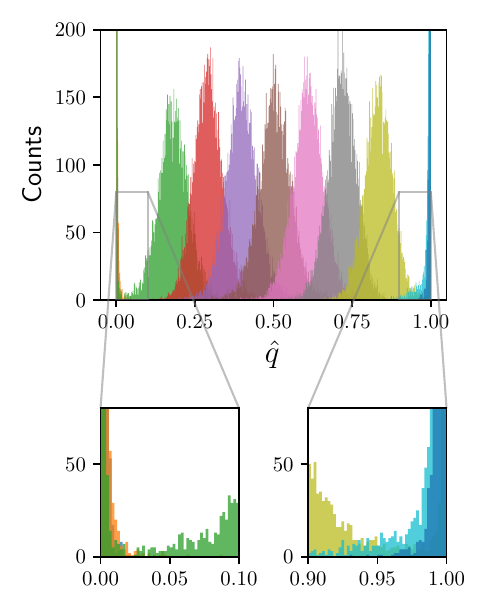}
\end{minipage}
\caption{(Left) Pseudo-code corresponding to the BKE--US method. Lines 2-6 are the sampling steps, Lines 7-9 are the optimization steps, and a feedback loop couples the sampling and optimization steps together. Note that the sampling of configuration $\* x^\alpha_m$ in Line 4 utilizes a fixed simulation length to obtain uncorrelated samples in the batch $\mathcal{M}_k^\alpha$---a convention used for all subsequent algorithms proposed in this work. (Right) Histograms of committor values from committor-based umbrella sampling. The histograms overlap near the transition state, with inset plots showing that the histograms are non-overlapping near the reactant and product states. See also \cref{fig:1dresults}(b, top) for the corresponding histograms in configuration space.}
\label{fig:basemethod}
\end{figure}

The algorithm requires two additional components.
First, the reactant and product batches $ \mathcal{A} $ and $ \mathcal{B} $ are generated using short MD/MC trajectories constrained in the reactant and product states, respectively.
Second, a formula for $\nabla_{\thetab} \hat{L}$ is needed for the optimizer and is obtained using a reweighting procedure \cite{thiede2016eigenvector} to compute the unbiased sample averages from biased samples. This yields
\begin{align}
\nabla_{\thetab} \hat{L}\left(\thetab_k;  \{ (\mathcal{M}^\alpha_k, z_\alpha) \}, \mathcal{A}_k, \mathcal{B}_k \right) =&  \dfrac{\sum\limits_{\alpha=1}^M  \dfrac{z_{\alpha}}{|\mathcal{M}^\alpha_k|}\sum\limits_{\* x \in \mathcal{M}^\alpha_k}  \left[\dfrac{\nabla_{\thetab} \ell(\* x; \thetab_{k})}{c(\* x; \thetab_k)} \right]  }{ \sum\limits_{\alpha=1}^M \dfrac{z_{\alpha}}{|\mathcal{M}^\alpha_k|}\sum\limits_{\* x \in \mathcal{M}^\alpha_k} \left[\dfrac{1}{c(\* x; \thetab_k) } \right]  } +\frac{\lambda_\mathrm{A}}{|\mathcal{A}_k|} \sum\limits_{\* x \in \mathcal{A}_k} \nabla_{\thetab} \ell_\mathrm{A}(\* x; \thetab_{k}) \nonumber
\\
                                                                                                                        &+\frac{\lambda_\mathrm{B}}{|\mathcal{B}_k|} \sum\limits_{\* x \in \mathcal{B}_k}\nabla_{\thetab} \ell_\mathrm{B}(\* x; \thetab_{k}) \,, \label{eq:basemethod_grad}
\end{align}
where $\mathcal{A}_k \subset \mathcal{A}$ and $\mathcal{B}_k \subset \mathcal{B}$ are mini-batches obtained from random sub-sampling of the reactant and product batches, respectively, and $c(\* x; \thetab) = \sum_{\alpha=1}^M e^{-\beta W_\alpha(\* x; \thetab)}$. Here, $z_\alpha$ is a reweighting factor given by the relative partition function
\begin{equation}
    z_\alpha = \frac{Z_\alpha}{\sum_{\alpha^\prime=1}^M Z_{\alpha^\prime}}=\dfrac{\int \diff \* x \ e^{-\beta[V(\* x)+W_\alpha (\* x; \thetab)]}}{\sum_{\alpha^\prime=1}^M \int \diff \* x \ e^{-\beta[V(\* x)+W_{\alpha^\prime}(\* x; \thetab)]}} \,, \label{eq:zalphadef}
\end{equation}
where $Z_\alpha$ is the partition function of the $\alpha$-th replica.
Given the batches of samples $ \{ \mathcal{M}^\alpha_{k} \}_{\alpha=1}^M$, various free-energy methods \cite{stoltz2010free} can be used to compute $Z_\alpha$ via the free-energy $F_\alpha = -\frac{1}{\beta} \ln Z_\alpha$. In this work, we use free-energy perturbation (FEP) \cite{zwanzig1954high} where the estimator for $z_\alpha$ is derived from the following exact identity: 
\begin{equation}
    \frac{z_\alpha}{z_{\alpha^\prime}}=e^{-\beta \Delta F_{\alpha, \alpha^\prime}} = \left\langle  \dfrac{\phi_{\alpha}(\* x;\thetab)}{\phi_{\alpha^\prime}(\* x;\thetab)} \right\rangle_{\alpha^\prime} \,, \label{eq:basicexpidentity}
\end{equation}
where $\langle \ldots \rangle_{\alpha^\prime}$ is an ensemble average over the distribution $\rho_{\alpha^\prime} \propto e^{-\beta [V(\* x)+W_{\alpha^\prime}(\* x; \thetab)]}$ obeyed by the $\alpha^\prime$-th replica, $\Delta F_{\alpha, \alpha^\prime} = F_{\alpha}- F_{\alpha^\prime}$ is the relative free-energy difference, and $\phi_\alpha(\*x; \thetab) = e^{-\beta W_\alpha(\*x; \thetab)}$. 
Given a batch $\mathcal{M}_k^{\alpha^\prime}$ from the $\alpha^\prime$-th replica, \cref{eq:basicexpidentity} can be estimated as
\begin{equation}
\frac{z_\alpha}{z_{\alpha^\prime}} \approx \frac{1}{|\mathcal{M}_k^{\alpha^\prime}|}  \sum_{\* x \in \mathcal{M}^{\alpha^\prime}_k} \dfrac{\phi_{\alpha}(\* x;\thetab_k)}{\phi_{\alpha^\prime}(\* x;\thetab_k)} \,. \label{eq:basicexpestimator}
\end{equation}
The accuracy of \cref{eq:basicexpestimator} quickly deteriorates if samples obtained between the $\alpha$-th and $\alpha^\prime$-th replicas do not overlap \cite{wu2005}.
To mitigate this issue, we can employ a strategy called stratification \cite{pohorille2010good}, where the forward and backward free-energy differences per \cref{eq:basicexpestimator} between adjacent replicas are used to compute the overall free-energy difference of replica $ \alpha $ in reference to replica $ \gamma $.
This strategy yields the following formula:
\begin{gather}
z_\alpha = \frac{z_\alpha^\star}{\sum_{\alpha=1}^M z_\alpha^\star}; \quad z_{\alpha }^\star = \begin{cases}
\prod\limits_{i=\gamma}^{\alpha-1}  e^{-\beta \Delta F_{(i+1),i}}  \approx \prod\limits_{i=\gamma}^{\alpha-1} \left(\dfrac{1}{|\mathcal{M}_k^i|}  \mathlarger{\sum}\limits_{\* x \in \mathcal{M}_k^i} \dfrac{\phi_{i+1}(\* x;\thetab_{k})}{\phi_i(\* x;\thetab_{k})}\right) & \alpha > \gamma
\\
\prod\limits_{i=\alpha}^{\gamma-1} e^{-\beta \Delta F_{(i-1),i}} \approx \prod\limits_{i=\alpha}^{\gamma-1} \left(\dfrac{1}{|\mathcal{M}_k^i|}  \mathlarger{\sum}\limits_{\* x \in \mathcal{M}_k^i} \dfrac{\phi_{i-1}(\* x;\thetab_{k})}{\phi_i(\* x;\thetab_{k})} \right) & \alpha < \gamma
\\
1 & \alpha = \gamma
\end{cases} \label{eq:expestimator} \,,
\end{gather}
where $\gamma \sim \mathrm{unif}\{1,M\}$ is randomly chosen at every iteration. 
In what follows, we shall refer to this complete algorithm as the BKE--US method, whose pseudocode is described in Algorithm~\ref{alg:base} (\cref{fig:basemethod}, left).
Note that Ref.~\cite{rotskoff2020learning} recommends choosing a different set of biasing potentials such that $ c ( \* x ; \thetab_{k} ) \approx 1 $, which corresponds to a special case of \cref{eq:basemethod_grad}. Additionally, Ref.~\cite{rotskoff2020learning} uses replica exchange, where configurations are exchanged between neighboring replicas to alleviate issues with metastability, which is not used here.

The challenge in the BKE--US method lies in selecting the bias potential parameters $\{ (\kappa_\alpha, q_\alpha) \}_{\alpha=1}^M$ such that the average loss functions and their gradients are  accurately estimated with low variance. 
Since these estimates are obtained by reweighting procedures their accuracy depends severely on obtaining an accurate estimate of the free-energy differences $ \Delta F_{\alpha, \alpha^\prime} $, and hence the reweighting factors $ z_{\alpha} $.
If one follows the procedures common to umbrella sampling and free-energy calculations, this is achieved by ensuring overlap in the histograms of the biased $\hat{q}(\* x; \thetab)$ values \cite{pohorille2010good}. 
One may choose as initial guess $q_\alpha = (\alpha - 1)/(M-1) $ with equal biasing strengths, which is the setting recommended in Ref.~\cite{rotskoff2020learning}, to obtain such overlap. However, since the committor varies rapidly near the transition state in the presence of high-energy barriers, this setting may lead to inadequate sampling of regions between the transition state and reactant/product state. 
This reduces the overlap between histograms, thereby reducing the accuracy as well as increasing the variance of the estimated average loss functions obtained from reweighting. 
Figure~\ref{fig:basemethod}(right) shows such behavior in the histograms of $\hat{q}$-values, with the replicas near the edges having progressively worse overlaps than the replicas biased towards the transition state. 
Such a non-overlapping behavior is even more apparent in the configuration space, as shown in \cref{fig:1dresults}(b) for a one-dimensional system, where large gaps in the histograms between the reactant/product basins and the transition states can be observed.
It may be plausible that further importance sampling near the edges increases the overlap, but this requires further fine-tuning of the bias parameters to focus more heavily on regions where $ \hat{q}(\* x ; \thetab ) \approx 0 $ and $ \hat{q}(\* x ; \thetab ) \approx 1 $; a non-trivial procedure to perform in high-dimensional systems. 
Alternatively, one may also increase the batch size to improve the chances of obtaining samples in the poorly targeted regions, but this task may require prohibitively long simulations.
Altogether, these issues motivate us to construct modifications to the BKE--US method, described in the next sections.

\subsection{Adding Elements of Supervised Learning} \label{sec:basesl}

To begin with, the accuracy of the BKE--US method (Algorithm~\ref{alg:base}) can be improved by adding supervised learning elements, where one can train the neural network to fit $\hat{q}(\* x; \thetab)$ to known estimates of $ q ( \* x ) $.
It has been found that supervised learning elements in the context of training neural network models achieve better performance by finding global minima in problems originally devoid of such elements \citep{du2018gradient, du2019gradient,hardtrecht}.
In our case, supervised learning can be implemented by evaluating an estimate of $ q ( \* x) $ denoted as the empirical committor function $ q_\mathrm{emp}(\* x) $ using short trajectories that start from a configuration $\* x$. The quantity $ q_\mathrm{emp}(\* x) $ can be obtained from a sample-mean estimator of  \cref{eq:forwardqdef}:
\begin{equation}
    q_\mathrm{emp}(\* x) = \frac{1}{H} \sum_{i=1}^{H} h_{B} \left(\* x_{\tau} ; \ \* x_0 = \* x \right) \,, \label{eq:empiricalcommittor}
\end{equation}
where the averaging is performed over $H$-many trajectories that are conditioned upon starting at $\* x_0 = \* x$, and ending at the first-passage time $\tau$.
This estimator obeys the binomial distribution and its variance scales as $ \frac{1}{H} $ \cite{peters2006}. 
It is important to note that supervised learning of committor functions without importance sampling is ineffective since it is necessary for the neural network to be trained on empirical committor values corresponding to rare events, i.e., configurations along the transition tube including the transition state. To this end, one may use either umbrella sampling as described before or the FTS method, which will be introduced in \cref{sec:basefts}, to target the transition tube.

At this stage, an objective function must be formulated to inform $ \hat{q}(\* x; \thetab) $ with the empirical committor function $ q_\mathrm{emp}(\* x) $.
To this end, a loss function in supervised learning is typically postulated as the squared error for every configuration $\* x$: 
\begin{equation}
    \ell_{\mathrm{MSE}}( q_\mathrm{emp}, \* x; \thetab) = \frac{1}{2}( \hat{q}(\* x; \thetab) - q_\mathrm{emp} )^2 \,. \label{eq:slse}
\end{equation}
Suppose that $q_\mathrm{emp}(\* x)$ is computed from configurations sampled by different replicas during importance sampling. 
For every $\alpha$-th replica, this allows us to generate a batch of samples $\mathcal{C}^\alpha$, which is a set of pairs of empirical committor function and its corresponding configuration. 
Denoting the collection of batches as $\{ \mathcal{C}^\alpha\}_{\alpha=1}^M$,  and given \cref{eq:slse}, the objective function as a mean-squared error has the form 
\begin{equation}
    \hat{L}_{\mathrm{MSE}}(\thetab; \{ \mathcal{C}^\alpha \}) = \frac{\lambda_{\mathrm{MSE}}}{M} \sum_{\alpha=1}^M \frac{1}{|\mathcal{C}^\alpha|} \sum_{(q_{\mathrm{emp}}, \* x) \in \mathcal{C}^\alpha} \ell_{\mathrm{MSE}}(q_\mathrm{emp}, \* x; \thetab)  \label{eq:loss_mse} \,,
\end{equation}
where $ \lambda_{\mathrm{MSE}} $ is the penalty strength. 
In practice, an optimizer to train the neural network requires the gradient $\nabla_{\thetab} \hat{L}_\mathrm{MSE}$ as additional input, which can be computed using a collection of mini-batches $\{ \mathcal{C}^\alpha_k \}$ with $\mathcal{C}^\alpha_k \subset \mathcal{C}^\alpha$ generated via random sub-sampling of the original batch $\mathcal{C}^\alpha$ similar to the sub-sampling procedure in \cref{eq:basemethod_grad}.

\begin{figure}[t]
\centering
\begin{minipage}{\linewidth}
\begin{algorithm}[H]
\caption{The BKE--US+SL Method \label{alg:basesl}}
\KwData{Initial conditions $\thetab_0$. Reactant and product batches $\mathcal{A}$ and  $\mathcal{B}$. Hyperparameters for optimizer $\etab$. Bias potential parameters $\{(\kappa_\alpha, q_\alpha)\}_{\alpha=1}^M$. Penalty strengths $\lambda_\mathrm{A}$,  $\lambda_\mathrm{B}$, and $\lambda_\mathrm{SL}$. Starting and ending iteration index, $k_\mathrm{emp,s}$ and $k_\mathrm{emp,e}$, and sampling period $\tau_\mathrm{emp}$ for supervised learning.}
{
\For {$k=0,\dots,K$}{
    \For {$\alpha=1,\dots,M$ in parallel}{
        \For {$m=1,\dots,|\mathcal{M}_k^\alpha|$}{
            Sample $\* x^{\alpha}_{m} \sim \rho_\alpha(\* x; \thetab_k)$ with MD/MC simulation, e.g., \cref{eq:ussamplinglangevin}.
        
            Store $\* x^{\alpha}_{m}$ in batch $\mathcal{M}_k^\alpha$.
            
            }
            \uIf{$ k \geq k_\mathrm{emp,s} $ and $ k < k_\mathrm{emp,e} $ and $ k \Mod{ \tau_\mathrm{emp} } = 0 $}{
                Evaluate $ q_\mathrm{emp} $ at $\* x^{\alpha} \in \mathcal{M}_k^\alpha $ with \cref{eq:empiricalcommittor}.
                
                Store $ ( q_\mathrm{emp} , \* x^{\alpha} ) $ in batch $\mathcal{C}^\alpha$.
            }
            
        }
    
        Sample mini-batch $\mathcal{A}_k \subset \mathcal{A}$, $\mathcal{B}_k \subset \mathcal{B}$, and $\mathcal{C}_k^{\alpha} \subset \mathcal{C}^\alpha $.
    
    Compute $z_\alpha$ with a free-energy method, e.g., FEP \cref{eq:expestimator}.
    
    Compute $\nabla_{\thetab} \hat{L}(\thetab_k; \{ (\mathcal{M}^\alpha_k, z_\alpha) \}, \mathcal{A}_k, \mathcal{B}_k)+\nabla_{\thetab}  \hat{L}_\mathrm{SL}(\thetab_k; \{\mathcal{C}_k^\alpha \})$ with \cref{eq:basemethod_grad,eq:loss_sl}.
    
    Update $\thetab_k \to \thetab_{k+1}$ with optimizer.
}
}
\end{algorithm}
\end{minipage}
\caption{Pseudo-code for the BKE--US+SL method.}
\label{fig:baseslmethod}
\end{figure}

Note that a finite number of trajectories are used to obtain estimates of committor values for each configuration $\* x$, resulting in a statistically noisy variation of $q_\mathrm{emp}(\* x)$. Therefore, using the objective function \cref{eq:loss_mse} to train the neural network may lead to overfitting issues and loss in accuracy. 
To alleviate this problem, we introduce a modified form of the objective function where we first evaluate the squared mean error for a batch of samples $\mathcal{C}^{\alpha}$ corresponding to the $\alpha$-th replica: 
\begin{equation}
    \ell_{\mathrm{ME}}(\mathcal{C}^{\alpha}; \thetab) = \frac{1}{2}\Bigg[\frac{1}{|\mathcal{C}^{\alpha}|} \sum_{ ( q_\mathrm{emp}, \* x ) \in \mathcal{C}^{\alpha}}( \hat{q}(\* x; \thetab) - q_\mathrm{emp} ) \Bigg]^2 \,.
\end{equation}
This is then reduced across all replicas, yielding the modified supervised learning objective function
\begin{equation}
    \hat{L}_\mathrm{SL}(\thetab; \{ \mathcal{C}^\alpha \}) = \frac{\lambda_{\mathrm{SL}}}{ M} \sum_{\alpha=1}^M  \ell_{\mathrm{ME}}(\mathcal{C}^{\alpha}; \thetab) \label{eq:loss_sl} \,,
\end{equation}
where $ \lambda_{\mathrm{SL}} $ is the penalty strength. 
\Cref{eq:loss_sl} indicates the neural network is trained on committor errors that are locally-averaged over a single replica. Such an averaging smears out the statistical error in $ q_\mathrm{emp}( \* x) $, alleviates the issue of overfitting, and further helps the neural network generalize to regions outside of the ones covered by sampling. A more detailed discussion, which shows results comparing the standard (\cref{eq:loss_mse}) and modified (\cref{eq:loss_sl}) objective functions for a two-dimensional system can be found in \cref{app:optimizer_dimer}

To incorporate the supervised learning strategy in the BKE--US method, each replica computes $q_\mathrm{emp}(\* x^{\alpha} )$ between the sampling and optimization steps of the algorithm, where $\* x^\alpha $ is the current configuration of replica $ \alpha $. 
The committor evaluation can be initiated at a chosen iteration $ k=k_\mathrm{emp,s}$ until $k= k_\mathrm{emp,e} $, after which no more $ q_\mathrm{emp}(\* x^{\alpha} )$ values are computed.
Since each $q_\mathrm{emp}(\* x^{\alpha} )$ requires the initiation of $H$-many trajectories starting at $\* x_0 = \* x^\alpha$, the committor is evaluated infrequently every $ \tau_\mathrm{emp}$ iterations to reduce the computational cost.
The pseudocode combining supervised learning with the BKE--US method is described in Algorithm~\ref{alg:base} (\cref{fig:baseslmethod}), and is herein referred to as the BKE--US+SL method.

\subsection{Replacing Feedback Loops with the Finite-Temperature String Method} \label{sec:basefts}
For methods employing umbrella sampling, it is important to ensure sufficient overlap in samples obtained from neighboring replicas, since the overlap guarantees accurate computation of reweighting factors $z_\alpha$, and further controls the accuracy in the estimator for the average loss functions, e.g., the average BKE loss function, which sets the reaction rate.
As mentioned before, this may require exhaustive fine-tuning of the algorithm parameters, or long simulations to obtain a larger number of samples. 
On the other hand, the framework of TPT already provides an algorithm called the finite-temperature string (FTS) method \cite{weinan2005finite,vanden2009revisiting}, which can homogeneously sample overlapping regions across the transition tube with few control parameters. 
The FTS method also yields the transition path $\varphib(s)$ without needing to compute the committor function $q(\*x)$. 
Therefore, if we replace the committor-based umbrella sampling with the FTS method, we eliminate the feedback loop between importance sampling and the neural network training in learning $q(\* x)$. Furthermore, it is also possible to obtain a low-variance estimate of the reaction rate due to the overlaps in samples obtained from the FTS method.
In what follows, we review the FTS method in \cref{sec:ftsmethod} and describe new algorithms for solving the BKE in \cref{sec:bkeftsdetail}; see also Ref.~\cite{vanden2009revisiting} for additional details on the FTS method. 
Readers who are familiar with the FTS method may skip \cref{sec:ftsmethod} and read \cref{sec:bkeftsdetail} directly for details on solving the BKE with the FTS method.

\subsubsection{Review of the Finite-Temperature String Method} \label{sec:ftsmethod}
\begin{figure}[t]
  \centering
  \includegraphics[width=0.75\linewidth]{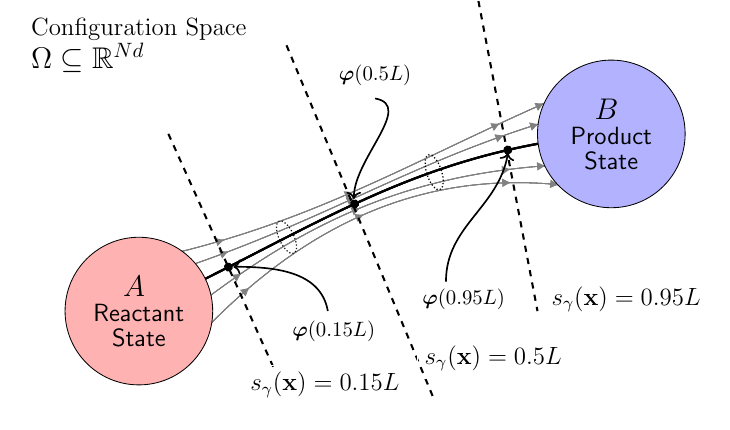}
  \caption{The local approximation of isocommittor surfaces as hyper-planes, which also correspond to the level sets of $s_\gamma(\* x)$. The normal vector of each hyper-plane is the tangent vector $\frac{\diff \varphib(s)}{\diff s }$.}
  \label{fig:tptillus3}
\end{figure}

The FTS method is an algorithm for obtaining a transition path $\varphib(s)$, as defined in \cref{eq:pathdef}, using sampling and optimization techniques. 
It emerges from an approximation of the committor function $q(\* x)$, which is locally built around the transition path $\varphib(s)$. 
This local approximation is achieved by constructing suitable functions $s_\gamma(\* x)$, which represent isocommittor surfaces as hyperplanes centered around $\varphib(s)$. 
If $\varphib(s)$ follows an arc-length parameterization, where $s$ is the arc-length, the approximation for $q(\* x)$ and the formula for $s_\gamma(\* x)$ can be written as
\begin{align}
    q(\*x) &\approx f(s_\gamma(\*x)) \,, \label{eq:qsgamma}
\\
    s_\gamma(\*x) &\equiv \argmin_{s \in [0,L]} \frac{1}{2}|\* x-\varphib(s)|^2 \,, \label{eq:sgammaprojection}
\end{align}
where $L$ is the total arc-length of the path, and $f:[0,L] \to [0,1]$ is an invertible scalar function. 
To see that the function $s_\gamma(\* x)$ approximates isocommittor surfaces as hyper-planes, one may perform the minimization in \cref{eq:sgammaprojection} to obtain the following equation:
\begin{equation}
\frac{\diff \varphib(s)}{\diff s } \cdot (\* x-\varphib(s)) = 0 \,, \label{eq:hyperplaneeq}
\end{equation}
which is a linear equation in $\* x$, indicating the set of all configurations satisfying \cref{eq:hyperplaneeq} for fixed value of $s \in [0,L]$ is a hyperplane; see \cref{fig:tptillus3} for illustration. 
On the other hand, the operation of fixing a configuration $\* x$, and finding $s$ that satisfies \cref{eq:hyperplaneeq} defines a mapping between configurations $\* x \in \Omega$ and the variable $s \in[0,L]$. This mapping is what we denote as $s_\gamma(\* x)$.

Given $s_\gamma(\* x)$ in \cref{eq:sgammaprojection}, the problem of finding $\varphib(s)$ can be posed as an optimization problem. 
To this end, using \cref{eq:qsgamma}, 
\cref{eq:pathdef} can be approximated as an integral over the hyperplane defined by $s_\gamma(\* x)$:
\begin{equation}
    \varphib(s) \approx \frac{\int_{\tilde{P}} \diff S \rho(\* x)f^\prime(s_\gamma(\* x))|\nabla_{\* x} s_\gamma(\* x)| \* x}{\int_{\tilde{P}} \diff S  \rho(\* x) f^\prime(s_\gamma(\* x))|\nabla_{\* x} s_\gamma(\* x)| } \,, \label{eq:pathdef_sgamma}
\end{equation}
where $\tilde{P}$ is a level set of the function $s_\gamma(\* x)$ given by $\tilde{P}= \{ \* x \in \Omega : s_\gamma(\* x)= s \}$.
Since $f^\prime(s_\gamma(\* x)) $ is constant over the level set $\tilde{P}$, \cref{eq:pathdef_sgamma} can be rewritten as 
\begin{equation}
\varphib(s) \approx \frac{\int_{\tilde{P}} \diff S \rho(\* x) |\nabla_{\* x} s_\gamma(\* x)| \* x}{\int_{\tilde{P}} \diff S \rho(\* x) |\nabla_{\* x} s_\gamma(\* x)| } \,. \label{eq:pathdef_sgammasetapprox} 
\end{equation} 
Using the identity \cite{Osher2003}
\begin{equation}
\int_{\tilde{P}} \diff S = \int_\Omega \diff \* x \delta(s_\gamma(\* x)-s) |\nabla_{\* x} s_\gamma(\* x)| \,,
\end{equation}
with $\delta(s_\gamma(\* x)-s)$ as the Dirac delta function, \cref{eq:pathdef_sgammasetapprox} can be rewritten as  
\begin{equation}
\varphib(s) \approx \frac{\int_\Omega \diff \* x \rho(\* x) \delta(s_\gamma(\* x)-s) |\nabla_{\* x} s_\gamma(\* x)|^2 \* x}{\int_\Omega \diff \* x \rho(\* x) \delta(s_\gamma(\* x)-s) |\nabla_{\* x} s_\gamma(\* x)|^2} = \frac{\langle \delta(s_\gamma(\* x)-s) |\nabla_{\* x} s_\gamma(\* x)|^2  \* x \rangle }{\langle \delta(s_\gamma(\* x)-s) |\nabla_{\* x} s_\gamma(\* x)|^2  \rangle} \label{eq:pathdef_deltafunc} \,. 
\end{equation}
Furthermore, assuming the path's curvature to be small, which implies that $|\nabla_{\* x} s_\gamma(\* x)|^2 \approx 1$ (see Appendix~A of Ref.~\cite{vanden2009revisiting} for a proof), \cref{eq:pathdef_deltafunc} can be simplified into a conditional average given by
\begin{equation}
\varphib(s) \approx \frac{\langle \delta(s_\gamma(\* x)-s) \* x \rangle }{\langle \delta(s_\gamma(\* x)-s) \rangle} = \langle \* x \mid s_\gamma(\* x) = s \rangle \label{eq:princcurve} \,.
\end{equation}
Lastly, one may use variational techniques to show that \cref{eq:princcurve} is the result of extremizing the following functional \citep{hastie1989principal,vanden2009revisiting}:
\begin{equation}
C[\varphib] =  \int_0^L \diff s \left\langle  \frac{1}{2} |\varphib(s)-\* x|^2 \delta(s_\gamma(\* x)-s) \right\rangle \label{eq:lsq_princcurve}
\end{equation}
such that 
\begin{equation}
\left|\frac{\diff \varphib(s)}{\diff s} \right|=1  \,. \label{eq:arc_param}
\end{equation}
\Cref{eq:arc_param} is the definition of arc-length parameterization, which sets a constraint on the possible paths that extremize \cref{eq:lsq_princcurve}. 

\Cref{eq:lsq_princcurve,eq:arc_param} form the starting points for developing the FTS method, with several discretization and approximation steps leading to a solvable optimization problem. 
To this end, discretizing $\varphib(s)$ into a set of equidistant nodal points $\{ \varphib^\alpha \}_{\alpha=1}^{M}$, satisfying \cref{eq:arc_param}, i.e., $|\varphib^{\alpha+1}-\varphib^{\alpha}|=|\varphib^{\alpha}-\varphib^{\alpha-1}|, \ \forall \alpha \in \{1,\ldots,M\}$, 
\cref{eq:lsq_princcurve} can be approximated as
\begin{equation}
C(\{\varphib^\alpha\}) =  \sum_{\alpha=1}^M \Delta s \left\langle  \frac{1}{2} |\varphib^\alpha-\* x|^2 \delta(s_\gamma(\* x)-s_\alpha) \right\rangle 
\label{eq:lsq_princcurve_disc1} \,,
\end{equation}
where $s_\alpha = \left(\frac{\alpha-1}{M-1}\right)L$ is the arc-length of the path up to node $\varphib^\alpha$, and $\Delta s$ is the arc-length between any two nodes. 
Furthermore, the Dirac delta function $\delta(s_{\gamma}(\* x)-s_\alpha)$ can be approximated with an indicator function (see Appendix~B of Ref.~\cite{vanden2009revisiting}):
\begin{equation}
h_{R_{\alpha}}(\* x) = \begin{cases}
\frac{1}{\Delta s} &  \* x \in R_\alpha(\{ \varphib^\alpha \}) = \{\* x \in \Omega : |\* x-\varphib^\alpha| < | \* x-\varphib^{\alpha^\prime}| \ \ \forall \alpha^\prime \neq \alpha \}
\\
0 & \text{otherwise} 
\end{cases}
\,, 
\label{eq:indicatorfunc}
\end{equation}
where $R_\alpha$ denotes a Voronoi cell centered at node $\varphib^\alpha$. With these steps,
\cref{eq:lsq_princcurve_disc1} can then be expressed as a least-squares function: 
\begin{equation}
C(\{\varphib^\alpha\}) =  \sum_{\alpha=1}^M \Delta s \left\langle \frac{1}{2} |\varphib^\alpha-\* x|^2 h_{R_\alpha}(\* x) \right\rangle = \sum_{\alpha=1}^M \left\langle \frac{1}{2} |\varphib^\alpha-\* x|^2 \right\rangle_{R_\alpha(\{ \varphib^\alpha \})} \label{eq:empriskfunc_fts} 
\end{equation}
where $\langle \ldots \rangle_{R_\alpha(\{\varphib^\alpha\})}$ is an ensemble average constrained inside a Voronoi cell. 

\begin{figure}[t]
\centering
\begin{minipage}{0.45\linewidth}
\includegraphics[width=0.95\linewidth]{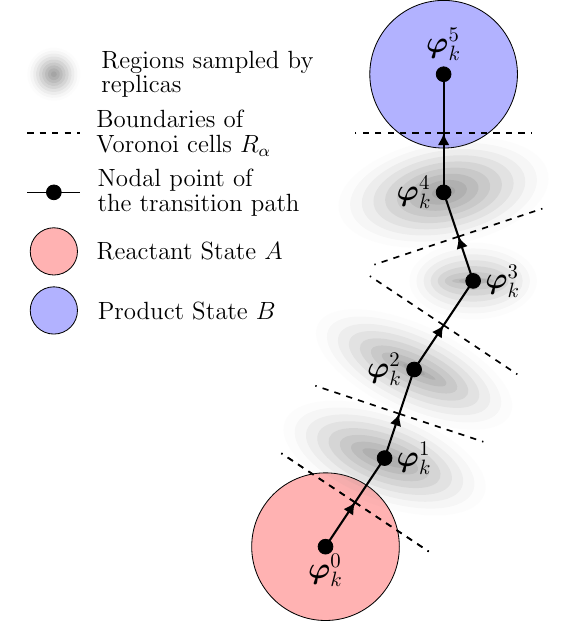}
\end{minipage}
\begin{minipage}{.525\linewidth}
\begin{algorithm}[H]
\caption{The BKE--FTS(ME) Method \label{alg:basefts}}
\KwData{Initial conditions $\thetab_0$, $\{ \varphib^\alpha_0 \}$. Reactant and product batches $\mathcal{A}$ and  $\mathcal{B}$. Hyperparameters for optimizers $\etab$. The FTS Method step size $\Delta \tau$ and penalty strength $\lambda_\mathrm{S}$. Penalty strengths $\lambda_\mathrm{A}$ and $\lambda_\mathrm{B}$.}
\For {$k=0,\dots,K$}{
    \For {$\alpha=1,\dots,M$ in parallel}{
        \For {$m=1,\dots,|\mathcal{R}_k^\alpha|$}{
            Sample $\* x^{\alpha}_{m}$ with MD/MC simulation constrained in the Voronoi cell $R_\alpha(\{\varphib^\alpha_k\})$, e.g., \cref{eq:ftssamplinglangevin,eq:ftsrejection}.
        
            Store $\* x^{\alpha}_{m}$ in batch $\mathcal{R}_k^\alpha$.
            
            }
            
        }
    $\varphib^{k+1}_\alpha \assign$ \cref{eq:sgd_fts,eq:projection_fts}.
    
    Sample mini-batch $\mathcal{A}_k \subset \mathcal{A}$ and $\mathcal{B}_k \subset \mathcal{B}$.
    
    Compute $z_\alpha$ by solving the master equation \cref{eq:ftsbalanceeq}. 
    
    Compute $\nabla_{\thetab} \hat{L}(\thetab_k; \{ (\mathcal{R}^\alpha_k, z_\alpha) \}, \mathcal{A}_k, \mathcal{B}_k)$ with \cref{eq:basefts_grad}.
    
    Update $\thetab_k \to \thetab_{k+1}$ with optimizer.

}
\end{algorithm}
\end{minipage}
\caption{(Left) An illustration of the FTS method, where each replica samples configurations inside a Voronoi cell. (Right) Pseudo-code for the BKE--FTS(ME) method. Note that the path is updated concurrently with the neural network at the $k$-th iteration.}
\label{fig:baseftsmethod}
\end{figure}

The ensemble averages in \cref{eq:empriskfunc_fts} can be estimated as averages over samples obtained from molecular simulations, which are constrained to be inside the Voronoi cells and are initiated with the configuration of the corresponding node. 
As illustrated in \cref{fig:baseftsmethod}(left), this step involves introducing $M$-many replicas of the system to sample configurations within each of the $M$-many Voronoi cells, where
each replica can evolve according to discrete overdamped Langevin dynamics with a rejection rule: 
\begin{align}
\* x^{\alpha}_{\star} &=    \*x^{\alpha}_{t}-\gamma^{-1}\nabla_{\* x} V(\* x^{\alpha}_{t}) \Delta t + \sqrt{2 \Delta t k_\mathrm{B} T\gamma^{-1}} \* w^{\alpha}_{t} \,, \label{eq:ftssamplinglangevin}
\\
\* x_{t+1}^\alpha &= \begin{cases}
    \*x^{\alpha}_{\star} & \text{if} \ \* x^{\alpha}_{\star} \in R_\alpha 
    \\
    \* x^{\alpha}_{t} & \text{otherwise} 
    \end{cases}
    \label{eq:ftsrejection} \,,
\end{align}
where $\* w_t^\alpha $ is a random variable with zero-mean and unit variance. Note that \cref{eq:ftssamplinglangevin} can be replaced with an MC step.
Introducing $\mathcal{R}^\alpha$ as the batch of samples obtained from the $\alpha$-th replica, \cref{eq:empriskfunc_fts} can be estimated as 
\begin{gather}
\hat{C}(\{\varphib^\alpha\}; \{\mathcal{R}^\alpha\}) =  \sum_{\alpha=1}^M \frac{1}{|\mathcal{R}^\alpha|} \sum_{\* x \in \mathcal{R}^\alpha}  \frac{1}{2} |\varphib^\alpha-\* x|^2 \,. 
\label{eq:empiricalriskfunc_fts} 
\end{gather}
To avoid large displacements in neighboring nodal points, a penalty function is added to \cref{eq:empiricalriskfunc_fts}, which yields 
\begin{gather}
\hat{C}(\{\varphib^\alpha\}; \{\mathcal{R}^\alpha\}) =  \sum_{\alpha=1}^M \frac{1}{|\mathcal{R}^\alpha|} \sum_{\* x \in \mathcal{R}^\alpha}  \frac{1}{2} |\varphib^\alpha-\* x|^2  + \frac{\lambda_\mathrm{S}}{2} \sum_{\alpha=1}^{M-1} |\varphib^{\alpha+1}-\varphib^{\alpha}|^2  \label{eq:empiricalriskfunc_fts_soft}  \,,
\\
\text{s.t.} \quad |\varphib^{\alpha+1}-\varphib^{\alpha}|=|\varphib^{\alpha}-\varphib^{\alpha-1}| \,, \label{eq:ftsconstraint}
\end{gather}
where $\lambda_\mathrm{S}$ is the penalty strength.

The FTS method minimizes \cref{eq:empiricalriskfunc_fts_soft} using a closed feedback loop between the replica dynamics, e.g., \cref{eq:ftssamplinglangevin,eq:ftsrejection}, and a modified gradient-descent step. 
At the $k$-th iteration of the loop, replicas generate a collection of batches $\{ \mathcal{R}^\alpha_k \}_{\alpha=1}^M$, where the batch $\mathcal{R}^\alpha_k$ consists of a short MD/MC trajectory run from the $\alpha$-th replica.
This data is then used in a two-part gradient descent update, where the first part corresponds to the following update:
\begin{equation}
\varphib_{\star}^\alpha = \varphib^\alpha_k-\Delta \tau \nabla_{\varphib^\alpha} \hat{C}(\{\varphib^\alpha_k\}; \{\mathcal{R}^\alpha_k\}) \,,
\label{eq:sgd_fts}
\end{equation}
with $\Delta \tau$ the step size. 
Note that one can replace \cref{eq:sgd_fts} with an implicit update for increased stability or a momentum-variant, such as the Heavy-Ball \cite{polyak1964some} and the Nesterov method \cite{nesterov1983method}, for accelerated convergence.
The second part enforces the constraint \cref{eq:ftsconstraint} with a reparameterization of the path using linear interpolation:
\begin{equation}
\varphib^{\alpha}_{k+1} = 
    \varphib^{a(\alpha)-1}_{\star}+\left(L_M\frac{\alpha-1}{M-1}-L_{a(\alpha)-1}\right) \frac{\varphib^{a(\alpha)}_{\star}-\varphib^{a(\alpha)-1}_{\star}}{\left|\varphib^{a(\alpha)}_{\star}-\varphib^{a(\alpha)-1}_{\star}\right|} \,, 
    \label{eq:projection_fts}
\end{equation}
where $L_\alpha = \sum_{\alpha^\prime=2}^\alpha |\varphib^{\alpha^\prime}_\star-\varphib^{\alpha^\prime-1}_\star|$ is the length of the path up to node $\varphib^\alpha_{\star}$, and $a(\alpha) \in \{1,\ldots,M\}$ is an index such that $L_{a(\alpha)-1} < \left(\frac{\alpha-1}{M-1}\right) L_M < L_{a(\alpha)}$. 
This process is repeated until convergence is achieved, yielding the transition path $\varphib(s)$.

\subsubsection{Solving the BKE with the Finite-Temperature String Method} \label{sec:bkeftsdetail} 
With the FTS method described in \cref{sec:ftsmethod}, we now proceed to construct new algorithms for minimizing the loss in \cref{eq:erm_bke}. 
The key idea  behind all subsequent new algorithms is to replace the committor-based umbrella sampling in the BKE--US method with the FTS method. 
This allows the replicas to generate samples that homogeneously cover the transition tube with little fine-tuning, and enables accurate low-variance estimation of the average loss functions and their gradients. 
As mentioned before, since the average BKE loss function is proportional to the chemical reaction rate, the FTS method also enables accurate estimation of reaction rates. 

\textbf{The FTS method with master equation:} The first algorithm that we construct involves updating the transition path, represented as a set of nodal points, simultaneously with the neural network training. In particular, the replicas from the FTS method generate batches of sampled configurations $\{ \mathcal{R}^\alpha_k \}_{\alpha=1}^M$ to update the current path $\{\varphib^\alpha_k \}$ via \crefrange{eq:sgd_fts}{eq:projection_fts}, as well as the neural network parameters $\thetab_k$ by computing the gradient of the loss in \cref{eq:erm_bke}. 
Note that, in this algorithm, there is no feedback loop between the neural network and updates to the path.
In this case, the loss gradient $\nabla_{\thetab} \hat{L}$ can be calculated using modified versions of \crefrange{eq:basemethod_grad}{eq:zalphadef}, where the bias potentials $W_\alpha$ are replaced with hard-wall potentials constraining each replica to its Voronoi cell, i.e., 
\begin{equation}
W_\alpha(\* x; \{\varphib^\alpha\}) = \begin{cases}
0 &  \* x \in R_\alpha
\\
\infty & \text{otherwise} 
\end{cases}
\,.
\label{eq:hardwall-fts}
\end{equation}
This yields 
\begin{align}
\nabla_{\thetab} \hat{L}\left(\thetab_k; \{ (\mathcal{R}^\alpha_k, z_\alpha) \}, \mathcal{A}_k, \mathcal{B}_k\right) =&  \sum\limits_{\alpha=1}^M \frac{z_{\alpha} }{|\mathcal{R}^\alpha_k|} \sum_{\* x \in \mathcal{R}^\alpha_k} \nabla_{\thetab} \ell(\* x; \thetab_{k}) +\frac{\lambda_\mathrm{A}}{|\mathcal{A}_k|} \sum\limits_{\* x \in \mathcal{A}_k} \nabla_{\thetab} \ell_\mathrm{A}(\* x; \thetab_{k}) \nonumber
\\
                                                                                                                      &+\frac{\lambda_\mathrm{B}}{|\mathcal{B}_k|} \sum\limits_{\* x \in \mathcal{B}_k}\nabla_{\thetab} \ell_\mathrm{B}(\* x; \thetab_{k}) \,, \label{eq:basefts_grad}
\end{align}
where the reweighting factors $z_\alpha$ are
\begin{equation}
z_\alpha = \frac{ \int_{R_\alpha} \diff \* x \ e^{-\beta V(\* x)}}{\int_{\bigcup_{\alpha=1}^M R_\alpha}  \diff \* x \ e^{-\beta V(\* x)}} = \frac{ \int_{R_\alpha} \diff \* x \ e^{-\beta V(\* x)}}{\int_{\Omega}  \diff \* x \ e^{-\beta V(\* x)}} = \int_{R_\alpha} \diff \* x \ \rho(\* x) \,.  \label{eq:zalphafts}
\end{equation}

\Cref{eq:zalphafts} indicates $z_\alpha$ is the equilibrium probability of finding $\* x$ to be in a Voronoi cell $R_\alpha$. 
This set of equilibrium probabilities can be computed as a solution to a steady-state master equation, whose form is found by identifying the instantaneous rates (or fluxes) between neighboring Voronoi cells \cite{vanden2009revisiting}. 
To this end, let $N_{\alpha \alpha^\prime}$ be the number of times that the $\alpha$-th replica attempts to exit its Voronoi cell $R_{\alpha}$ and enter a neighboring Voronoi cell $R_{\alpha^\prime}$, e.g., the number of times that $\* x_\star^\alpha \in R_{\alpha^\prime}$ for the replica dynamics given by \crefrange{eq:ftssamplinglangevin}{eq:ftsrejection}. 
Let $k_{\alpha \alpha^\prime}$ be the rate at which the system transitions between $ R_\alpha $ to $ R_{\alpha^\prime} $. 
Denoting $N_\mathrm{steps}^\alpha$ as the total simulation length of the $\alpha$-th replica, the previous rate can be evaluated as $k_{\alpha \alpha^\prime} \approx N_{\alpha \alpha^\prime}/N_\mathrm{steps}^\alpha$. 
The steady-state master equation is then given by a balance between the total rate of leaving and entering the Voronoi cell $R_\alpha$:
\begin{equation}
    \sum_{\alpha^\prime=1}^M z_{\alpha^\prime} k_{\alpha^\prime \alpha} = \sum_{\alpha^\prime=1}^M z_{\alpha} k_{\alpha \alpha^\prime }, \quad \forall \alpha \in \{1,\ldots,M\} \,, \label{eq:ftsbalanceeq} 
\end{equation}
which can be solved to obtain $z_\alpha$; see \cref{app:freenergymethods} for more details, 
and also Section~III of Ref.~\cite{vanden2009markovian} for a more detailed discussion of \cref{eq:ftsbalanceeq}. 
\Cref{eq:basefts_grad,eq:ftsbalanceeq} constitute the new algorithm, and will herein be referred to as the BKE--FTS(ME) method, whose 
pseudocode is described in Algorithm~\ref{alg:basefts} (\cref{fig:baseftsmethod}, right).

\begin{figure}[t]
\centering
\begin{minipage}{\linewidth}
\begin{algorithm}[H]
\caption{The BKE--FTS(US) Method \label{alg:baseftsus}}
\KwData{Initial conditions $\thetab_0$. Nodal points of the transition path $\{ \varphib^\alpha \}$ obtained from the FTS method. Reactant and product batches $\mathcal{A}$ and  $\mathcal{B}$. Hyperparameters for optimizers $\etab$. Penalty strengths $\lambda_\mathrm{A}$ and $\lambda_\mathrm{B}$.}
\For {$k=0,\dots,K$}{
    \For {$\alpha=1,\dots,M$ in parallel}{
        \For {$m=1,\dots,|\mathcal{M}_k^\alpha|$}{
            Sample $\* x^{\alpha}_{m} \sim \rho_\alpha(\* x; \{\varphib^\alpha \})$ with MD/MC simulation, e.g., \cref{eq:ussamplinglangevin} and \cref{eq:umbrella-fts}.
        
            Store $\* x^{\alpha}_{m}$ in batch $\mathcal{M}_k^\alpha$.
            
            }
        }
    
    Sample mini-batch $\mathcal{A}_k \subset \mathcal{A}$ and $\mathcal{B}_k \subset \mathcal{B}$.
    
   Compute $z_\alpha$ with a free-energy method, e.g., FEP \cref{eq:expestimator}.
    
    Compute $\nabla_{\thetab} \hat{L}(\thetab_k; \{ (\mathcal{M}^\alpha_k, z_\alpha) \}, \mathcal{A}_k, \mathcal{B}_k)$ with \cref{eq:basemethod_grad}.
    
    Update $\thetab_k \to \thetab_{k+1}$ with optimizer.
    
}
\end{algorithm}
\end{minipage}
\caption{Pseudo-code for the BKE--FTS(US) method.}
\label{fig:baseftsusmethod}
\end{figure}

\begin{figure}[t]
\centering
\begin{minipage}{\linewidth}
\begin{algorithm}[H]
\caption{The BKE--FTS(ME)+SL Method \label{alg:baseftssl}}
\KwData{Initial conditions $\thetab_0$, $\{ \varphib^\alpha_0 \}$. Reactant and product batches $\mathcal{A}$ and  $\mathcal{B}$. Hyperparameters for optimizers $\etab$. The FTS Method step size $\Delta \tau$ and penalty strength $\lambda_\mathrm{S}$. Penalty strengths $\lambda_\mathrm{A}$,  $\lambda_\mathrm{B}$, and $\lambda_\mathrm{SL}$. Starting and ending iteration index, $k_\mathrm{emp,s}$ and $k_\mathrm{emp,e}$, and sampling period $\tau_\mathrm{emp}$ for supervised learning.}
{
\For {$k=0,\dots,K$}{
    \For {$\alpha=1,\dots,M$ in parallel}{
        \For {$m=1,\dots,|\mathcal{R}_k^\alpha|$}{
            Sample $\* x^{\alpha}_{m}$ with MD/MC simulation constrained in the Voronoi cell $R_\alpha(\{\varphib^\alpha_k\})$, e.g., \crefrange{eq:ftssamplinglangevin}{eq:ftsrejection}.
        
            Store $\* x^{\alpha}_{m}$ in batch $\mathcal{R}_k^\alpha$.
            
            }
            \uIf{$ k \geq k_\mathrm{emp,s} $ and $ k < k_\mathrm{emp,e} $ and $ k \Mod{ \tau_\mathrm{emp} } = 0 $}{
                Evaluate $ q_\mathrm{emp} $ at $\* x^{\alpha} \in \mathcal{R}_k^\alpha $ with \cref{eq:empiricalcommittor}.
                
                Store $ ( q_\mathrm{emp} , \* x^{\alpha} ) $ in batch $\mathcal{C}^\alpha$.
            }
            
        }
    
        $\varphib^{k+1}_\alpha \assign$ \crefrange{eq:sgd_fts}{eq:projection_fts}.

        Sample mini-batch $\mathcal{A}_k \subset \mathcal{A}$, $\mathcal{B}_k \subset \mathcal{B}$, and $ \mathcal{C}_k^{\alpha} \subset \mathcal{C}^\alpha $.
    
    Compute $z_\alpha$ by solving the master equation \cref{eq:ftsbalanceeq}. 
    
    Compute $\nabla_{\thetab} \hat{L}(\thetab_k; \{ (\mathcal{R}^\alpha_k, z_\alpha) \}, \mathcal{A}_k, \mathcal{B}_k)+\nabla_{\thetab}  \hat{L}_\mathrm{SL}(\thetab_k; \{\mathcal{C}_k^\alpha \})$ with \cref{eq:basefts_grad,eq:loss_sl}.
    
    Update $\thetab_k \to \thetab_{k+1}$ with optimizer.
    
}
}
\end{algorithm}
\end{minipage}
\caption{Pseudo-code for the BKE--FTS(ME)+SL method.}
\label{fig:ftsslmethod}
\end{figure}
\begin{figure}[t]
\vspace{24pt}
\centering
\begin{minipage}{\linewidth}
\begin{algorithm}[H]
\caption{The BKE--FTS(US)+SL Method \label{alg:baseftsussl}}
\KwData{Initial conditions $\thetab_0$. Nodal points of the transition path $\{ \varphib^\alpha \}$ obtained from the FTS method. Reactant and product batches $\mathcal{A}$ and  $\mathcal{B}$. Hyperparameters for optimizers $\etab$. Penalty strengths $\lambda_\mathrm{A}$,  $\lambda_\mathrm{B}$, and $\lambda_\mathrm{SL}$. Starting and ending iteration index, $k_\mathrm{emp,s}$ and $k_\mathrm{emp,e}$, and sampling period $\tau_\mathrm{emp}$ for supervised learning.}
{

\For {$k=0,\dots,K$}{
    \For {$\alpha=1,\dots,M$ in parallel}{
        \For {$m=1,\dots,|\mathcal{M}_k^\alpha|$}{
            Sample $\* x^{\alpha}_{m} \sim \rho_\alpha(\* x; \{\varphib^\alpha\})$ with MD/MC simulation, e.g., \cref{eq:ussamplinglangevin,eq:umbrella-fts}.
        
            Store $\* x^{\alpha}_{m}$ in batch $\mathcal{M}_k^\alpha$.
            
            }
            \uIf{$ k \geq k_\mathrm{emp,s} $ and $ k < k_\mathrm{emp,e} $ and $ k \Mod{ \tau_\mathrm{emp} } = 0 $}{
                Evaluate $ q_\mathrm{emp} $ at $\* x^{\alpha} \in \mathcal{M}_k^\alpha $ with \cref{eq:empiricalcommittor}.
                
                Store $ ( q_\mathrm{emp} , \* x^{\alpha} ) $ in batch $\mathcal{C}^\alpha$.
            }
            
        }
    
        Sample mini-batch $\mathcal{A}_k \subset \mathcal{A}$, $\mathcal{B}_k \subset \mathcal{B}$, and $\mathcal{C}_k^{\alpha} \subset \mathcal{C}^\alpha $.
    
    Compute $z_\alpha$ with a free-energy method, e.g., FEP \cref{eq:expestimator}.
    
    Compute $\nabla_{\thetab} \hat{L}(\thetab_k; \{ (\mathcal{M}^\alpha_k, z_\alpha) \}, \mathcal{A}_k, \mathcal{B}_k)+\nabla_{\thetab}  \hat{L}_\mathrm{SL}(\thetab_k; \{\mathcal{C}_k^\alpha \})$ with \cref{eq:basemethod_grad,eq:loss_sl}.
    
    Update $\thetab_k \to \thetab_{k+1}$ with optimizer.
    
}
}
\end{algorithm}
\end{minipage}

\caption{Pseudo-code for the BKE--FTS(US)+SL method.}
\label{fig:ftsusslmethod}
\end{figure}

\begin{figure}[t]
    \centering
    \includegraphics[width=\linewidth]{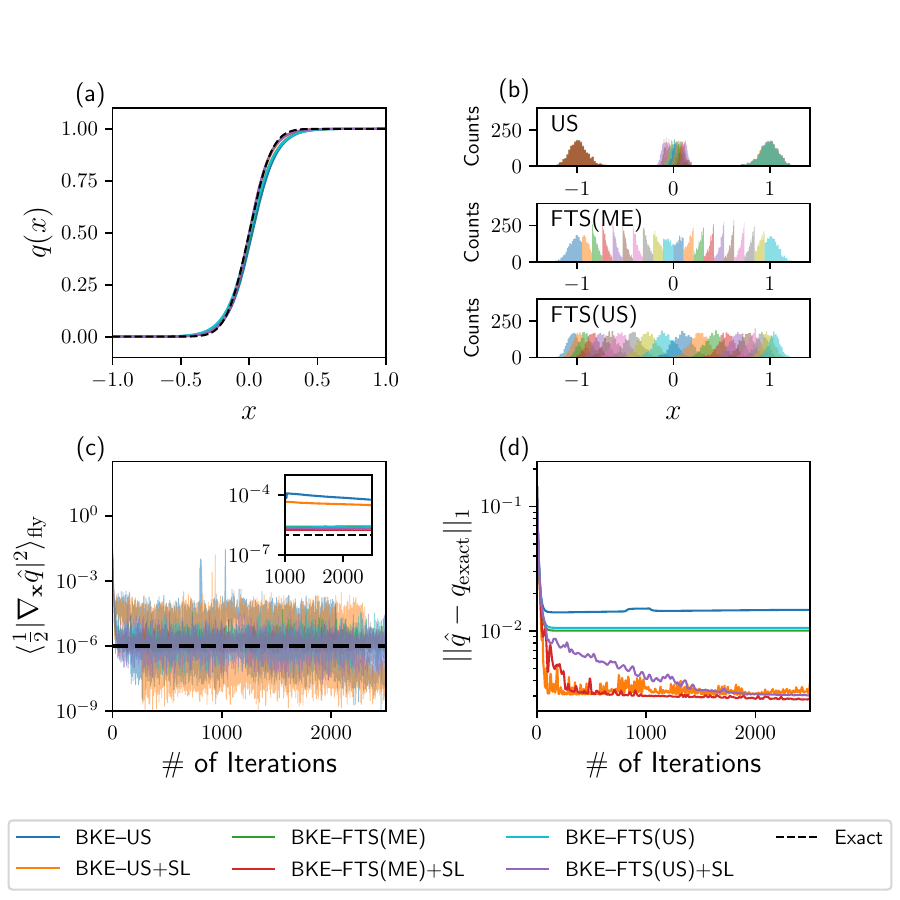}
    \caption{(a) Committor function obtained from all methods compared with the exact solution. (b) Histograms of samples obtained from the BKE--US method (top), the BKE--FTS(ME) method (middle), and the BKE--FTS(US) method (bottom). 
    (c) On-the-fly estimates of the average BKE loss obtained at every iteration and computed using a batch size of 16, with an inset plot showing their cumulative averages over the last 1500 iterations. (d) The $L_1$-norm error as a function of iterations. }
    \label{fig:1dresults}
\end{figure}

\textbf{The FTS method with umbrella sampling:} As mentioned before, given a sufficient number of nodes, the BKE--FTS(ME) method guarantees homogeneous sampling across the transition path (see also \cref{fig:1dresults}(b)), which better ensures low-variance estimation from reweighting. Accuracy can also be improved by running longer simulations, i.e., larger $N_\mathrm{steps}^\alpha$, since they lead to more accurate estimates of the rates $k_{\alpha \alpha^\prime}$,
thereby reducing the error in the estimated reweighting factor $z_\alpha$. 
Despite this, the error in $z_\alpha$ is difficult to study as it involves the error propagation of $k_{\alpha \alpha^\prime}$, which forms a random matrix in the master equation. 
On the other hand, $z_\alpha$ computed from umbrella sampling is amenable to error analysis \cite{thiede2016eigenvector,shirts2005comparison}, which makes it feasible to determine the error in the estimates computed from reweighting as a function of batch size. 
This motivates us to construct a modification to the BKE--FTS(ME) method where the computation of $z_\alpha$ is based on umbrella sampling and FEP (\cref{eq:expestimator}).  
The modified algorithm consists of running the FTS method before the neural network training to obtain the transition path $\{ \varphib^\alpha \}_{\alpha=1}^M$, which is then used as a basis for umbrella sampling across the transition tube to subsequently train the neural network. 

The path-based umbrella sampling requires new bias potentials that can lead to better overlaps between adjacent replicas, as well as sufficient exploration of regions transverse to the path. The latter is necessary to ensure the neural network representing the committor function is also accurate in regions away from the transition path. To this end, we construct new bias potentials such that different bias strengths can be specified in directions parallel and transverse to the path. Let $\* t^\alpha$ be the unit tangent vector at node $\varphib^\alpha$, evaluated using finite differences. We then form the projection matrices $\* P_\alpha^\parallel = \* t^\alpha \otimes \* t^\alpha$ and $\* P_\alpha^\bot = \* I-\* t^\alpha \otimes \* t^\alpha$ to decompose a vector into a component that is parallel and transverse to $\* t^\alpha$, respectively. The bias potential for the $\alpha$-th replica can be written as 
\begin{equation}
W_\alpha(\* x; \{ \varphib^\alpha \}) = \frac{1}{2}\kappa_\alpha^\parallel (\* x-\varphib^\alpha)\* P_\alpha^\parallel (\* x-\varphib^\alpha)+\frac{1}{2}\kappa_\alpha^\bot (\* x-\varphib^\alpha)\* P_\alpha^\bot(\* x-\varphib^\alpha)\,,
\label{eq:umbrella-fts}
\end{equation}
where $\kappa_\alpha^\parallel$ and $\kappa_\alpha^\bot$ are the bias strengths for the parallel and transverse direction, respectively. To promote exploration of regions transverse to the path, the bias strengths are set such that $\kappa_\alpha^\bot< \kappa_\alpha^\parallel$. For sufficiently strong bias, this results in every replica exploring an oblate ellipsoidal region, where the center of the ellipsoid is located at node $\varphib^\alpha$, and its axis of rotation is parallel to the tangent vector $\* t^\alpha$. Note that a similar bias potential has also been used in Ref.~\cite{zinovjev2017} but defined with respect to a low-dimensional collective-variable space.

The loss gradient  $\nabla_{\thetab} \hat{L}$ needed for this algorithm can be computed with  \cref{eq:basemethod_grad} and \cref{eq:expestimator} from the BKE--US method, using samples obtained from biased MD/MC simulations. As in the BKE--FTS(ME) method, there exists no feedback loop between the neural network and umbrella sampling because the bias potentials are based on the transition path, which remains static during training. This modification to the BKE--FTS(ME) method shall be referred to as the BKE--FTS(US) method, whose pseudocode is described in Algorithm~\ref{alg:baseftsus} (\cref{fig:baseftsusmethod}). The algorithm shares similar advantages as the BKE--FTS(ME) method, since homogeneous sampling across the transition tube and overlap in configuration space is readily achieved for large enough bias strengths. Unlike the master-equation approach, the bias and variance in the reweighting factors $z_\alpha$ estimated from FEP are amenable to error analysis \citep{shirts2005comparison}. As shown later in \cref{sec:lognorm}, we provide an error analysis of the estimated average loss function, and a procedure where the bias in the estimator can be removed, thereby enabling accurate estimation of reaction rates with smaller batch sizes. 

\textbf{The FTS method with supervised learning:} Both the BKE--FTS(ME) and BKE--FTS(US) methods can be combined with the supervised learning methodology developed in \cref{sec:basesl} to further improve the accuracy of the committor function. 
Since the samples obtained by either method homogeneously cover the transition tube, they provide access to configurations that can be used for computing empirical committor function $q_\mathrm{emp}(\* x)$ necessary for supervised learning. The empirical committor function $q_\mathrm{emp}(\* x)$ may be evaluated by the replicas between the sampling and optimization step of the algorithms. 
Similar to the procedure described in \cref{sec:basesl}, it can be evaluated at a rate $ \tau_\mathrm{emp} $ between a starting iteration $ k_\mathrm{emp,s} $ and an ending iteration $ k_\mathrm{emp,e} $. 
Given these estimates, the supervised-learning loss in \cref{eq:loss_sl} can be used to compute the compound loss gradient to update the neural network. We shall call these composite algorithms as the BKE--FTS(ME)+SL and BKE--FTS(US)+SL method,  whose pseudo-codes are described in Algorithm~\ref{alg:baseftssl} (\cref{fig:ftsslmethod}) and Algorithm~\ref{alg:baseftsussl} (\cref{fig:ftsusslmethod}), respectively.

\textbf{Limitations of the FTS Method:} The proposed methods for solving the BKE with the FTS method inherit the limitations of the FTS method itself. For instance, the application of the FTS method to molecular systems may fail since the distance metrics defining the Voronoi cells are not invariant with respect to rigid-body transformations. As a result, replicas can escape from their respective Voronoi cells without any structural change via rotations and/or translations alone. To resolve this issue, the FTS method is typically applied in the space of collective variables (CVs), which are invariant under translation and rotation by construction. While a solution independent of CVs remains an open problem, the work in Ref.~\cite{vanden2009revisiting} proposes a sufficiently general CV, denoted as $\Theta$, if the system configuration $\* x$ can be divided into a sub-system configuration $\* x_\mathrm{S}$ that undergoes the structural change and solvent degrees of freedom $\* x_\mathrm{E}$ that make up the surrounding environment. This CV takes $\* x_\mathrm{S}$ and a string nodal point $\varphib^\alpha$ as input, and it can be written as 
\begin{align}
    \Theta ( \* x_\mathrm{S} ; \* R^*, \* b^*) &= \* R^* \left( \* x_\mathrm{S} - \* b^* \right) \,, \label{eq:cv}
    \\
    (\* R^*, \* b^*) &= \argmin_{(\* R, \* b)} | \* R \left( \* x_\mathrm{S} - \* b \right) - \varphib^\alpha | \,,  \label{eq:opt_cv}
\end{align}
where $ \* R^* $ and $\* b^*$ are a rotation matrix and translation vector, respectively, that form a rigid body transformation of the sub-system.
By minimizing the distance metric in \cref{eq:opt_cv}, the chosen rigid transformation has the effect of matching the center-of-mass and orientation axis of $\* x_\mathrm{S}$ to that of  $\varphib^\alpha$. This results in a CV that not only retains some of the original molecular degrees of freedom, but also removes the degeneracy due to translations and rotations. The transformation defined by \cref{eq:opt_cv} can also be done at a relatively low computational cost by translating the sub-system to match its center of mass with the center of mass of $\varphib^\alpha$ and subsequently rotating the sub-system via the Kabsch algorithm \cite{kabsch1976solution}. Other CVs are also possible and may be needed when dealing with rare-event problems  where the system cannot be subdivided, e.g., nucleation and self-assembly. 

Despite the generality of \cref{eq:cv}, it may not be sufficient at high densities where the solvent molecules/particles move in a highly correlated fashion during the transition, i.e., solvent reorganization.
In this situation, the BKE--FTS methods can still use the FTS method with the CV as given in \cref{eq:cv} to train neural networks that are implicitly aware of the solvent reorganization, since each replica samples the solvent configurations that participate in the transition. Such a strategy of utilizing the FTS method with the CV in \cref{eq:cv} is used in \cref{sec:dimer_solvent} to compute committor functions and reaction rates in a solvated dimer system with relatively high accuracy.

The FTS method is also ill-suited for problems involving multiple reaction pathways. 
This problem can possibly be addressed by evolving multiple independent strings that are repulsive with respect to each other, as is done in an extension of the string method in the CV space termed the climbing multistring method \cite{shrivastav2019string}, but it remains to be extended to the FTS method.
Other methods more amendable to studying processes with multiple reaction pathways, such as Markov State Models \cite{schutte1999msm,schutte2011markov,bowman2013introduction}, could also be considered in future work.

\section{Computational Studies in Low-Dimensional Systems} \label{sec:results}
In this section, we test Algorithms~\ref{alg:base}--\ref{alg:baseftsussl} to two model systems consisting of a single particle diffusing in non-convex potential energies in one dimension (1D) and two dimensions (2D), respectively. 
Reference solutions can be obtained in 1D and 2D via analytical method and the finite element method (FEM), respectively, which will be used to ascertain the relative accuracy of the algorithms. 
Before we introduce these two systems, we elaborate on the choice of the neural network, optimizer, and initial conditions. 
For both systems, we use a single-hidden layer neural network with ReLU activation functions and a sigmoidal output layer \cite{hardtrecht}:
\begin{equation}
\hat{q}(\* x;\thetab=\{\* W_1, \* w_2, \* b\}) = \sigma \left(\* w_2 \cdot \mathrm{ReLU}(\* W_1 \* x + \* b_1) \right) \,, \label{eq:simplenn}
\end{equation}
where $\mathrm{ReLU}(s) = \max(0,s)$, $\sigma(s) = \frac{1}{1+e^{-s}}$, $\* W_1$ is an $m$-by-$d$ matrix of weights of the hidden layer, $\* w_2$ and $\* b_1$ are $m$-dimensional vectors of weights of the output layer and biases of the hidden layer, respectively, and the number of neurons is $m=200$. 
The chosen optimizer is the Heavy-Ball method \cite{polyak1964some} and Adam \cite{kingma2014adam} for the 1D and 2D system, respectively; see \cref{app:optimizer} for a brief review of each optimizer and associated hyperparameters for each study. 

The neural network parameters are initialized randomly and subsequently updated by minimizing the following mean-squared error function
\begin{equation}
    I(\thetab; \{ \* x_0^\alpha \} ) = \frac{1}{M} \sum_{\alpha =1}^M \left(\hat{q}(\* x_0^\alpha; \thetab)-\frac{\alpha-1}{M-1} \right)^2 \,, \label{eq:choice-initial-minimize}
\end{equation}
where a gradient descent algorithm is used with a stepsize of $ 0.001 $ until $ I (\thetab) \leq 10^{-3} $. Here, $\* x_0^\alpha$ is the initial configuration of the $\alpha$-th replica, and  
is chosen to be the linear interpolation between a known energy-minimizing configuration at the reactant state $\* x_{0}^\mathrm{A}$ and product state $\* x_{0}^\mathrm{B} $:
\begin{equation}
    \* x_0^\alpha = \left( 1 - \frac{\alpha-1}{M-1}\right) \* x_{0}^{\mathrm{A}} + \left(\frac{\alpha-1}{M-1}\right)\* x_{0}^{\mathrm{B}}\,.
\end{equation}
For the BKE--FTS(ME) and BKE--FTS(ME)+SL methods, the initial nodal points of the path are chosen as $\varphib_0^\alpha = \* x_0^\alpha$. For the BKE--FTS(US) and the BKE--FTS(US)+SL method, since the FTS method is run before the neural network training, $\* x_0^\alpha$ is set to the nodal point $\varphib^\alpha$ of the converged path.
The choice in \cref{eq:choice-initial-minimize} ensures an initial guess of $\thetab$ that results in a monotonic increase of the committor function from the reactant to the product states. It also provides an initial value of the committor function that is compatible with the target value of the committor-based umbrella sampling, avoiding large force evaluations for MD simulations. 
Additional details pertaining to individual studies such as sampling schemes generating mini-batches for optimization, choices of penalty strengths, and parameters controlling the FTS method can be found in the \cref{app:optimizer}.

The accuracy of the algorithms is measured using both an $L_{1}$ norm measuring error in $ \hat{q}( \* x ; \thetab ) $, and the ensemble average of the BKE loss function given by \cref{eq:bkeloss}.
The latter is proportional to the reaction rate in \cref{eq:rxnrate}.
The $ L_{1} $-norm error is defined over the region spanned by the transition tube, $ T_{\Lambda} = \{ \* x \in \Omega : |\* J(\* x)| \geq \Lambda \}$ where $\Lambda$ is a cut-off value, and normalized by the volume of the region. This yields
\begin{equation}
    ||\hat{q}-q||_{1}= \frac{1}{\int_{T_{\Lambda}} \diff \* x }\int_{T_{\Lambda}} \diff \* x |\hat{q}(\* x; \thetab)-q(\* x)| \,. \label{eq:l1norm_avg}
\end{equation}
In all algorithms, an on-the-fly estimate of the ensemble average of \cref{eq:bkeloss} is computed at the $k$-th iteration with the following formula: 
\begin{gather}
\left\langle \frac{1}{2} |\nabla_{\* x} \hat{q}(\* x; \thetab_{\mathrm{k}}) |^2\right\rangle_\mathrm{fly} = \begin{cases}
\frac{\mathlarger{\sum}\limits_{\alpha=1}^M  \dfrac{z_{\alpha}}{|\mathcal{M}^\alpha_k|}\mathlarger{\sum}\limits_{\* x \in \mathcal{M}^\alpha_k}  \left[\dfrac{\ell(\* x; \thetab_{k})}{c(\* x; \thetab_k)} \right]  }{ \mathlarger{\sum}\limits_{\alpha=1}^M \dfrac{z_{\alpha}}{|\mathcal{M}^\alpha_k|}\mathlarger{\sum}\limits_{\* x \in \mathcal{M}^\alpha_k} \left[\dfrac{1}{c(\* x; \thetab_k) } \right]  } & \text{for umbrella sampling}
\\
\sum\limits_{\alpha=1}^M \dfrac{z_{\alpha} }{|\mathcal{R}^\alpha_k|} \sum\limits_{\* x \in \mathcal{R}^\alpha_k} \ell(\* x; \thetab_{k}) & \text{for the master equation}
\end{cases}  \,, \label{eq:flyloss}
\end{gather}
where the reweighting factors $ z_{\alpha} $ are evaluated using \cref{eq:expestimator} for umbrella sampling and \cref{eq:ftsbalanceeq} for the master equation, respectively. The estimate in \cref{eq:flyloss} is then compared to the average BKE loss function that is evaluated using reference solutions. 

\subsection{First Study: 1D Quartic Potential} \label{sec:1dbrownian}

In this section we study a 1D particle diffusing in a quartic potential $V(\* x)=(1-\* x^2)^2$ with $k_\mathrm{B} T=1/15$. 
This potential has two minima at $\* x=-1,1$ with a saddle point at $\* x=0$, which is the transition state of the model. 
Setting the reactant state $A=(-\infty,-1]$ and product state $B=[1,\infty+)$, the exact solution for the committor function $q_\mathrm{exact}(\* x)$ can be obtained as 
\begin{align}
q_\mathrm{exact}(\* x) &= \frac{\int_{-1}^{\* x} \diff \* x^\prime e^{15V(\* x^\prime)}}{\int_{-1}^1 \diff \* x^\prime e^{15 V (\* x^\prime)}} \label{eq:1dexactsol} \,.
\end{align}
Using \cref{eq:1dexactsol}, the average of the BKE loss function $\left\langle \frac{1}{2} |\nabla_{\* x} q_\mathrm{exact}(\* x) |^2\right\rangle$ can be computed as 
\begin{align}
\left\langle \frac{1}{2} |\nabla_{\* x}  q_\mathrm{exact}(\* x)|^2 \right\rangle &= \frac{1}{2}\left(Z \left(\int_{-1}^1 \diff \* x \ e^{\beta V(\* x)}\right)\right)^{-1} \!\! \!\! \!\! \approx  10^{-6} \,.  
\end{align}
To compute the $L_1$-norm error, we set the transition tube region $ T_{\Lambda}=\Omega \setminus A \cup B = (-1,1)$. 

Figure~\ref{fig:1dresults}(a) shows that the neural network approximations $ \hat{q}( \* x ; \thetab ) $ obtained from all methods converge to the exact solution. 
However, the histograms of sampled configurations obtained from committor-based umbrella sampling lack overlap between the reactant/product states and the transition state (\cref{fig:1dresults}(b), top). 
As discussed in \cref{sec:base}, this lack of overlap indicates that on-the-fly estimates of the average BKE loss, and thus the chemical reaction rates, may not be accurate and are subject to large variance/noise. 
On the other hand, the histograms from algorithms that use the FTS method (\cref{fig:1dresults}(b), middle and bottom) show homogeneous sampling across the transition tube with sufficient overlaps, which should translate to accurate low-variance estimates of reaction rates.
Indeed, \cref{fig:1dresults}(c) shows that the on-the-fly estimates from the BKE--US and BKE--US+SL methods exhibit large fluctuations, spanning six orders in magnitude for a batch size of 16, while the algorithms that use the FTS method can reduce this variance by approximately one order of magnitude for the same batch size. When these on-the-fly estimates are cumulatively averaged, as shown in the inset of \cref{fig:1dresults}(c), we also see that the BKE--US and BKE--US+SL methods yield inaccurate estimates of the average BKE loss when compared to the algorithms employing the FTS method, as these estimates are off from the exact value by two orders of magnitude. 
Irrespective of the sampling method, the addition of supervised learning elements can yield an order-of-magnitude increase in the accuracy of the committor function, as seen from the $ L_1 $-norm error in \cref{fig:1dresults}(d). 
Based on these results, we may conclude that the addition of the FTS method and SL elements yields accurate committor functions and low-variance estimates of the reaction rates.

\subsection{Second Study: 2D M{\"u}ller-Brown Potential} \label{sec:2dmullerbrown}

\begin{figure}[t]
    \centering
    \includegraphics[width=0.95\linewidth]{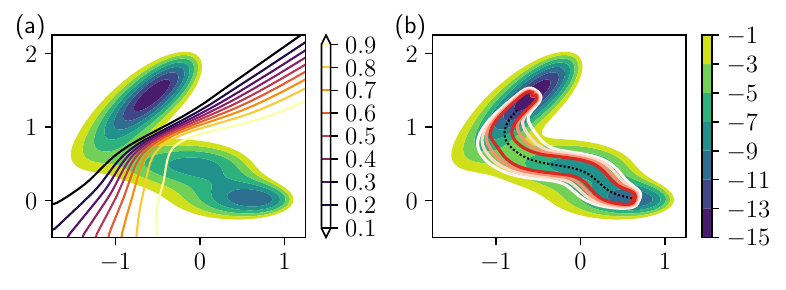}
    \caption{(a) MB potential along with isocommittor lines from the FEM solution. Note the MB contours correspond to $ \beta V_{\rm MB} $. (b) MB potential along with contours indicating the lines of increasing flux $ \* J(\* x) $ from white to red along with the transition path in black, computed using the FEM solution via \cref{eq:pathdef}.}
    \label{fig:mb_potential}
\end{figure}

Although the 1D system already showcases the salient advantages of incorporating SL elements and the FTS method, it only serves as a check to ensure that all algorithms can converge in a setting where an exact solution is available. The advantages and disadvantages of all algorithms can be observed with a more complex problem involving a 2D potential energy landscape, where the transition path is curved. To this end, we now study a particle subject to the 2D M{\"u}ller-Brown (MB) potential \cite{muller1979location}, which is a Gaussian mixture potential given by
\begin{gather}
 \label{eq:v_mb}
 V_{\mathrm{MB}}(\* x) = \sum_{k=1}^4 A_k \exp \left( a_{i} \left( x - \bar{x}_{i} \right)^{2} + b_{i} \left( x - \bar{x}_{i} \right) \left( y - \bar{y}_{i} \right) + c_{i} \left( y - \bar{y} \right)^{2} \right) \,,
 \\
 \begin{aligned}
     A &= (-200,-100,-170,15), \quad  a = (-1, -1, -6.5, -0.7) \,, \\
     b &= (0,0,11,0.6), \quad  c = (-10, -10, -6.5, 0.7) \,, \\
     \bar{x} &= (1, 0, -0.5, -1), \quad  \bar{y} = (0, 0.5, 1.5, 1) \,.
 \end{aligned} \nonumber
\end{gather}
It has two minima at $ \* x_{0}^\mathrm{A} \approx (-0.558, 1.442) $ and $ \* x_{0}^\mathrm{B} \approx (0.623, 0.028) $. In what follows, we study this model at a temperature where $ k_{\mathrm{B}} T = 10 $.
While an analytical form of $ q ( \* x ) $ for the MB potential is unknown, we use FEM to numerically solve the BKE (\cref{eq:bke}) via FEniCS \cite{logg2012automated,alnaes2015fenics}, and obtain a solution to the committor function $ q_{\mathrm{FEM}}(\* x) $.
This is done on the domain $ \Omega = \left[ -1.75 , 1.25 \right] \times \left[ -0.5, 2.25 \right] $, with the reactant and product states defined by $ A = \lbrace \* x \in \Omega : | \* x - \* x_{0}^\mathrm{A} | < 0.025 \rbrace $ and $ B = \lbrace \* x \in \Omega : | \* x - \* x_{0}^\mathrm{B} | < 0.025 \rbrace $, respectively.
The FEM solution is obtained by applying Dirichlet boundary conditions as per \cref{eq:bkebc} along with a zero-flux Neumann boundary condition on $ \partial \Omega $, and a mesh of roughly $3 \cdot 10^5$ elements.
Contours of the MB potential along with isocommittor lines of $ q_{\mathrm{FEM}}( \* x ) $ are shown in \cref{fig:mb_potential}(a), along with contours of increasing flux and the transition path in \cref{fig:mb_potential}(b).

The ensemble-averaged BKE loss with $ q_{\mathrm{FEM}}(\* x) $ over $ \Omega $ is obtained by evaluating the variational objective function in \cref{eq:variational_estimate}:
\begin{equation}
    \left\langle \frac{1}{2} | \nabla_{\* x} q_\mathrm{FEM} |^2 \right\rangle \approx 2.46 \cdot 10^{-4} \,. \label{eq:femloss}
\end{equation}
To compute the $ L_{1}$-norm error, we select the transition tube domain to be $ T_{\Lambda} = \lbrace \* x \in \Omega : | \* J (\* x) | > \Lambda=1.61 \cdot 10^{-4}  \rbrace $, which corresponds to the outermost white line in \cref{fig:mb_potential}(b). 
In addition to on-the-fly estimates, the ensemble average of the BKE loss from the neural network representation $ \hat{q}( \* x; \thetab ) $ can be evaluated by numerically integrating over the entire domain, and is given by
\begin{equation}
    \left\langle \frac{1}{2} |\nabla_{\* x} \hat{q}(\* x; \thetab_{\mathrm{k}}) |^2\right\rangle_\mathrm{full} = \int_{\Omega} \diff \* x \rho(\* x) \ell(\* x; \thetab_{\mathrm{k}}) = \langle \ell(\* x; \thetab_{\mathrm{k}}) \rangle \,. \label{eq:fullloss}
\end{equation}
\Cref{eq:fullloss} provides an additional metric for evaluating accuracy;
in particular, comparing \cref{eq:fullloss} with the on-the-fly estimates allows us to evaluate the sampling error that arises from the choice of estimator, while comparing \cref{eq:fullloss} with the FEM value (\cref{eq:femloss}) allows us to evaluate the error inherent to the neural network.

\begin{figure}[t]
    \centering
    \includegraphics[width=\linewidth]{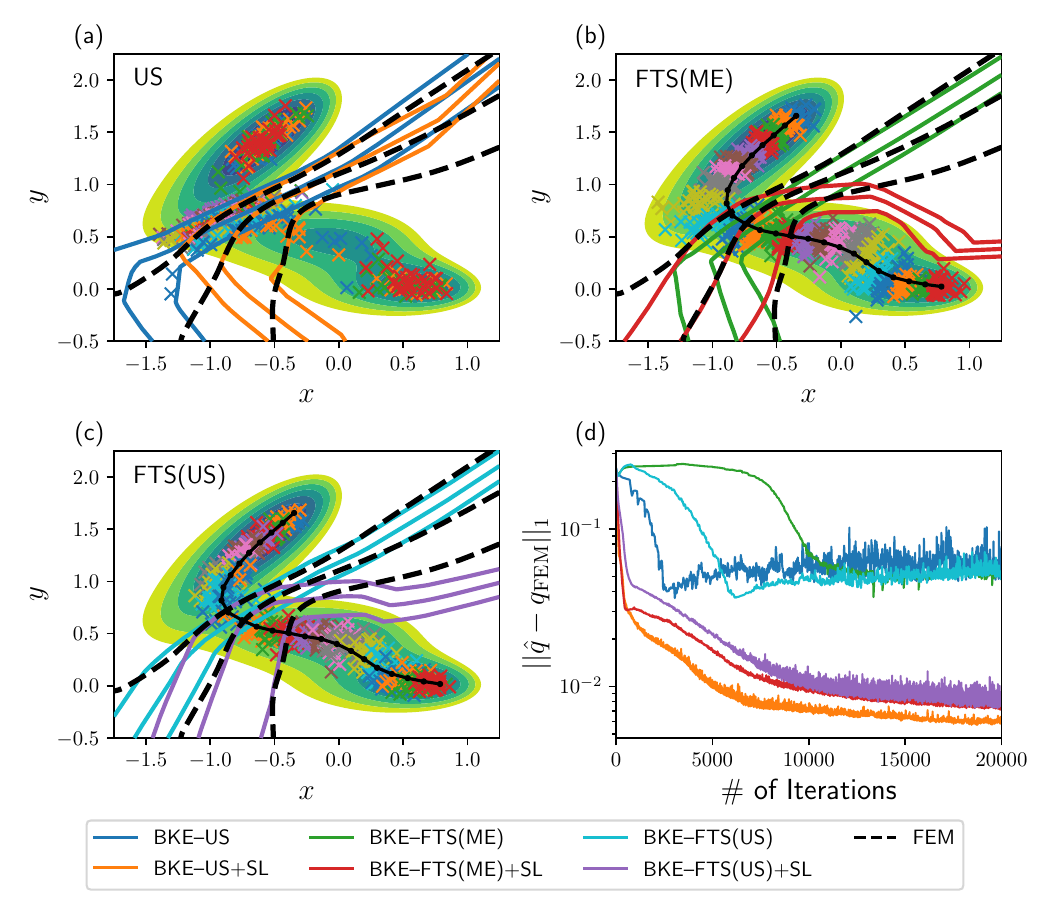}
    \caption{Isocommittor lines for $ q = 0.1 $, $ 0.5 $, and $ 0.9 $ (left to right) from (a) the BKE--US and BKE--US+SL method, (b) the BKE--FTS(ME) and BKE--FTS(ME)+SL method, and (c) the BKE--FTS(US) and BKE--FTS(US)+SL method. $\times$ markers denote representative samples obtained from algorithms without supervised learning. Dotted lines are the transition paths obtained from the FTS method. (d) The $L_1$-norm error of the committor function as a function of iterations. }
    \label{fig:2dresults}
\end{figure}

\begin{figure}[t]
    \centering
    \includegraphics[width=\linewidth]{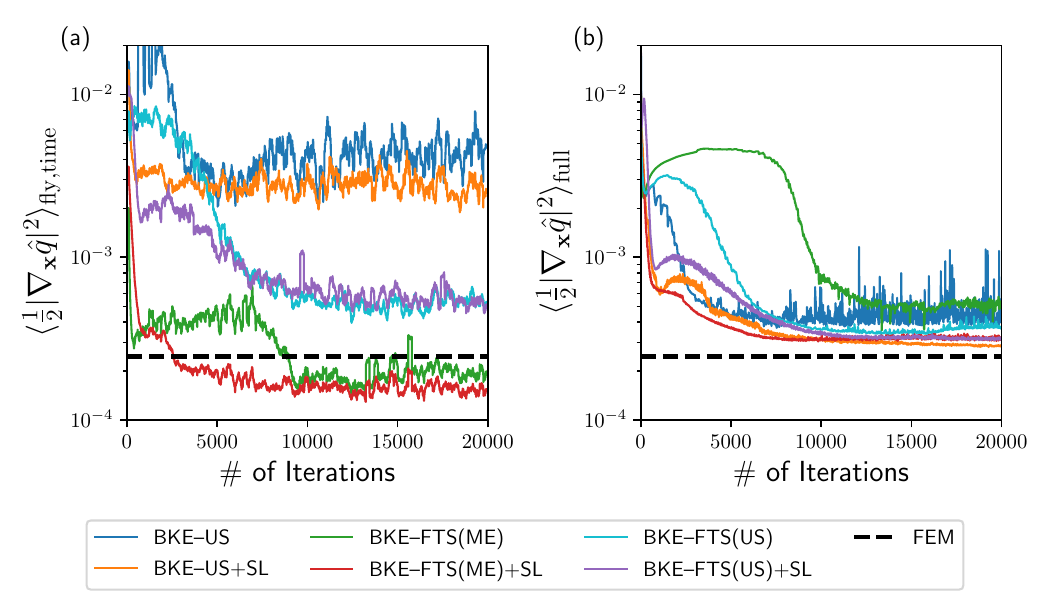}
    \caption{(a) The filtered on-the-fly estimate of the BKE loss obtained at every iteration, with the filtering window set to 200 iterations. (b) The ensemble-averaged loss per \cref{eq:fullloss} obtained at every iteration.}
    \label{fig:mb_metric}
    \centering
    \includegraphics[width=0.825\linewidth]{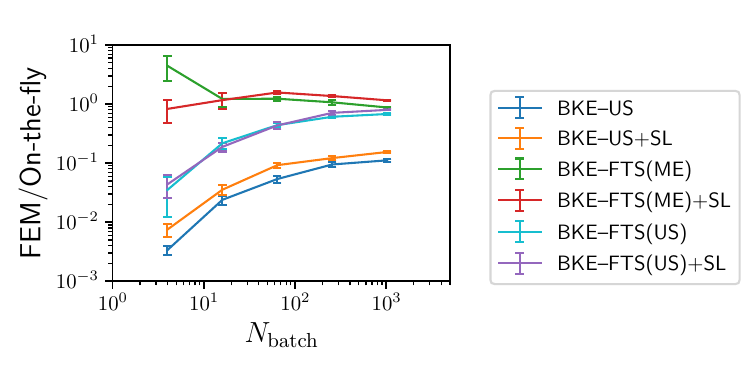}
    \caption{The ratio between the average BKE loss from FEM solution (\cref{eq:femloss}) and on-the-fly estimates, where the latter is cumulatively averaged over the last 3000 iterations of the neural network training.}
    \label{fig:errorallmethods}
\end{figure}

Figures~\ref{fig:2dresults}(a-c) show the isocommittor lines and sampled configurations obtained from all algorithms. We see from the isocommittor lines that methods employing supervised learning elements improve the accuracy of the committor functions both in and outside the transition tube, as these surfaces follow the FEM solution far more closely than the ones without such elements. 
This increase in accuracy is also reflected in the $L_{1}$-norm error shown in \cref{fig:2dresults}(d), where the error from methods with supervised learning is reduced by an order of magnitude regardless of the chosen sampling method. 
Furthermore, similar to the 1D system, committor-based umbrella sampling yields samples that are focused near the transition state with little overlap between the reactant/product basins and the transition state region; see \cref{fig:2dresults}(a). As mentioned in \cref{sec:base}, this lack of overlap can negatively impact the accuracy of the estimated reaction rates due to inaccurate estimates of free energy differences between neighboring replicas and thereby the reweighting factors (\cref{fig:us_z_l_norm}).
Conversely, all algorithms using the FTS method yield overlapping samples that homogeneously cover the transition tube and hence accurate estimates of reweighting factors (\cref{fig:ftsme_z_l_norm,fig:ftsus_z_l_norm}), indicating that reaction rate estimates may be computed with higher accuracy and lower variance.

Figure~\ref{fig:mb_metric}(a) shows the on-the-fly estimates of the reaction rates or the average BKE loss from all methods, computed using a smaller batch size of 64 samples and filtered over the nearest $200$ iterations.  With the exception of the BKE--FTS(ME) and BKE--FTS(ME)+SL methods, these on-the-fly estimates converge towards values far from the FEM solution even though the ensemble-averaged BKE loss computed by numerical integration (\cref{eq:fullloss}) shows convergence towards the FEM value (\cref{fig:mb_metric}(c)). 
This shows the sampling error is still large, and larger batch sizes ($N_\mathrm{batch}$) are needed to obtain accurate on-the-fly estimates. Figure~\ref{fig:errorallmethods}(a) shows the ratio of the FEM and the on-the-fly estimates as a function of batch size, where all the methods employing the FTS methods converge towards the FEM value with the exception of the BKE--US and BKE--US+SL methods, which plateau to a ratio of $0.1$. As mentioned in \cref{sec:base}, this discrepancy is related to the lack of overlaps in the samples between the transition state and the reactant/product basins, resulting in the inaccurate estimates of $z_\alpha$ (\cref{fig:us_z_l_norm}). 
These results show that replacing the committor-based umbrella sampling with the FTS method results in more accurate estimates of the reaction rates.

Furthermore, the FTS method with path-based umbrella sampling is amenable to error analysis, allowing us to estimate the errors in the reaction rates. 
In what follows, we provide such an analysis for the BKE--FTS(US) and BKE--FTS(US)+SL methods, using which the sampling errors in the on-the-fly estimates can be eliminated. As will be shown later in \cref{fig:error_ftsus_final}, this allows accurate computation of the average BKE loss functions for the BKE--FTS(US) and BKE--FTS(US)+SL methods at any batch size. Lastly, although the average BKE loss computed by numerical integration may be closer to the FEM solution than the on-the-fly estimates, such computation is impractical for high-dimensional problems due to the increased cost of quadrature, necessitating the procedure constructed from error analysis to improve the accuracy in the on-the-fly estimates.

\subsection{Error Analysis of the Average BKE Loss Estimator}  
\label{sec:lognorm}

\begin{figure}[p]
    \centering
    \includegraphics[width=0.9\linewidth]{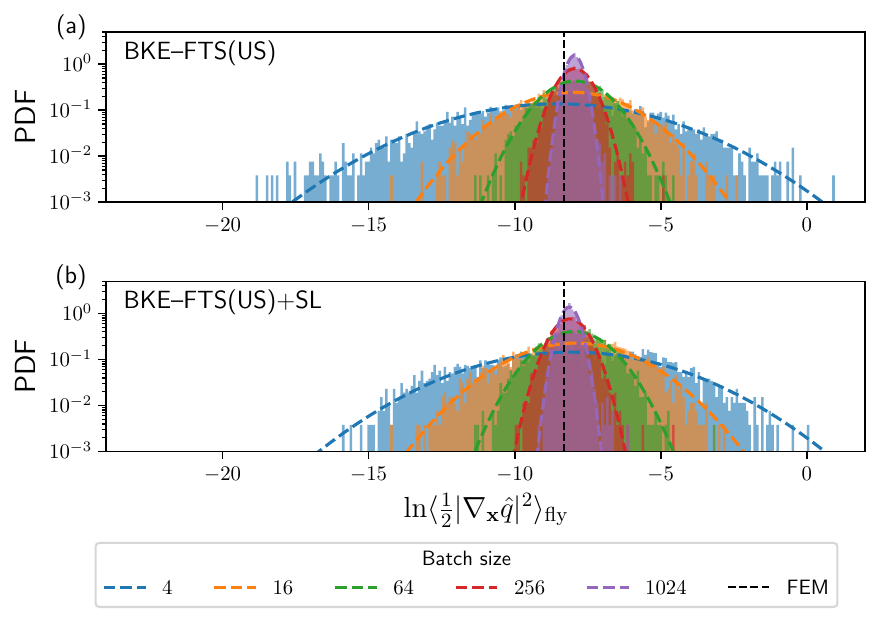}
    \caption{Histograms yielding the probability density functions for the on-the-fly estimate of the BKE loss at various batch sizes from the last $ 3000 $ iterations of training for the (a) BKE--FTS(US), and (b) BKE--FTS(US)+SL methods. Corresponding dashed lines are log-normal distributions fitted using the method of moments \cite{pearson1936method}, while the dashed vertical black line corresponds to the average BKE loss from the FEM solution.}
    \label{fig:energy_log_ftsus_histogram}
    \centering
    \includegraphics[width=0.9\linewidth]{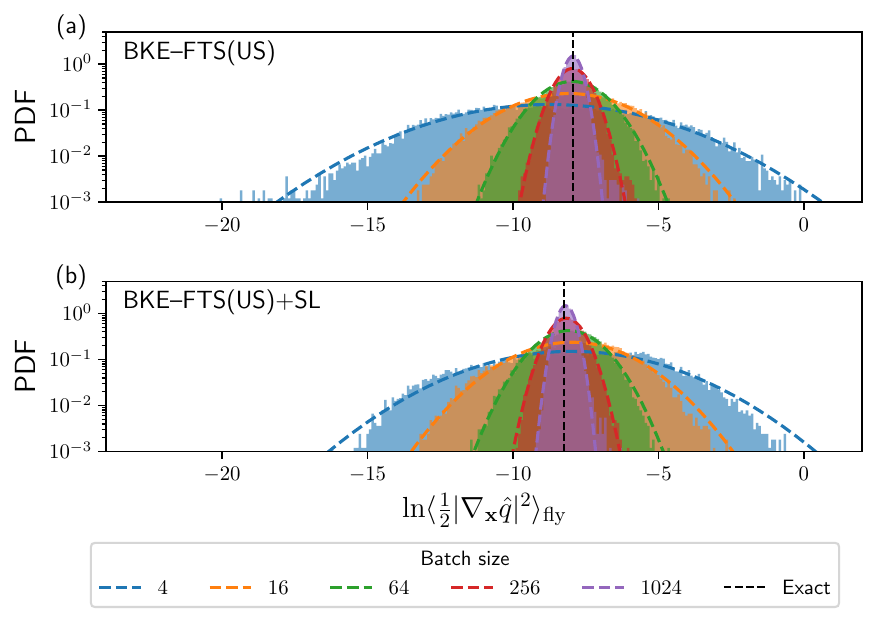}
    \caption{Histograms of the on-the-fly estimate of the average BKE loss at various batch sizes, with the neural network parameters fixed at every iteration for the (a) BKE--FTS(US), and (b) BKE--FTS(US)+SL methods. 
    The neural network configuration corresponds to the one obtained from training at batch size $ 4 $. 
    Corresponding dashed lines are log-normal distributions fitted using the method of moments \cite{pearson1936method}, while the dashed vertical black line corresponds to the average BKE loss  computed by numerical integration (\cref{eq:fullloss}).} 
    \label{fig:energy_log_ftsus_histogram_sampling}
\end{figure}

\begin{figure}[t]
    \centering
    \includegraphics[width=0.9\linewidth]{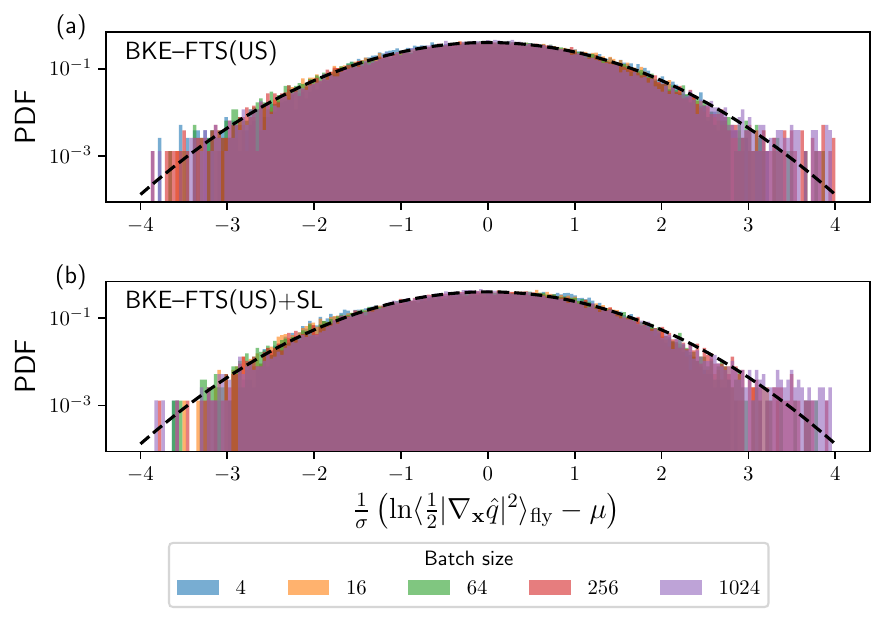}
    \caption{Histograms of the on-the-fly estimate of the average BKE loss shifted by the mean $ \mu $ and normalized by the standard deviation $ \sigma $ at various batch sizes, with the neural network parameters fixed at every iteration for the (a) BKE--FTS(US), and (b) BKE--FTS(US)+SL methods. 
    The neural network configuration corresponds to the one obtained from training with a batch size of $ 4 $. 
    The black dashed line is a log-normal distributions with $ \mu = 0 $ and $ \sigma = 1 $.}
    \label{fig:energy_log_ftsus_histogram_sampling_renormalized}
\end{figure}

Before we begin the error analysis, we first plot the normalized histograms, i.e., the empirical probability density functions (PDFs), of the logarithm of on-the-fly BKE loss for both the BKE--FTS(US) and BKE--FTS(US)+SL methods (\cref{fig:energy_log_ftsus_histogram}), which show that fluctuations of these estimates are centered around the FEM value. 
Furthermore, the resulting PDFs can be fitted to a log-normal distribution via the method of moments \cite{pearson1936method} with increasing agreement as the batch size is increased. 
The emergence of the log-normal distribution can be attributed to either the change in model parameters $\thetab_k$ during optimization or the nature of umbrella sampling when used in conjunction with the estimator given by \cref{eq:flyloss}. 
Since the log-normal statistics emerge when the neural network is already converged, it is more likely for sampling to be the chief cause of these statistics, rather than the optimization. 
This hypothesis can be tested by computing the on-the-fly BKE loss when the neural network parameters are fixed at every iteration, which has the effect of decoupling the influence of optimization from sampling. 
The histograms from this numerical experiment are shown in \cref{fig:energy_log_ftsus_histogram_sampling}, where log-normal distributions are produced as before, and their peaks are located precisely at the ensemble-averaged BKE loss computed by numerical integration (\cref{eq:fullloss}).
The logarithm of the average BKE loss can be shifted by the mean and normalized by the standard deviation of the corresponding distributions to produce approximate standard normal distributions as seen in \cref{fig:energy_log_ftsus_histogram_sampling_renormalized}, with increasing batch sizes having an increasing agreement with a standard normal distribution.

With the observation of log-normal statistics established, we now determine its origin by investigating each component that contributes to the computation of the on-the-fly BKE loss in \cref{eq:flyloss}. To this end, we provide a more concise notation for the estimator (\cref{eq:flyloss}) by re-writing it as
\begin{align}
    \left\langle \frac{1}{2} |\nabla_{\* x} \hat{q}(\* x; \thetab_{\mathrm{k}}) |^2 \right\rangle_{\mathrm{fly}} = \frac{\mathlarger{\sum}\limits_{\alpha=1}^M  z_{\alpha} \left(\dfrac{1}{|\mathcal{M}^\alpha_k|}\mathlarger{\sum}\limits_{\* x \in \mathcal{M}^\alpha_k}  \left[\dfrac{\ell(\* x; \thetab_{k})}{c(\* x; \thetab_k)} \right] \right) }{ \mathlarger{\sum}\limits_{\alpha=1}^M z_{\alpha}\left(\dfrac{1}{|\mathcal{M}^\alpha_k|}\mathlarger{\sum}\limits_{\* x \in \mathcal{M}^\alpha_k} \left[\dfrac{1}{c(\* x; \thetab_k) } \right] \right) } = \frac{\mathlarger{\sum}\limits_{\alpha=1}^M  z_{\alpha} \bar{\ell}^{*}_{\alpha} }{ \mathlarger{\sum}\limits_{\alpha=1}^M z_{\alpha} \bar{1}^{*}_{\alpha}  } \,, \label{eq:estimator}
\end{align}
where we define the division by $ c(\* x; \thetab_k) $ per sample with the $ * $ operator, and denote the standard sample mean using the bar operator. \Cref{eq:estimator} requires computing free energies through $z_\alpha$, and sample means from each replica through $\bar{\ell}^*_\alpha$ and $\bar{1}^*_\alpha$, which indicates that the origin of the log-normal statistics of the average BKE loss can be found once the statistics for $ z_\alpha $, $\bar{\ell}^*_\alpha$, and $\bar{1}^*_\alpha$ are determined individually.
In what follows, we first investigate the statistics of $z_\alpha$ as computed via FEP.

To begin, we write the free-energy difference $\Delta F_{\alpha, \alpha^\prime}=F_\alpha-F_{\alpha^\prime}$ per \cref{eq:basicexpestimator} as
\begin{equation}
    \beta \Delta F_{\alpha, \alpha^\prime} = -\log \left [ \frac{1}{|\mathcal{M}_k^{\alpha^\prime}|}  \sum_{\* x \in \mathcal{M}^{\alpha^\prime}_k} \exp ( - \beta \Delta W_{\alpha, \alpha^\prime}(\* x; \thetab_k ) ) \right] \,, \label{eq:flylossconcise}
\end{equation} 
where $ \Delta W_{\alpha, \alpha^\prime}(\* x; \thetab_k )=W_\alpha ( \* x ; \thetab_k )-W_{\alpha^\prime}(\* x; \thetab_k ) $. Note that free-energy differences are typically computed for adjacent replicas, so that $ \alpha = \alpha^\prime \pm 1 $. For sufficiently small $ \Delta W_{\alpha, \alpha^\prime}(\* x ; \thetab_k ) $, use of Taylor series expansions yields
\begin{align}
    \beta \Delta F_{\alpha, \alpha^\prime} &\approx -\log \left [ \frac{1}{|\mathcal{M}_k^{\alpha^\prime}|}  \sum_{\* x \in \mathcal{M}^{\alpha^\prime}_k} \left( 1  - \beta \Delta W_{\alpha, \alpha^\prime}(\* x ; \thetab_k ) \right) \right] \\
                                          &\approx -\log \left [ 1 - \frac{1}{|\mathcal{M}_k^{\alpha^\prime}|} \sum_{\* x \in \mathcal{M}^{\alpha^\prime}_k} \beta \Delta W_{\alpha, \alpha^\prime}(\* x ; \thetab_k ) \right] \\
                                          &\approx \frac{1}{|\mathcal{M}_k^{\alpha^\prime}|} \sum_{\* x \in \mathcal{M}^{\alpha^\prime}_k} \beta \Delta W_{\alpha, \alpha^\prime}(\* x ; \thetab_k ) \,.
\end{align}
According to the central limit theorem and assuming that the samples $ \* x  \in \mathcal{M}^{\alpha}_k $ are independent and identically distributed, the sample mean of $ \Delta W_{\alpha, \alpha^\prime} $ is normally distributed, and thus the free-energy differences $\Delta F_{\alpha, \alpha^\prime}$ are also normally distributed.
This argument only holds for small $ \Delta W_{\alpha, \alpha^\prime}(\* x ; \thetab_k) $, which can be achieved when there is overlap in configuration space---a condition that is ensured with a good choice of the bias strength parameters. Since $ \Delta F_{\alpha, \alpha^\prime} $ is normally distributed, its exponentiation $ e^{-\beta \Delta F_{\alpha, \alpha^\prime}} $ is log-normally distributed.
Using \cref{eq:expestimator}, for $ \alpha $ not equal to the reference index $ \gamma $, the un-normalized reweighting factor $ z_\alpha^\star $ obtained from FEP is also log-normally distributed, since it is computed from products of $ e^{-\beta \Delta F_{\alpha, \alpha^\prime}} $  factors that are log-normally distributed 
\cite{forbes2011statistical}.
Upon normalizing $ z_\alpha^\star $ to obtain $ z_\alpha $, we should observe approximately log-normal statistics for $ z_\alpha $, since the normalization requires dividing $ z_\alpha^\star $ with its sum, which is approximately log-normal \cite{marlow1967,barouch1986sums,beaulieu1996,mehta2007,asmussen2008}.

\begin{figure}[t]
    \centering
    \includegraphics[width=0.95\linewidth]{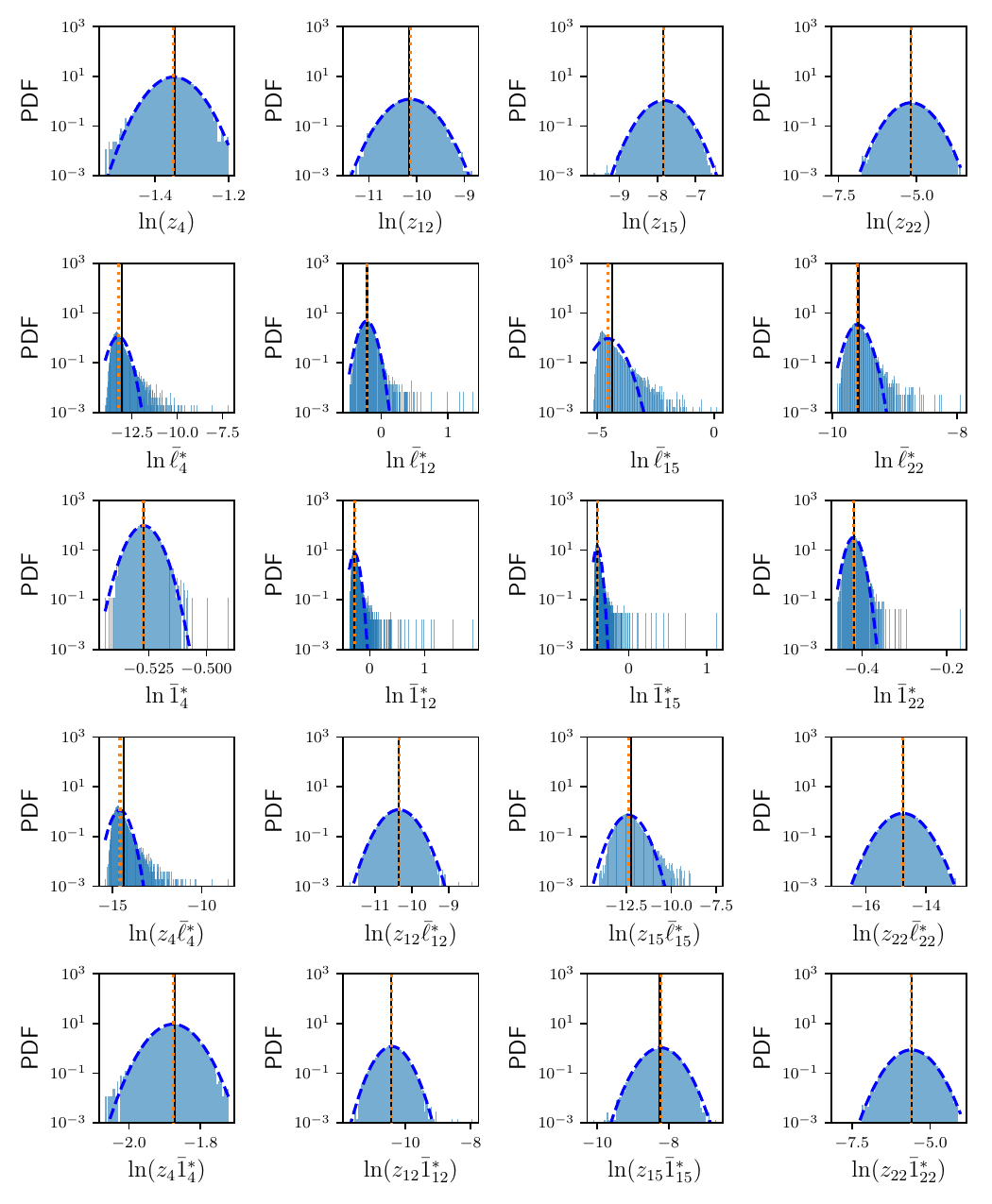}
    \caption{Probability density functions of the various quantities representative of Table ~\ref{tab:error_table}. Data is obtained from sampling with batch size $1024$, with a fixed neural network obtained from the BKE--FTS(US)+SL method at the same batch size. Dashed blue lines are log-normal distributions fitted  using the method of moments \cite{pearson1936method}, while the vertical dotted orange and solid black lines correspond to the mean of the histograms and the corresponding ensemble average computed via numerical integration, respectively.}
    \label{fig:summary_overall}
\end{figure}

The arguments we put forth for the statistics of $ \beta \Delta F_{\alpha, \alpha^\prime} $ and $ z_\alpha $ can be verified in simulations by evaluating the probability density functions for the quantities of interest.
For the forward free-energy differences $ \beta \Delta F_{(\alpha+1) , \alpha} $ and the backward free-energy differences $ \beta \Delta F_{(\alpha-1), \alpha} $, the observed distributions can be described by normal distributions (\cref{fig:summary_fwd,fig:summary_bwd}), which immediately imply that their exponentiation is log-normally distributed.
The resulting reweighting factors $ z_{\alpha} $ are found to be log-normally distributed, in agreement with our heuristic arguments, as seen from the PDFs of $ \ln z_{\alpha} $ in the first row of \cref{fig:summary_overall} for representative replicas, and \cref{fig:summary_z} for all replicas.
Note that there exist free-energy differences, such as $ \beta \Delta F_{9,8} $ and $ \beta \Delta F_{10,9} $,  that have a slight deviation in the tails due to the presence of higher-moment terms.
These effects are mostly removed when evaluating the PDFs for $ \ln z_{\alpha} $, 
and it is expected that these tails disappear as the batch size is increased since this leads to free-energy differences that further obey a normal distribution.  
To summarize the statistics observed in all replicas, we group replicas with similar behaviors into four groups, corresponding to the reactant (1-10), transition (11-13), metastable (14-18), and product (19-24) states. The results for $ z_{\alpha} $ for these groups are shown in the second column of Table~\ref{tab:error_table}. 

\begin{figure}[t]
    \centering
    \begin{tabular}{|l|c|c|c|c|c|c|c|}
        \hline
        Replicas & $ z_\alpha $ & $ \bar{\ell}_{\alpha}^{*} $ & $ \bar{1}_{\alpha}^{*} $ & $ z_\alpha \bar{\ell}_{\alpha}^{*} $ & $ z_\alpha \bar{1}_{\alpha}^{*} $ & 
        $\begin{aligned}
        &\varop ( \ln z_{\alpha} ) > \\
        &\ \varop ( \ln \bar{\ell}_{\alpha}^{*} )
        \end{aligned}$ &
        $\begin{aligned}
        &\varop ( \ln z_{\alpha} ) > \\
        &\ \varop ( \ln \bar{1}_{\alpha}^{*} )
        \end{aligned}$ \\ \hline
        Reactant State (1-10) & \cmark & \xmark & \xmark & \xmark & \cmark & \xmark & \cmark \\ \hline
        Transition State (11-13) & \cmark & \cmark & \xmark & \cmark & \cmark & \cmark & \cmark \\ \hline
        Metastable State (14-18) & \cmark & \xmark & \xmark & \xmark & \cmark & \xmark & \cmark \\ \hline
        Product State (19-24) & \cmark & \cmark & \xmark & \cmark & \cmark & \cmark & \cmark \\ \hline
    \end{tabular}
    \captionof{table}{Summary of error analysis results. For columns two through six, \cmark\ indicates the distributions in all or most replicas are log-normal, while \xmark\ indicates the distributions in all or most replicas are approximately log-normal with a skew or slight deviations in the tails. For columns seven and eight, they indicate if the inequality holds. The corresponding histograms for columns two through six can be found in \cref{fig:summary_z,fig:summary_grad,fig:summary_norm,fig:summary_grad_z,fig:summary_norm_z}.}
    \label{tab:error_table}
 \end{figure}

With the sampling distributions of $z_\alpha$ understood, we now study the sampling distributions for $ \bar{\ell}_{\alpha}^{*} $ and $ \bar{1}_{\alpha}^{*} $. 
Assuming the values of $ \ell_{\alpha}^{*} $ and $ 1_{\alpha}^{*} $ are independent and identically distributed, one may expect the corresponding sample means $ \bar{\ell}_{\alpha}^{*} $ and $ \bar{1}_{\alpha}^{*} $ to be normally distributed according to the central limit theorem. 
However, we observe from simulations that these sample means are better described by log-normal distributions; see the second and third rows of \cref{fig:summary_overall} for representative histograms, and \cref{fig:summary_grad,fig:summary_norm} for all histograms. Since log-normality arises when normally-distributed random variables are exponentiated, its origin is likely due to the sums of exponentials in $c(\* x;\thetab_k)$ for $\bar{1}_{\alpha}^{*}$, and the neural network model $\hat{q}(\* x; \thetab_k)$ for $ \bar{\ell}_{\alpha}^{*} $, where the output layer of $\hat{q}(\* x; \thetab_k)$ contains the sigmoidal function $\sigma(s) = 1/(1+e^{-s})$. 
Nevertheless, the distributions possess tails that render the log-normality only approximate in nature. 
We summarize these observations in the third and fourth columns of Table~\ref{tab:error_table}. 

Despite the approximate log-normality in $ \bar{\ell}_{\alpha}^{*} $ and $ \bar{1}_{\alpha}^{*} $, one need not understand accurately the distributions of $\bar{\ell}_{\alpha}^{*}$ and $\bar{1}_{\alpha}^{*}$, as the distributions obtained for the products $ z_{\alpha} \bar{\ell}_{\alpha}^{*} $ and $ z_{\alpha} \bar{1}_{\alpha}^{*} $, which are needed by the estimator in \cref{eq:flylossconcise}, are log-normal;  see the fourth and fifth rows of \cref{fig:summary_overall} for representative histograms, and \cref{fig:summary_grad_z,fig:summary_norm_z} for all histograms, as well as the fifth and sixth columns of Table~\ref{tab:error_table} for a concise summary. The only exceptions are the histograms for $ z_{\alpha} \bar{\ell}_{\alpha}^{*} $ at the reactant (1-10) and metastable state (14-18), which have slightly skewed log-normal behavior. However, these do not contribute significantly to the overall BKE loss when compared to the transition state.
To understand why log-normality emerges again for $ z_{\alpha} \bar{\ell}_{\alpha}^{*} $ and $ z_{\alpha} \bar{1}_{\alpha}^{*} $, let us convert the products into sums by taking the logarithm, so that 
$ \ln z_{\alpha} \bar{\ell}_{\alpha}^{*} = \ln z_{\alpha} + \ln \bar{\ell}_{\alpha}^{*} $ and $ \ln z_{\alpha} \bar{1}_{\alpha}^{*} = \ln z_{\alpha} + \ln \bar{1}_{\alpha}^{*} $. The distribution of the sum of two independent random variables, denoted more generally as $Y= X_{1} + X_{2} $, can be obtained from the distributions for $ X_{1}$ and $ X_{2}$ in terms of a convolution  
\begin{equation}
    \rho_{Y}( y) = \int_{-\infty}^{\infty} \diff x \  \rho_{X_{1}}(x ) \rho_{X_{2}}(y - x) \,.
\end{equation}
When one random variable, e.g., $ X_{2} $, possesses a much lower variance than the other random variable, we expect that the value of $ X_{2} $ will be constant relative to $ X_{1} $.
In this limit, we may approximate $ \rho_{X_{2}}( x) $ with a Dirac delta function to yield
\begin{equation}
    \rho_{Y}(y) \approx \int_{-\infty}^{\infty} \diff x \ \rho_{X_{1}}(x) \delta (y - x) = \rho_{X_{1}}(y) \,. \label{eq:product_argument}
\end{equation}
Thus, the distribution for the sum is solely determined by the distribution of the random variable with the highest variance. Although this argument is only a weak approximation, as the random variables involved in $ z_{\alpha} \bar{\ell}_{\alpha}^{*}$ and $z_{\alpha} \bar{1}_{\alpha}^{*}$ are correlated due to being processed from the same $ \* x $ values, it gives an insight as to why $ z_{\alpha} \bar{\ell}_{\alpha}^{*} $ and $ z_{\alpha} \bar{1}_{\alpha}^{*} $ are log-normally distributed. Note that the true distributions of $ \ln \bar{\ell}_{\alpha}^{*} $ and $ \ln \bar{1}_{\alpha}^{*} $ are not exactly known, but the distributions of $ \ln z_{\alpha} $ consist of normal distributions. If $ \ln z_{\alpha} $ possesses a larger variance than $ \ln \bar{\ell}_{\alpha}^{*} $ or $ \ln \bar{1}_{\alpha}^{*} $ we expect from \cref{eq:product_argument} that the distribution of the sum in $ \ln z_{\alpha} \bar{\ell}_{\alpha}^{*} $ and $ \ln z_{\alpha} \bar{1}_{\alpha}^{*} $ matches the normal distribution of $ \ln z_{\alpha}$.
This argument is verified in the seventh and eighth columns of Table~\ref{tab:error_table}, where we see that $ \ln z_{\alpha} \bar{\ell}_{\alpha}^{*} $ and $ \ln z_{\alpha} \bar{1}_{\alpha}^{*} $ are normally distributed whenever $\ln z_\alpha$ possess higher variance. 

\begin{figure}[t]
    \centering
    \includegraphics[width=\linewidth]{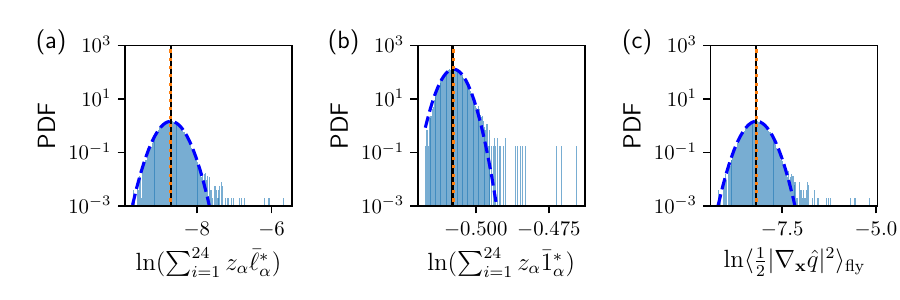}
    \caption{Probability density functions of sums of $ z_{\alpha} \bar{\ell}_{\alpha}^{*} $, $ z_{\alpha} \bar{1}_{\alpha}^{*} $, and the on-the-fly estimate of the BKE loss function in logarithmic space. Data is obtained from sampling with batch size $1024$, with a fixed neural network obtained from the BKE--FTS(US)+SL method at the same batch size. Dashed blue lines are log-normal distributions fitted using the method of moments \cite{pearson1936method}, while the vertical dotted orange and solid black lines correspond to the mean of the histograms and the corresponding ensemble average computed via numerical integration, respectively.}
    \label{fig:summary_process}
\end{figure}

With the log-normality of $z_{\alpha} \bar{\ell}_{\alpha}^{*} $ and $ z_{\alpha} \bar{1}_{\alpha}^{*} $ verified, we can examine the numerator $\sum_{\alpha=1}^M z_{\alpha} \bar{\ell}_{\alpha}^{*}$ and denominator $\sum_{\alpha=1}^M z_{\alpha} \bar{1}_{\alpha}^{*} $ of \cref{eq:estimator}, which make up the on-the-fly average BKE loss. 
Since the sum of log-normal random variables can be approximately described by a log-normal distribution \cite{marlow1967,barouch1986sums,beaulieu1996,mehta2007,asmussen2008}, both the numerator and denominator should be approximately log-normal.
From simulations, we find that the numerator is log-normally distributed (\cref{fig:summary_process}(a)) while the denominator is log-normally distributed with slight deviations in the tails (\cref{fig:summary_process}(b)). Since the ratio of two log-normal random variables is also log-normal, the resulting on-the-fly BKE loss should be log-normal, as shown in \cref{fig:summary_process}(c). This is also in agreement with what is observed during training  (\cref{fig:energy_log_ftsus_histogram}), and when the neural network is fixed (\cref{fig:energy_log_ftsus_histogram_sampling}). 
Although the log-normality of the denominator is only approximate, one can use the previous argument on sums of random variables, i.e., \cref{eq:product_argument}, to show that the sampling distribution of the on-the-fly BKE loss is still log-normal, since the numerator has higher variance than the denominator, thereby allowing the log-normality of the numerator to dominate in the on-the-fly BKE loss. Given these results, we conclude that the on-the-fly estimates of the average BKE loss obtained from the BKE--FTS(US) and BKE--FTS(US)+SL methods are approximately log-normal. 

\begin{figure}[t]
    \centering
    \includegraphics[width=0.7\linewidth]{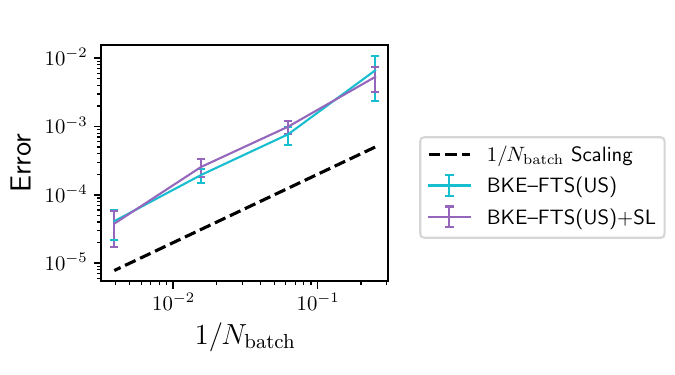}
    \caption{The absolute error in the on-the-fly BKE loss at different batch sizes, with respect to the largest batch size. All error bars are 95\ confidence intervals.}
    \label{fig:error_ftsus_scaling}
\end{figure}

\begin{figure}[p]
    \centering
    \includegraphics[width=0.9\linewidth]{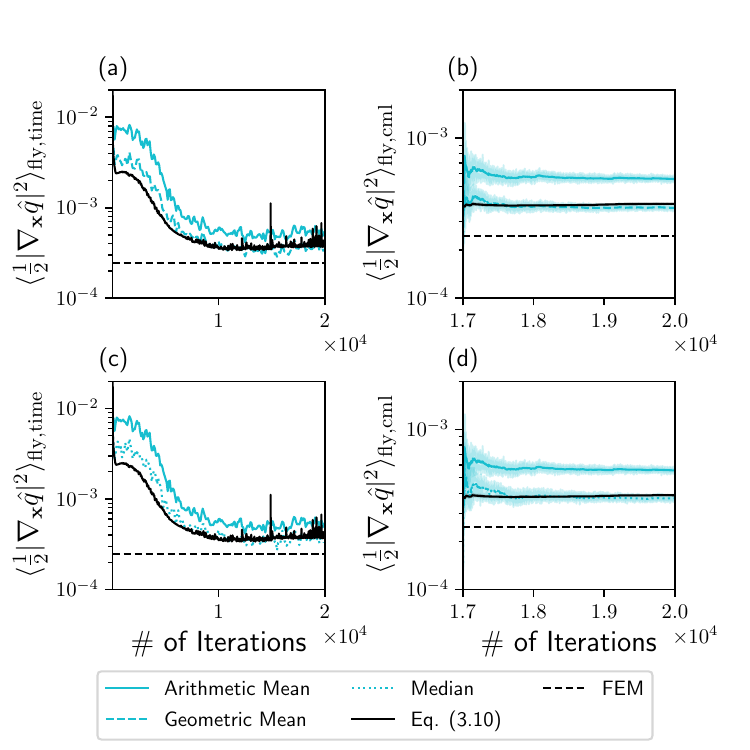}
    \caption{Using the geometric mean (a,b) and median (c,d) to remove the sampling error in the filtered on-the-fly estimates (left column) and cumulative average (right column) of the on-the-fly estimates in the BKE--FTS(US) method for batch size $64$. Note that the remaining error between the FEM value and the average BKE loss computed per \cref{eq:fullloss} is due to the inherent error of the chosen neural network. Cumulative mean and median are performed over the last 3000 iterations of the algorithm. Shaded colors in (b) and (d) are 95\ confidence intervals.}
    \label{fig:error_ftsus}
    \centering
    \includegraphics[width=0.95\linewidth]{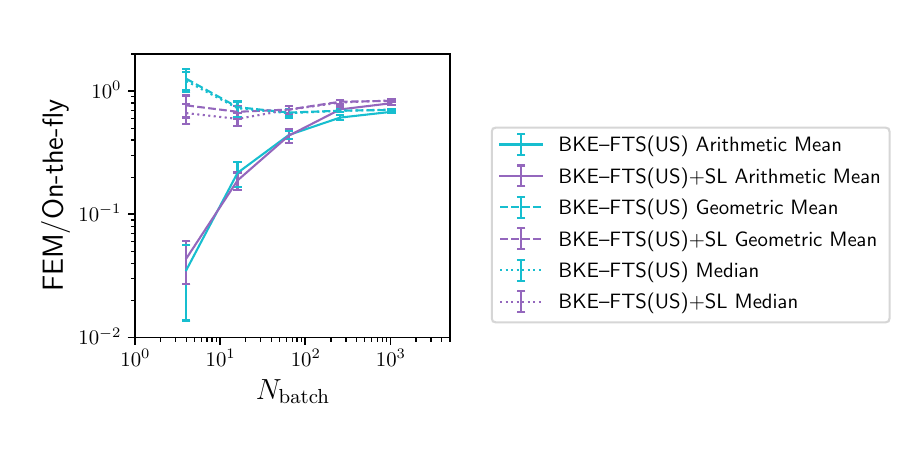}
    \caption{The ratio between FEM and on-the-fly estimates, after taking the geometric mean and median. Error bars are 95\ confidence interval.}
    \label{fig:error_ftsus_final}
\end{figure}

Using the log-normal distribution of the average BKE loss, one can determine the asymptotic behavior of the sampling error as a function of batch size $N_\mathrm{batch}$.
Denoting the mean and variance of the log-normal distribution as $ \mu $ and $ \sigma^{2} $, respectively, we expect that the cumulative mean of the on-the-fly BKE loss over iterations is given by \cite{forbes2011statistical}
\begin{equation}
    \frac{1}{K-k^\star+1} \sum_{k=k^\star}^{K} \left\langle \frac{1}{2} |\nabla_{\* x} \hat{q}(\* x; \thetab_{\mathrm{k}}) |^2 \right\rangle_{\mathrm{fly}} \approx \exp \left( \mu + \frac{1}{2} \sigma^{2} \right)  \label{eq:meanfromlognormal} \,,
\end{equation}
where $K$ is the final iteration index, and $k^\star$ is the iteration index when the on-the-fly estimates begin to fluctuate around a plateau.
\Cref{eq:meanfromlognormal} implies that the cumulative mean of on-the-fly estimates is always multiplied by a factor $ \exp \left(\frac{1}{2} \sigma^{2} \right) > 1$, since $\sigma^2 > 0$. This explains why the on-the-fly estimates in \cref{fig:2dresults}(a) from both the BKE--FTS(US) and BKE--FTS(US)+SL methods are larger than the FEM value, and why the ratio between the FEM value and the on-the-fly estimates in \cref{fig:mb_metric} is always less than one. Furthermore, $ \sigma^{2} \sim O(1/N_\mathrm{batch}) $, implying  for large $N_\mathrm{batch}$ that 
\begin{align}
\frac{1}{K-k^\star+1} \sum_{k=k^\star}^{K} \left\langle \frac{1}{2} |\nabla_{\* x} \hat{q}(\* x; \thetab_{\mathrm{k}})|^2 \right\rangle_{\mathrm{fly}} \sim \exp \left( \mu\right)(1+O(1/N_\mathrm{batch})) \,,  
\end{align}
thus showing the sampling error in the on-the-fly estimates scales as $O(1/N_\mathrm{batch})$. Defining the absolute error as the difference between the cumulative mean of the on-the-fly estimates obtained at smaller batch sizes and the one obtained at the largest batch size, we plot the absolute error as a function of $N_\mathrm{batch}$ in Figure~\ref{fig:errorallmethods} for both the BKE--FTS(US) and BKE--FTS(US)+SL methods, where the $O(1/N_\mathrm{batch})$ scaling can be observed.

The knowledge of the log-normal distribution can also be used to remove the sampling error between the on-the-fly estimates and the ensemble-averaged loss computed by numerical integration (\cref{eq:fullloss}). This can be achieved by taking the median and geometric mean of the on-the-fly estimates since they are equal to the true mean $ \exp ( \mu ) $  for log-normally distributed random variables \cite{forbes2011statistical}. We demonstrate this by applying the geometric mean (Figs.~\ref{fig:error_ftsus}(a,b)) and median (Figs.~\ref{fig:error_ftsus}(c,d)) to remove the sampling error in the filtered on-the-fly estimates and the cumulative mean from the BKE--FTS(US) method. 

Furthermore, the geometric mean or median can be used to obtain similar accuracy in the average BKE loss across all batch sizes, as seen in \cref{fig:error_ftsus_final} where we plot the ratio between the FEM value and the geometric mean and median of the on-the-fly estimates from the BKE--FTS(US) and BKE--FTS(US)+SL methods. Note that the ratio obtained from the BKE--FTS(US) method at the smallest batch size is larger than one, in contrast to the expected log-normal prediction that is less than one, but this result is consistent with the presence of the tails in the histograms for the smallest batch size; see Figs.~\ref{fig:energy_log_ftsus_histogram}(a) and \ref{fig:energy_log_ftsus_histogram_sampling}(a). Nevertheless, the accuracy obtained from the smallest batch size after applying the geometric mean and median is comparable to the accuracy obtained from the largest batch size. Thus, one can use the BKE--FTS(US) and BKE-FTS(US)+SL methods to train neural networks with smaller batch sizes, which results in cheaper simulation costs, without loss in the accuracy in the reaction rates estimated on-the-fly. 

\section{Computational Study of a Solvated Dimer System} \label{sec:dimer_solvent}
\begin{figure}[t]
    \centering
    \subfloat{\includegraphics[width=0.47\linewidth]{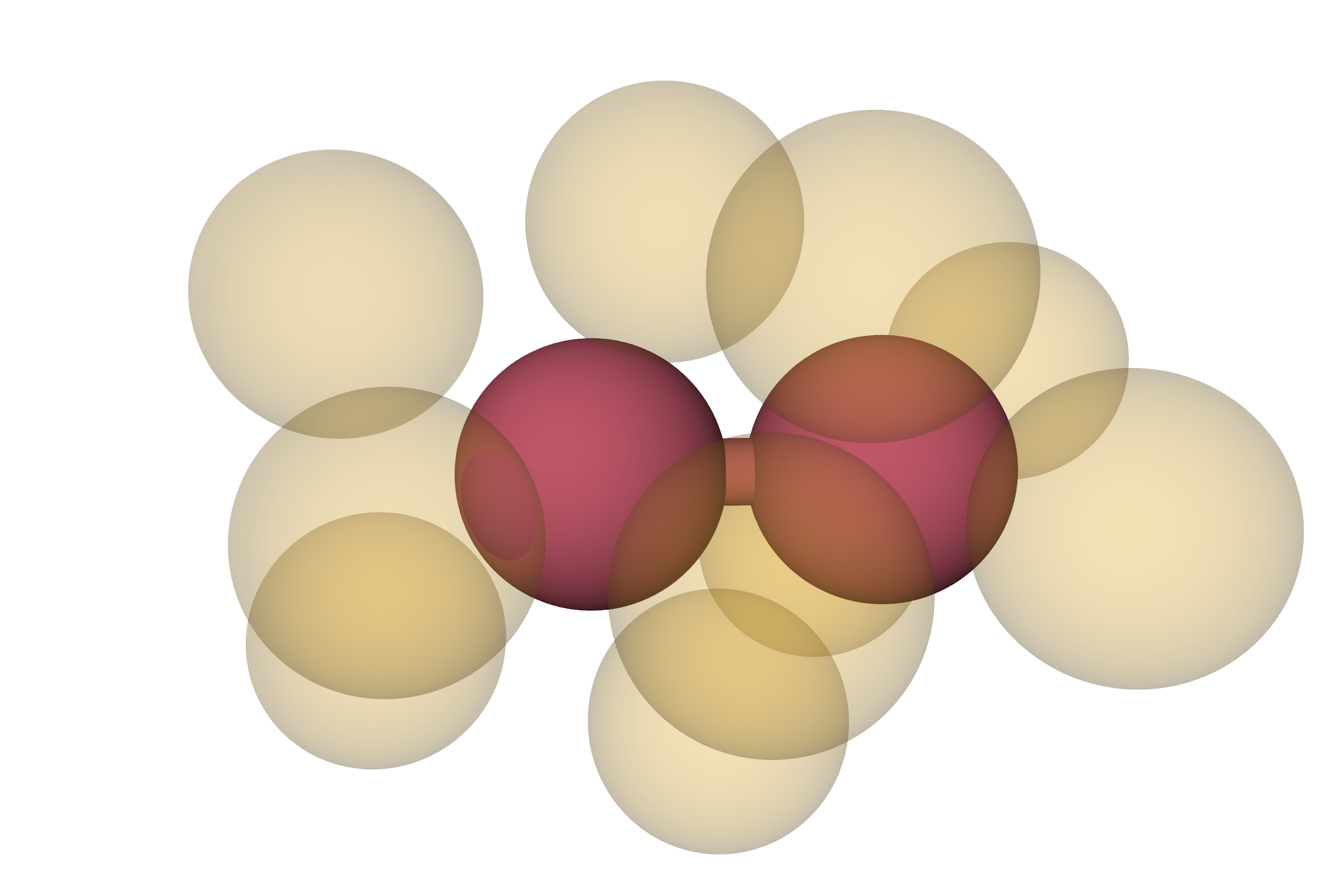}}
    \qquad
    \subfloat{\includegraphics[width=0.47\linewidth]{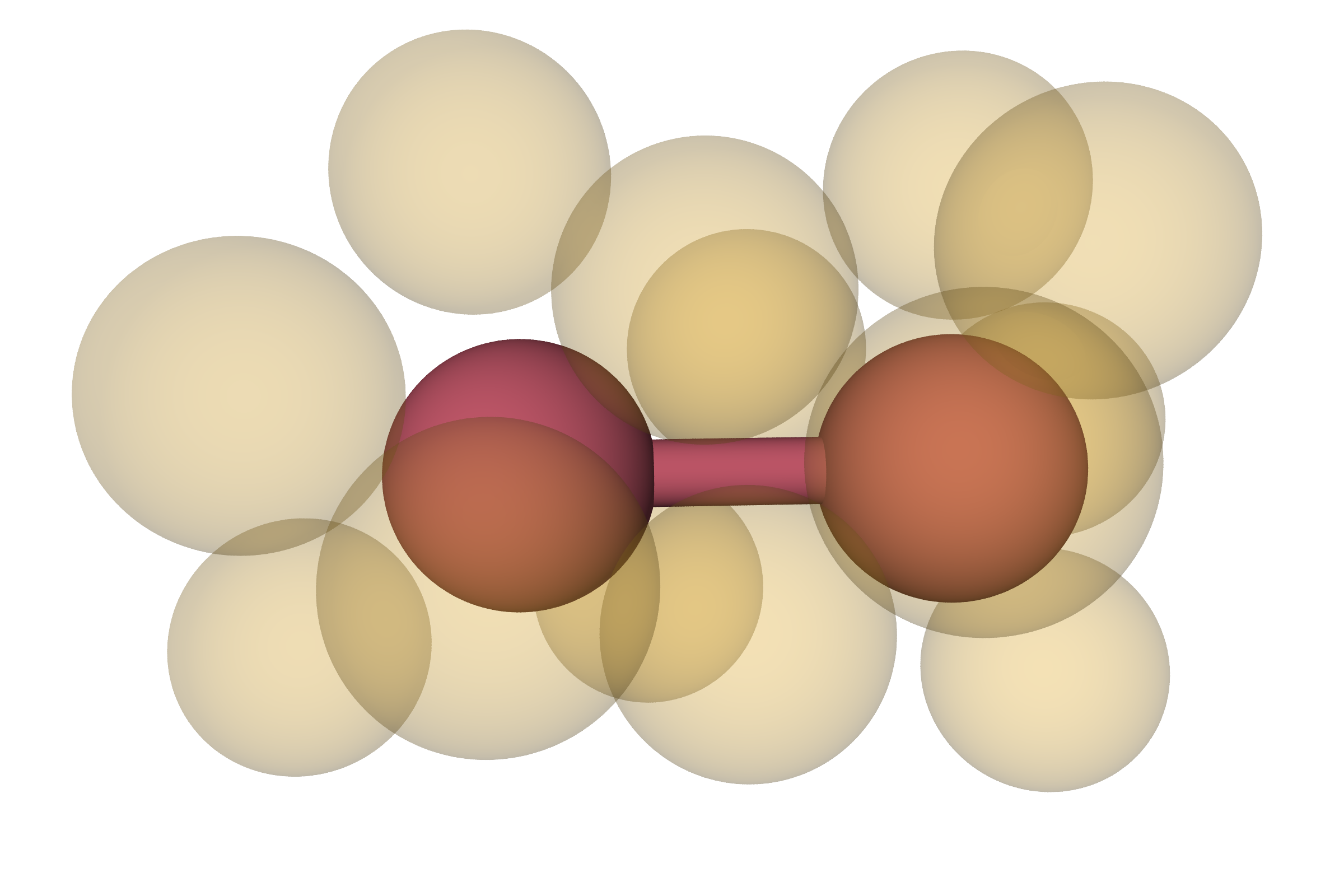}}
    \caption{Dimer particles in (left) compact and (right) extended states for $ r_{0} = 2^{1/6} $ and $ s = 0.25 $ for a system with $ \rho = 0.9 $. Only the seven nearest neighbors of each solvent particle are visualized and made transparent. Image created using Ovito \cite{ovito}.}
    \label{fig:dimer_pics}
\end{figure}

Until now, all previous studies correspond to a single particle diffusing in low-dimensional energy landscapes where a reference solution for $ q(\* x) $ is known through analytical or numerical methods, 
allowing us to understand the accuracy of the proposed methods.
However, the neural network representation of the committor function can also be employed in molecular systems with a high-dimensional configuration space with no reference solution, demonstrating the applicability of the proposed methods. 
To this end, we now test Algorithms~\ref{alg:base}--\ref{alg:baseftsussl} on a solvated dimer system \cite{dellago1999calculation}, where the dimer transitions between a compact and an extended state; see \cref{fig:dimer_pics}.
In what follows, we compute the committor function and reaction rate corresponding to the transition between the compact and the extended states of the dimer.

In this system, the dimer particles interact via a bond potential given by
\begin{align}
    V_{\text{dimer}} \left( r \right) = h \left[ 1 - \frac{\left( r - r_{0} - s \right)^{2}}{s^2} \right]^{2} \,,
\end{align}
where $ r $ is the distance between the particles, $ h = 5.0 \ k_\mathrm{B} T $ is the height of the barrier, $ r_{0} = 2^{1/6} $ sets the distance in the compact state, and $ s= 0.25 $ sets the distance in the extended state.
The distance in the compact state is $ r = r_{0} $, and the distance in the extended state is $ r = r_{0} + 2 s $ (\cref{fig:dimer_pics}).
The solvent particles interact between themselves and the dimer particles by the Weeks-Chandler-Andersen potential \cite{weeks1971WCA}
\begin{align}
    V_{\text{WCA}} \left( r \right) = \left( 4 \epsilon \left[ \left( \frac{1}{r} \right)^{12} - \left( \frac{1}{r} \right)^{6} \right] + \epsilon \right) \Theta \left( r_{\text{WCA}} - r \right) \,,
\end{align}
where $\epsilon=1.0$, $ r_{\text{WCA}} = 2^{1/6} $, and $ \Theta \left( x \right) $ is the Heaviside function.
We test all the methods on systems of densities $ 0.05 $, $ 0.4 $, and $ 0.7 $ with a dimer and $ 30 $ solvent particles, and a system of density $ 0.9 $ with a dimer and $ 46 $ solvent particles. For all systems, the temperature is maintained at  $k_\mathrm{B} T=1$.

In comparison to the low-dimensional systems, molecular systems may have many particles with different species identities.
To increase efficiency in training, the neural network should satisfy invariances with respect to translations, rotations, and permutations of the particle positions $ \* x $ and species identities $ \* z $.
To this end, we use a neural network of the form
\begin{equation}
    \hat{q}(\* x, \* z ;\thetab) = \sigma \left( f(\* x, \* z ; \thetab) \right) \,, \label{eq:schnetnn}
\end{equation}
where the species identities $ \* z $ correspond to $ z = 1 $ for a dimer particle and $ z = 0 $ for a solvent particle, and $ f (\* x, \* z ; \thetab) $ being the implementation of SchNet \cite{schutt2018schnet} available with PyTorch Geometric \cite{fey2019pyg}.
SchNet is a message-passing neural network that determines the contribution to the committor function for each particle, satisfying permutation invariance of the particle identities, using a scheme dependent only on the distances between particles, satisfying the aforementioned translational and rotational invariances.
SchNet first maps for each particle a high dimensional feature vector that is obtained from an embedding of the particle identities.
The feature vectors are then updated using continuous-filter convolutions over the relative distances of a particle to its neighboring particles, which incorporate information about the particle environment; these operations are termed interaction blocks.
The use of the feature vectors and interaction blocks allows for SchNet to learn the effect of particle environments on the per particle contribution to the committor function without the use of handcrafted descriptors.
The feature vectors are then reduced into a scalar per particle contribution to the committor function through a dense neural network, which are summed together and passed through a sigmoid to obtain the neural network representation of the committor function.
In this work, we use a feature vector size of $ 64 $ and $ 3 $ interaction blocks and perform the continuous-filter convolution for each particle over all other particles.
For details on the associated hyper-parameters for each study and parameters used for BKE--US, BKE--FTS(ME), and BKE--FTS(US), see \cref{app:optimizer_dimer}. See also Ref.~\cite{schutt2018schnet} for more details on the general architecture of SchNet and our code repository\footnote[2]{\url{  https://github.com/muhammadhasyim/tps-torch}} for its implementation in this work. 

We apply the same training procedure as done for the 1D and 2D systems with the BKE--US, BKE--FTS(ME), and BKE--FTS(US) methods plus their SL variants, where all methods use $ 24 $ replicas of a batch size of $ 8 $ samples collected every $ 25 $ steps. 
Initial configurations for sampling are obtained using umbrella sampling simulations with respect to the dimer bond distance $r$ with a potential of the form 
\begin{equation}
    W_{\alpha} = \frac{1}{2} \kappa_{\alpha} \left( r - r_{\alpha} \right)^{2} \,, \label{eq:us_bare}
\end{equation}
where $ \kappa_{\alpha} = 1200 \ k_\mathrm{B} T $ and $ r_{\alpha} = 0.75+\frac{1.9-0.75}{31} \left( \alpha - 1 \right) $ for $ \alpha \in \left[ 1 , 32 \right] $.
These simulations generate a set of equilibrium configurations corresponding to the reactant, product, and in-between states.
Furthermore, they are used to initialize the neural network and evaluate the quality of the trained neural network with a fixed data set.
This data set consists of $ 10^4 $ samples per umbrella sampling replica generated from simulations of length $ 10^7 $ time steps with a sampling period of $ 10^3 $ time steps.

The neural network initialization is done through a similar procedure as described in \cref{sec:results}.
The neural network parameters are initialized randomly, and updated by minimizing \cref{eq:choice-initial-minimize} using Adam with a stepsize of $ 1 \cdot 10^{-5} $ until $ I (\thetab) \leq 10^{-4} $.
The initial configurations $\* x_0^\alpha$ are chosen to be the configurations obtained using the above umbrella sampling procedure with bond distances closest to $ r_{\alpha} = 0.98 + \frac{1.75-0.98}{23} (\alpha - 1) $ for $\alpha \in \left[1,24 \right]$.
As in the previous 1D and 2D cases, the BKE--FTS(ME) and BKE--FTS(ME)+SL methods use $\varphib_0^\alpha = \* x_0^\alpha$, and the BKE--FTS(US) and the BKE--FTS(US)+SL methods sets $\* x_0^\alpha$ to be the nodal point $\varphib^\alpha$ of the converged path. 
All additional details related to sampling schemes generating mini-batches for optimization, penalty strengths, and parameters controlling the FTS method can be found in the \cref{app:optimizer_dimer}.

\begin{figure}[t]
    \centering
    \includegraphics[width=\linewidth]{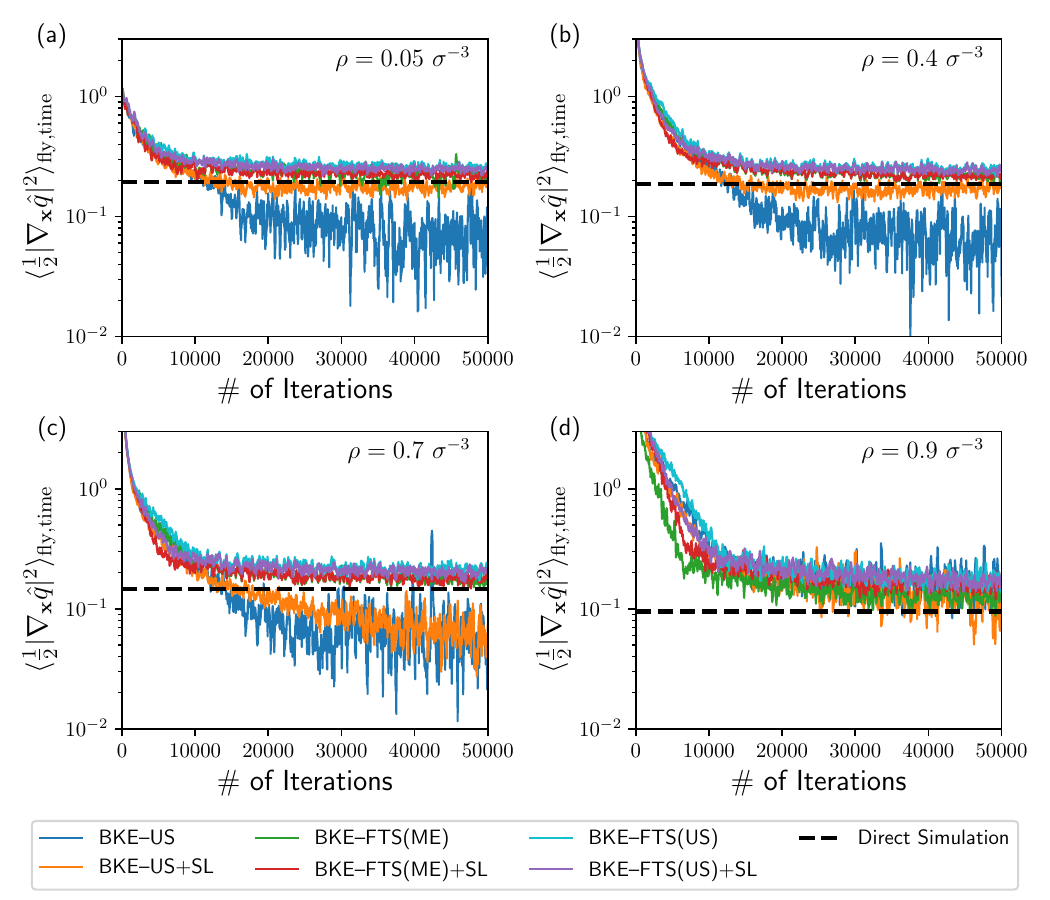}
    \caption{The filtered on-the-fly estimate of the BKE loss obtained at every iteration for the solvated dimer system, with the filtering window set to $ 200 $ iterations. A total of $ 10^{4} $ unbiased trajectories are used to compute a direct estimate of the reaction rate (dashed line) for comparison with the proposed methods.}
    \label{fig:dimer_losses}
\end{figure}

\begin{figure}[htbp!]
    \centering
    \includegraphics[width=\linewidth]{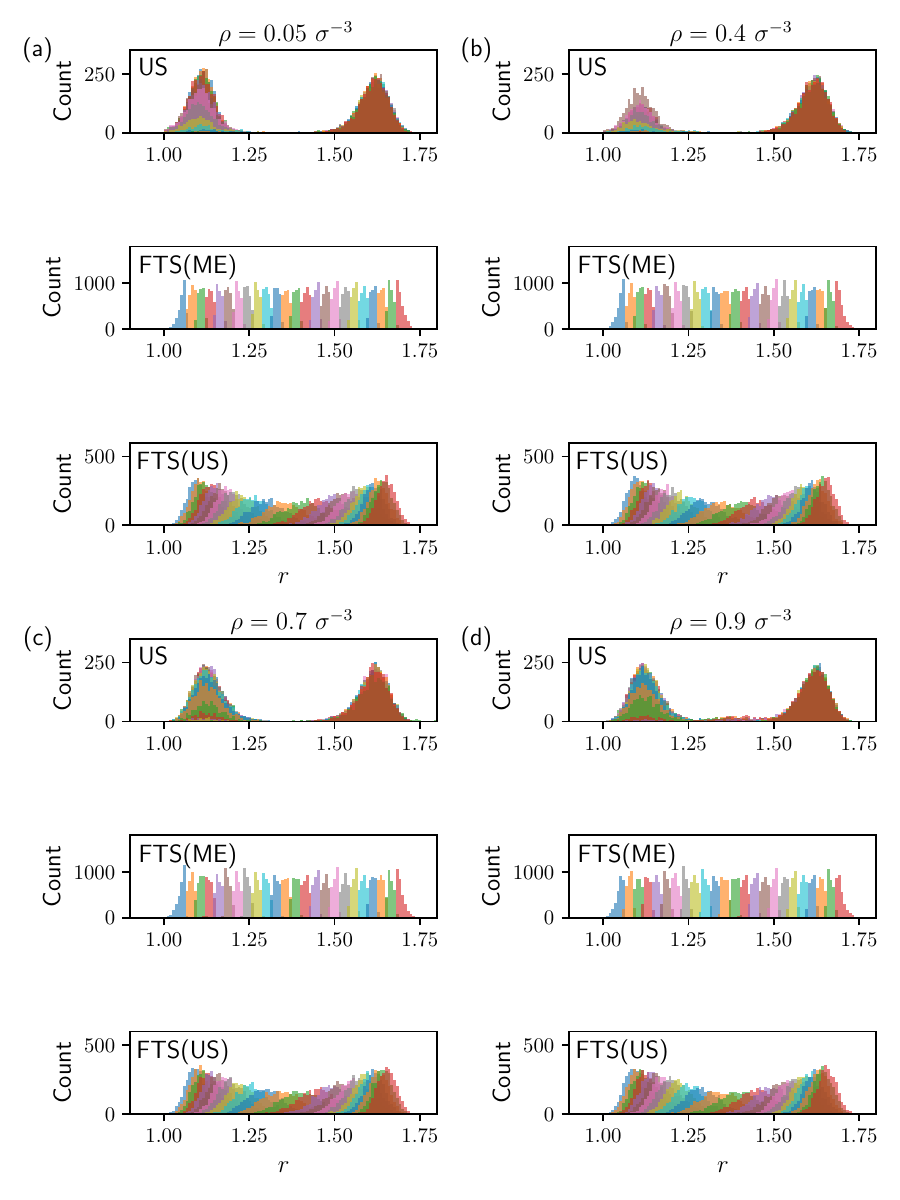}
    \caption{Histograms of dimer distances obtained from the BKE--US method (top), the BKE--FTS(ME) method (middle), and the BKE--FTS(US) method (bottom) for  (a) $ \rho = 0.05  $, (b) $ \rho = 0.4  $, (c) $ \rho = 0.7  $, and (d) $ \rho = 0.9  $.}
    \label{fig:dimer_histograms}
\end{figure}

Figure~\ref{fig:dimer_losses} shows the on-the-fly estimates of the reaction rates or the average BKE loss from all methods tested on various densities for a batch size of $ 8 $ samples.
For densities of $ 0.05 $, $ 0.4 $, and $ 0.7  $ (\cref{fig:dimer_losses}(a-c)), the BKE--FTS(ME) and BKE--FTS(US) estimates plateau around the same value near the estimate obtained from direct simulation, while BKE--US has high variance around a different plateau.
For a density of $ 0.9  $ all methods plateau around the same value.
As with the low-dimensional systems, the BKE--FTS(ME) and BKE--FTS(US) methods sample the reaction pathway, corresponding to dimer distances between the compact and extended states, homogeneously across all densities. In contrast, the BKE--US method does not homogeneously sample the reaction pathway although the transition state is better sampled at $ \rho = 0.9  $ compared to lower densities (\cref{fig:dimer_histograms}). This behavior results in slightly improved overlaps between samples from the reactant/product state and the transition state, which may explain why the reasonable agreement is obtained between the BKE--US method and the direct estimate at $\rho = 0.9$.

\begin{figure}[t]
    \centering
    \includegraphics[width=0.95\linewidth]{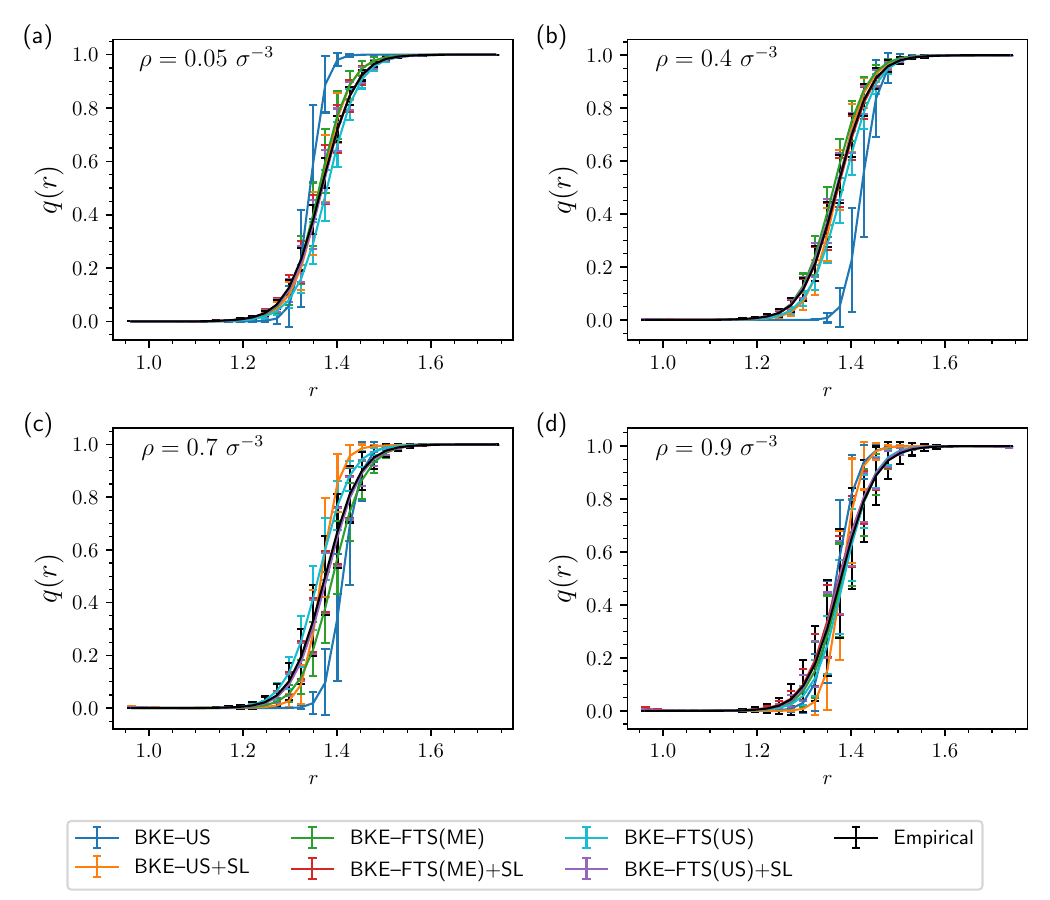}
    \caption{Average committor profiles for the methods compared to the empirical results for dimer in solvent systems. The values are binned for $ 31 $ windows between $ r_\mathrm{min} = 0.95 $ and $ r_\mathrm{max} = 1.75 $.}
    \label{fig:dimer_cp}
\end{figure}
\begin{figure}[t]
    \centering
    \includegraphics[width=0.95\linewidth]{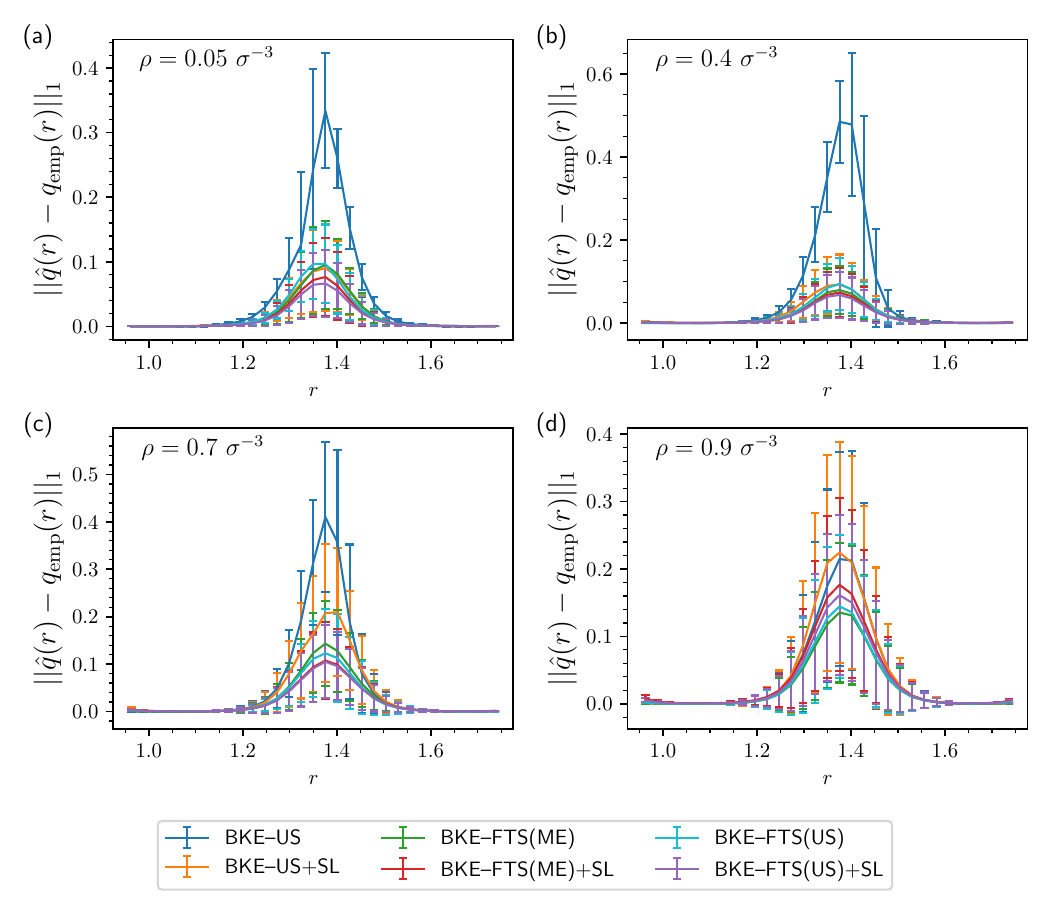}
    \caption{Mean absolute error profiles for the methods compared to the empirical results for dimer in solvent systems. The values are binned for $ 31 $ windows between $ r_\mathrm{min} = 0.95 $ and $ r_\mathrm{max} = 1.75 $.}
    \label{fig:dimer_mae}
\end{figure}

The accuracy of all methods can be assessed by comparing an empirical committor function $q_\mathrm{emp}(r)$ computed at a fixed value of bond length $ r $ with the corresponding value $\hat{q}(r, \* z; \thetab)$ obtained from the neural network. At fixed $r$, the committor values are spread across a distribution since the committor depends not only on $r$ but also on solvent configurations. Thus, both $q_\mathrm{emp}(r)$ and $\hat{q}(r, \* z; \thetab)$ represent estimates of the mean committor at fixed $r$. 
Given the full empirical committor function $q_\mathrm{emp} (\* x )$ (\cref{eq:empiricalcommittor}) and neural network $\hat{q} (\* x, \* z; \thetab)$, we can compute these means via a binning procedure. Letting $ \mathcal{Q}_i $ be a set of configurations such that every $ \* x \in \mathcal{Q}_{i} $ satisfies $ r \in ( r_{i-1} , r_{i} ] $, the binning procedure yields the following formulas: 
\begin{align}
    \hat{q} ( r_{i}, \* z ; \thetab ) &= \frac{1}{| \mathcal{Q}_{i} |} \sum_{\* x \in \mathcal{Q}_{i}} \hat{q} (\* x, \* z; \thetab) \label{eq:committor_nn_binned} \,, \\ 
    q_\mathrm{emp} ( r_{i}) &= \frac{1}{| \mathcal{Q}_{i} |} \sum_{\* x \in \mathcal{Q}_{i}} q_\mathrm{emp} (\* x ) \,, \label{eq:committor_emp_binned}
\end{align}
where every $\* x \in \mathcal{Q}_{i}$ is obtained from the configurations sampled via the umbrella potential in \cref{eq:us_bare} and $q_\mathrm{emp} (\* x) $ is computed using $ 1250 $ trajectories per configuration $\* x$.  
Figure~\ref{fig:dimer_cp} plots $q_\mathrm{emp}(r_i)$ and $\hat{q}(r_i, \* z; \thetab)$ with their respective variances, which represent the intrinsic spread of committor values around their mean at $r=r_i$. 
We see that the BKE--US and BKE--US+SL methods have a systematic difference between the average binned neural network and empirical values. Meanwhile, the
BKE--FTS(ME) and BKE--FTS(US) have a slightly lower systematic difference, which decreases further upon the use of supervised learning.

We further assess the accuracy of all methods by computing the mean of absolute error between the binned values of the neural network committor and the empirical committor, i.e.,
\begin{equation}
    ||\hat{q} ( r_{i} )-q_\mathrm{emp} ( r_{i} ) ||_{1}= \frac{1}{| \mathcal{Q}_{i} |} \sum_{\* x \in \mathcal{Q}_{i}} |\hat{q}(\* x, \* z; \thetab)-q_\mathrm{emp} (\* x)| \,. \label{eq:l1_dimer_binned}
\end{equation}
Figure~\ref{fig:dimer_mae} shows the mean of absolute errors for all densities, where we find that the error is the largest near $q(r)=1/2$. Furthermore, we observe a hierarchy in the reduction of errors. For densities $\rho$ of $ 0.05 \text{--} 0.7  $, the order of methods with increasing accuracy goes as BKE--US < BKE--FTS(ME) < BKE--FTS(US), and the addition of supervised learning improves the accuracy of each respective method. 

\begin{figure}[t]
    \centering
    \includegraphics[width=\linewidth]{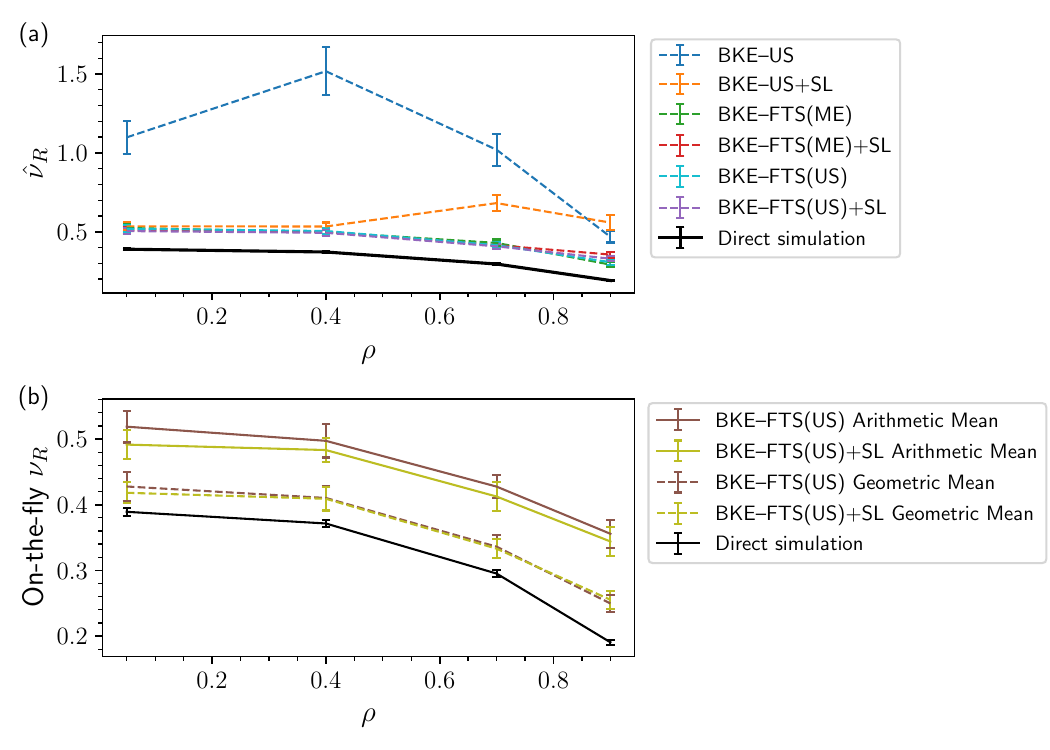}
    \caption{(a) The reaction rate $ \hat{\nu}_{R} $ of the neural network per \cref{eq:rr_dimer} as a function of density. (b) Comparison between the arithmetic mean and geometric mean applied to the last $ 3000 $ samples from training to direct simulation as a function of density. Error bars are 95\ confidence interval.}
    \label{fig:dimer_stats}
\end{figure}

We now assess the accuracy of the methods through the average BKE loss, and thereby the reaction rates.
Unlike the low-dimensional studies, where the average  BKE loss of the neural network can be evaluated via quadrature (\cref{eq:fullloss}), numerically exact calculation is not possible in high-dimensional problems and a new scheme is needed. 
To this end, we choose umbrella sampling with a reweighting procedure to compute the average BKE loss with minimal sampling error. 
This new scheme utilizes the earlier dataset obtained for the initialization of the neural network as a validation dataset, where umbrella sampling with respect to \cref{eq:us_bare} was used to obtain $10^4$ configurations from all $ 32 $ replicas. 
Given this dataset, we compute the reweighting factors $ z_{\alpha} $ using the multistate Bennett acceptance ratio (MBAR) method. Note that MBAR is used instead of FEP since it yields estimates of $z_\mathrm{\alpha}$ with lower error than FEP, albeit at a higher computational cost \cite{shirts2008statistically}.
Once the MBAR reweighting factors $ z_{\alpha}^{\mathrm{MBAR}}$ are computed, the reaction rate from the neural network can be estimated from a modification of \cref{eq:flyloss} for umbrella sampling,
\begin{equation}
    \hat{\nu}_R = \left\langle |\nabla_{\* x} \hat{q}(\* x; \thetab) |^2\right\rangle = \frac{\mathlarger{\sum}\limits_{\alpha=1}^{32}  \dfrac{2 z_{\alpha}^{\mathrm{MBAR}}}{|\mathcal{M}^\alpha|}\mathlarger{\sum}\limits_{\* x \in \mathcal{M}^\alpha}  \left[\dfrac{\ell(\* x; \thetab)}{c(\* x; \thetab)} \right]  }{ \mathlarger{\sum}\limits_{\alpha=1}^{32} \dfrac{z_{\alpha}^{\mathrm{MBAR}}}{|\mathcal{M}^\alpha|}\mathlarger{\sum}\limits_{\* x \in \mathcal{M}^\alpha} \left[\dfrac{1}{c(\* x; \thetab) } \right]  } \,. \label{eq:rr_dimer}
\end{equation}
Evaluating \cref{eq:rr_dimer} produces the results seen in Fig.~\ref{fig:dimer_stats}(a), which are compared to the true reaction rate as estimated by direct molecular simulation.
The results in Fig.~\ref{fig:dimer_stats}(a) mirror the trends seen in \cref{fig:dimer_mae}.

As established by the error analysis in \cref{sec:lognorm}, we may avoid costly computation in \cref{eq:rr_dimer} for the BKE--FTS(US) and BKE--FTS(US)+SL methods via the geometric-mean estimate to eliminate sampling error at low batch sizes. The comparison between the arithmetic and geometric mean on the on-the-fly estimates, taken from the last portion of training, is shown in \cref{fig:dimer_stats}(b). 
Similar to the low-dimensional case, the geometric mean is able to recover estimates of the reaction rate closer to the true reaction rate than the arithmetic mean, demonstrating the generality of the results from the error analysis. 
Furthermore, the trend between the geometric mean agrees reasonably well with the true reaction rate across all densities. This result supports the points made in \cref{sec:bkeftsdetail} that the BKE--FTS methods are able to account for solvent effects despite using a CV that ignores solvent configurations and thereby predicting the correct trend of the reaction rate as a function of density.

\section{Conclusion \& Future Work}

In summary, building on the work of Ref.~\cite{rotskoff2020learning}, we have introduced and discussed a set of ML-based algorithms for computing accurate and precise committor functions and reaction rates. 
Accuracy in computing committor functions is improved by adding elements of supervised learning, where committor values obtained from short molecular trajectories are used to improve the neural network training. 
On the other hand, accuracy in the estimated reaction rates is significantly improved by incorporating the FTS method, which allows homogeneous sampling across the transition tube necessary for obtaining accurate free energies and reweighting factors. 
Furthermore, for the FTS method via path-based umbrella sampling as in the BKE--FTS(US) and BKE--FTS(US)+SL method, we provide an error analysis, which shows that the on-the-fly estimates of the average BKE loss obey log-normal statistics. 
This analysis also shows that the sampling error in the on-the-fly estimates of reaction rates can be removed by computing its geometric mean or median. 
The different combinations of supervised learning and the FTS method yield five additional algorithms, which were tested against three model systems. 
Out of the six algorithms, we recommend the BKE--FTS(US)+SL method, which combines all the strengths of supervised learning and the FTS method, in conjunction with the geometric mean/median procedure that allows accurate and precise computation of reaction rates with a small number of samples, e.g., batch size of $O(10^1)$.

Future work involves investigating ways of  further increasing the accuracy of the methods on molecular systems.
The accuracy could likely be increased through the use of an equivariant neural network \cite{thomas2018tensor}, with neural networks satisfying equivariance throughout the hidden layers having been shown to yield increased accuracy in predictions of molecular properties over SchNet \cite{batzner2021e3equivariant}.
Future work should also explore other model systems ranging from ionic association/dissociation in $\mathrm{Na Cl}$ solution, where the transition pathway involves the association/dissociation of $\mathrm{Na}^+$--$\mathrm{Cl}^-$ ionic pairs \cite{Laio12562,geissler1999kinetic,geissler2001autoionization,ballard2012toward}, to excitation events in glassy systems, where the transition state is known to have elastic signatures that are crucial for the structural relaxation \cite{hasyim2021theory}. 



\section*{Acknowledgments}

We thank Professors Benjamin Recht and Moritz Hardt for helpful discussions about machine learning methodologies, and the use of supervised learning elements in reinforcement learning. 
We also thank Chloe Hsu for insightful comments.
This work is supported by Director, Office of Science, Office of Basic Energy Sciences, of the U.S. Department of Energy under contract No. DEAC02-05CH11231.

\section*{Data Availability}
The data that support the findings of this study are available
from the corresponding author upon reasonable request

\printbibliography

\newpage 

\appendix

\section{Computing Reweighting Factors in the Master-Equation Approach} \label{app:freenergymethods}

Recall that the reweighting factor $z_\alpha$ in the BKE--FTS(ME) and BKE--FTS(ME)+SL methods are computed by solving the master equation \cref{eq:ftsbalanceeq}. 
One can re-write \cref{eq:ftsbalanceeq} as a matrix equation:
\begin{equation}
 \* K \* z = \* 0 \,, \label{eq:ftsbalancematrix}
\end{equation}
where $\* z = z_\alpha \* e_\alpha$, $\* K = K_{\alpha \alpha^\prime} \* e_\alpha \otimes \* e_{\alpha^\prime}$, and $K_{\alpha \alpha^\prime} = k^T_{\alpha \alpha^\prime}$ for $\alpha \neq \alpha^\prime$ and $K_{\alpha \alpha} = -\sum_{\alpha^\prime} k_{\alpha \alpha^\prime}$. 
Since \cref{eq:ftsbalancematrix} defines $\* z$ as the basis vector of the null-space of $\* K$, one can use singular value decomposition (SVD) to factorize $\* K = \* U \bm \Sigma \* V^T$, and set the solution $\* z$ as the column vector of $\* V$ corresponding to the zero singular value. 
One can then normalize the vector $\* z$ to satisfy the constraint $\sum_{\alpha=1}^M z_\alpha = 1$. 

In extremely short simulation runs, the off-diagonals of $N_{\alpha \alpha^\prime}$ can be zero due to the absence of rejection counts, which may result in estimates of $z_\alpha$, i.e., elements of column vector of $\* V$, that are not strictly positive. 
To ensure that the algorithm computes the correct column vector, we shift the off-diagonals $k_{\alpha (\alpha+1)}$ and  $k_{(\alpha-1)\alpha}$ by a tolerance value of $2 \cdot 10^{-9}$, i.e., slightly lower than the machine epsilon of single-precision floats, and set the tolerance for zero singular-value detection to be $10^{-6}$. For this choice of tolerance values, the estimated $z_\alpha$ converge in the limit of large batch sizes to the $z_\alpha$ computed by numerical integration of \cref{eq:zalphafts}; see \cref{fig:ftsme_z_l_norm}. Note that a range of tolerance values $10^{-11}$--$10^{-8}$ have also been used with no change to the results. 

\section{Computational Details on Optimization and Sampling} \label{app:optimizer}

In this section, we provide additional details relevant to both the sampling and optimization steps of all algorithms. 
The simulation of multiple replicas are distributed with MPI and interfaced with PyTorch \cite{NEURIPS2019_9015} for performing optimization\footnote[2]{\url{  https://github.com/muhammadhasyim/tps-torch}}.

\subsection{First Study: 1D Quartic Potential}

In the first study, umbrella sampling is performed with $M=20$ replicas with dynamics described by the overdamped Langevin dynamics in \cref{eq:ussamplinglangevin}. 
For committor-based umbrella sampling, bias potential parameters for each replica are set to $\kappa_\alpha= 50$ and $q_\alpha = \frac{\alpha-1}{M-1}$. For the path- or string-based umbrella sampling, the bias strength $\kappa_\alpha^\parallel=5$, and the choice of  $\kappa_\alpha^\bot$ is irrelevant since there is no perpendicular direction in 1D. The transition path used as input for the path-based umbrella sampling is obtained by running the FTS method up to 100 iterations.
Note that the FTS method is also performed with the same number of replicas, but with dynamics described by Eqs.~\eqref{eq:ftssamplinglangevin}-\eqref{eq:ftsrejection}. 
In all algorithms, the friction coefficient $\gamma=1$, and step size $\Delta t = 0.005$. 
The size of $\alpha$-th batch at every iteration is set to $|\mathcal{M}_k^\alpha|=16$ and $|\mathcal{R}_k^\alpha|=16$ for methods employing umbrella sampling and the FTS method, respectively. 
Each sample $\* x \in \mathcal{M}_k^\alpha$ and $\* x \in \mathcal{R}_k^\alpha$ is collected every 25 timesteps.

For the supervised learning component, the penalty strength $\lambda_\mathrm{SL} = 100$ at all iterations, and empirical committor values are collected at every 40 iterations of the algorithm, i.e., $\tau_\mathrm{emp} = 40$. 
The initial and final iteration index are set to $k_{\mathrm{emp},s}=10$ and $k_{\mathrm{emp},f}=2500$, respectively. The number of trajectories for each replica is $H = 100$. The size of $\alpha$-th mini-batch is $|\mathcal{C}_k^\alpha| = 0.5 |\mathcal{C}^\alpha|$, and thus the size of mini-batch during iterations grows as more samples are stored into $\mathcal{C}^\alpha$

For the boundary conditions, the penalty strengths are $\lambda_\mathrm{A}=\lambda_\mathrm{B}=10^4$. 
The reactant and product batches $\mathcal{A}$ and $\mathcal{B}$ are collected prior to the start of each algorithm using dynamics given by Eqs.~\eqref{eq:ftssamplinglangevin}-\eqref{eq:ftsrejection}, but with $R_\alpha$ replaced with $A$ and $B$, respectively. 
The size $|\mathcal{A}|=|\mathcal{B}|=250M$ and each sample is also collected every 100 timesteps. 
During optimization, the mini-batch is randomly sampled without replacement from the original batch $\mathcal{A}$ and $\mathcal{B}$, where the size $|\mathcal{A}_k|=|\mathcal{B}_k|=125M$. 

The chosen optimizer to train the neural network is the Heavy-Ball method \cite{polyak1964some}, which takes in two hyper-parameters as inputs. 
The first is the step size/learning rate $\eta$, while the second is the momentum coefficient $\mu$. 
Given any function $f(\thetab)$  to minimize, the Heavy-Ball method updates model parameters $\thetab_{k}$ with the following equation:
\begin{align}
    \* m_{k+1} &= \mu \* m_{k}+\nabla_{\thetab} f(\thetab_k) \,, \label{eq:hbupdate} 
\\
    \thetab_{k+1} &= \thetab_k-\eta \* m_{k+1} \,,
\end{align}
where $\* m_0 = \* 0$. 
Note that our notation is consistent with PyTorch's implementation of the Heavy-Ball method. 
For all methods, $\eta = 5 \cdot 10^{-4}$ and $\mu = 0.95$.
The gradient $\nabla_{\thetab} f(\thetab)$ in \cref{eq:hbupdate} corresponds to, e.g., \cref{eq:basemethod_grad} for the BKE--US and BKE--FTS(US) method, and \cref{eq:basefts_grad} for the BKE--FTS(ME) method, with additional mini-batches used as inputs to the gradient computation. 

For the FTS method, the penalty strength is set to $\lambda_\mathrm{S} = 0.1M$, where $M=20$ is the number of replicas. 
In addition, we replace the SGD step in \cref{eq:sgd_fts} with a momentum-variant called the Nesterov's method \cite{nesterov1983method}. 
As implemented in PyTorch, which follows the simplified version in \cite{bengio2013advances}, the Nesterov's update can be written as
\begin{align}
    \* m_{k+1} &= \mu^2 \* m_k +(1+\mu) \nabla_{\varphib^\alpha} \hat{C}(\{ \varphib_{k}^\alpha \}) \,,
\\
    \varphib_{\star}^\alpha &= \varphib_{k}^\alpha - \Delta \tau \* m_{k+1} \,, 
\end{align}
where $\* m_0 = \* 0$. 
We set the step size/learning rate $\Delta \tau= 10^{-2}$ and momentum coefficient $\mu = 0.9$. 

\subsection{Second Study: Muller-Brown Potential} \label{app:optimizer_mb}

In the second study, umbrella sampling is performed with $M=24$ replicas with dynamics given by Metropolis Monte Carlo \cite{metropolis1949monte,frenkel2001understanding}.
The particle is displaced in both directions by a random value between $ - \Delta r $ and $ \Delta r $ to yield a new position $ \* x' $, which is accepted with probability given by $P_{\text{acc}} = \min \left[ 1 , \exp \left( -\beta ( V_\mathrm{MB}(\* x') - V_\mathrm{MB}(\* x) ) \right)  \right] $.
The value of $ \Delta r $ is $ 0.05 $ when generating the batches $ \mathcal{M}_{k}^{\alpha} $ for umbrella sampling, $\mathcal{R}_{k}^{\alpha} $ for the FTS method, and $ \mathcal{C}_{k}^{\alpha} $ for supervised learning, while it is set to $ 0.01 $ for sampling the reactant and product states. For committor-based umbrella sampling, $q_\alpha $ is set to be $ \frac{\alpha-1}{M-1}$ and $ \kappa_\alpha = 10000 $ for all $ \alpha $. For the path- or string-based umbrella sampling, we choose bias strength $\kappa_\alpha^\parallel =1100$ and $\kappa_\alpha^\bot=600$ and the transition path used as input is obtained by running the FTS method up to 100 iterations.
Note the FTS method also uses the same amount of replicas as umbrella sampling, and the Monte Carlo method to sample configurations inside the Voronoi cells.
For \cref{fig:2dresults,fig:mb_metric}, the size of $\alpha$-th batch at every iteration is set to $|\mathcal{M}_k^\alpha|=16$ and $|\mathcal{R}_k^\alpha|=4$ for methods employing umbrella sampling and the FTS method, respectively. For \crefrange{fig:errorallmethods}{fig:summary_norm_z}, we use a list of batch sizes $[4,16,64,256,1024]$, where each sample $\* x \in \mathcal{M}_k^\alpha$ and $\* x \in \mathcal{R}_k^\alpha$ is collected every $ 25 $ timesteps.

For the supervised learning component, the penalty strength is set to $\lambda_\mathrm{SL} = 100$ initially.
Beginning at iteration $ 300 $, $ \lambda_\mathrm{SL} $ is increased linearly to $ 25000 $ at iteration $ 10000 $.
Empirical committor values are collected at every 10 iterations of the algorithm, i.e., $\tau_\mathrm{emp} = 10$.
The initial and final iteration index where we start and end supervised learning is set to $k_{\mathrm{emp},s}=10$ and $k_{\mathrm{emp},f}=1000$.  
The number of trials for every window $H = 100$. 
The size of $\alpha$-th mini-batch is $|\mathcal{C}_k^\alpha| = 0.5 |\mathcal{C}^\alpha|$, and thus the size of mini-batch we use during iterations again grows as more samples are stored into $\mathcal{C}^\alpha$ as in the 1D case.

For the boundary conditions, the penalty strengths $\lambda_\mathrm{A}=\lambda_\mathrm{B}=10^4$. 
The reactant and product batches $\mathcal{A}$ and $\mathcal{B}$ are collected before the start of each algorithm with dynamics confined to regions $A$ and $B$, respectively. The size of the number of samples is $|\mathcal{A}|=|\mathcal{B}|=100M$ and each sample is also collected every 10 timesteps. The mini-batch is randomly sampled without replacement from the original batch $\mathcal{A}$ and $\mathcal{B}$ with $|\mathcal{A}_k|=|\mathcal{B}_k|=50M$. 

The optimizer used to train the neural network in MB systems is Adam \cite{kingma2014adam}, which takes in four hyper-parameters as inputs: the first is the step size/learning rate $\eta$, the second and third are momentum coefficients $\beta_1 $ and $ \beta_2$ that control the change in the momentum and momentum squared respectively, and the fourth parameter $ \epsilon $ is a term added to improve numerical stability. For any function $f(\thetab)$  being optimized, the Adam update of model parameters $\thetab_{k}$ can be written as
\begin{align}
    \* m_{k+1} &= \beta_1 \*  m_k+(1-\beta_1) \nabla_{\thetab} f(\thetab_k) \,,
\\ \label{eq:adam_begin}
    \* v_{k+1} &= \beta_2  \* v_k+(1-\beta_2) \left[ \nabla_{\thetab} f(\thetab_k) \odot \nabla_{\thetab} f(\thetab_k) \right] \,,
\\
    \hat{\*  m}_{k+1} &= \frac{\*  m_k}{1-(\beta_1)^k} \,,
\\
    \hat{\* v}_k &= \frac{\* v_k}{1-(\beta_2)^k}  \,,
\\
    \* H_{k+1} &= \mathrm{diag}\left[\sqrt{\hat{\* v}_k } \right]+\epsilon \*  \,, \label{eq:hkmatrix}
\\
    \thetab_{k+1} &= \thetab_{k}-\eta (\* H_{k+1})^{-1}\hat{\* m}_{k+1}  \,, \label{eq:adam_end}
\end{align}
where $\odot$ is the element-wise product between two vectors that yields a new vector of the same dimension, the square root in \cref{eq:hkmatrix} is applied element-wise to the vector, $\mathrm{diag}\left[\ldots \right]$ is a diagonal matrix obtained from elements of an input vector. Initially $\* m_0 = \* 0$ and $\* v_0 = \* 0$. Note that our notation is consistent with PyTorch's implementation of Adam, and for all methods, $\eta = 1 \cdot 10^{-3}$, $\beta_1 = 0.9 $, $\beta_2 = 0.999 $, and $ \epsilon = 10^{-8} $. 

For the FTS method, the penalty strength for the M{\"u}ller-Brown potential is set to $\lambda_\mathrm{S} = 0.1M$. The previously mentioned Nesterov's update scheme is also used here, with the step size/learning rate $\Delta \tau= 0.05$ and momentum coefficient $\mu = 0.9$. 

\subsection{Third Study: Solvated Dimer} \label{app:optimizer_dimer}

In the third study, umbrella sampling is performed with $M=24$ replicas with dynamics described by the overdamped Langevin dynamics in \cref{eq:ussamplinglangevin}.
For committor-based umbrella sampling, bias potential parameters for each replica are set to $\kappa_\alpha= 100$ for $ \rho = 0.05 $, $ 0.4 $, and $ 0.7 $ and $\kappa_\alpha = 50$ for $ \rho = 0.9$,  and $q_\alpha = \frac{\alpha-1}{M-1}$ for all densities. 
For the path- or string-based umbrella sampling, the bias strength $\kappa_\alpha^\parallel=\kappa_\alpha^\bot=1200$.
The transition path used as input for the path-based umbrella sampling is obtained by running the FTS method to 20000 iterations.
Note that the FTS method is also performed with the same number of replicas, but with dynamics described by Eqs.~\eqref{eq:ftssamplinglangevin}-\eqref{eq:ftsrejection}. 
In all algorithms, the friction coefficient $\gamma=1$, and step size $\Delta t = 0.0001$. 
The size of $\alpha$-th batch at every iteration is set to $|\mathcal{M}_k^\alpha|=8$ and $|\mathcal{R}_k^\alpha|=8$ for methods employing umbrella sampling and the FTS method, respectively. 
Each sample $\* x \in \mathcal{M}_k^\alpha$ and $\* x \in \mathcal{R}_k^\alpha$ is collected every 25 timesteps.

For the supervised learning component, the penalty strength is set to $\lambda_\mathrm{SL} = 100$ initially.
Beginning at iteration $ 200 $, $ \lambda_\mathrm{SL} $ is increased linearly to $ 1000 $ at iteration $ 10000 $.
Empirical committor values are collected at every 10 iterations of the algorithm, i.e., $\tau_\mathrm{emp} = 10$.
The initial and final iteration index where we start and end supervised learning is set to $k_{\mathrm{emp},s}=10$ and $k_{\mathrm{emp},f}=5000$.  
The number of trials for every window $H = 100$. 
The size of $\alpha$-th mini-batch is $|\mathcal{C}_k^\alpha| = 0.5 |\mathcal{C}^\alpha|$, and thus the size of mini-batch we use during iterations again grows as more samples are stored into $\mathcal{C}^\alpha$ as in the 1D and 2D cases.

For the boundary conditions, the penalty strengths $\lambda_\mathrm{A}=\lambda_\mathrm{B}=10^4$. 
The reactant and product batches $\mathcal{A}$ and $\mathcal{B}$ are collected before the start of each algorithm with dynamics confined to regions $A$ and $B$, respectively. The size of the number of samples is $|\mathcal{A}|=|\mathcal{B}|=100M$ and each sample is also collected every 10 timesteps. The mini-batch is randomly sampled without replacement from the original batch $\mathcal{A}$ and $\mathcal{B}$ with $|\mathcal{A}_k|=|\mathcal{B}_k|=50M$. 

The optimizer used to train the neural network in dimer systems is Adam as previously described in \cref{app:optimizer_mb}.
Initially $\* m_0 = \* 0$ and $\* v_0 = \* 0$. 
For all densities and methods, $\eta = 1 \cdot 10^{-5} $, $\beta_1 = 0.9 $, $\beta_2 = 0.999 $, and $ \epsilon = 10^{-8} $.

For the FTS method, the penalty strength for the M{\"u}ller-Brown potential is set to $\lambda_\mathrm{S} = 0.1M$. 
The previously mentioned Nesterov's update scheme is also used here, with the step size/learning rate $\Delta \tau= 0.001$ and momentum coefficient $\mu = 0.9$. 

\section{Comments on the Supervised Learning Loss Function} \label{app:sl_loss}

In this section, we compare the results from the supervised learning scheme used in this work with that of the more standard scheme seen in the literature \cite{hardtrecht}, which utilizes the mean-squared error (MSE) loss function given by \cref{eq:loss_mse} instead of the supervised-learning loss given by \cref{eq:loss_sl}. Switching the supervised-learning loss yields new algorithms denoted as the BKE--US+MSE, BKE--FTS(ME)+MSE, and the BKE--FTS(US)+MSE methods.
The procedure for training the neural network follows that described in \cref{app:optimizer_mb}, except for the BKE--US+MSE method where $ \lambda_\mathrm{MSE} $ is increased linearly to $ 2500 $. 

\begin{figure}[t]
    \centering
    \includegraphics[width=0.975\linewidth]{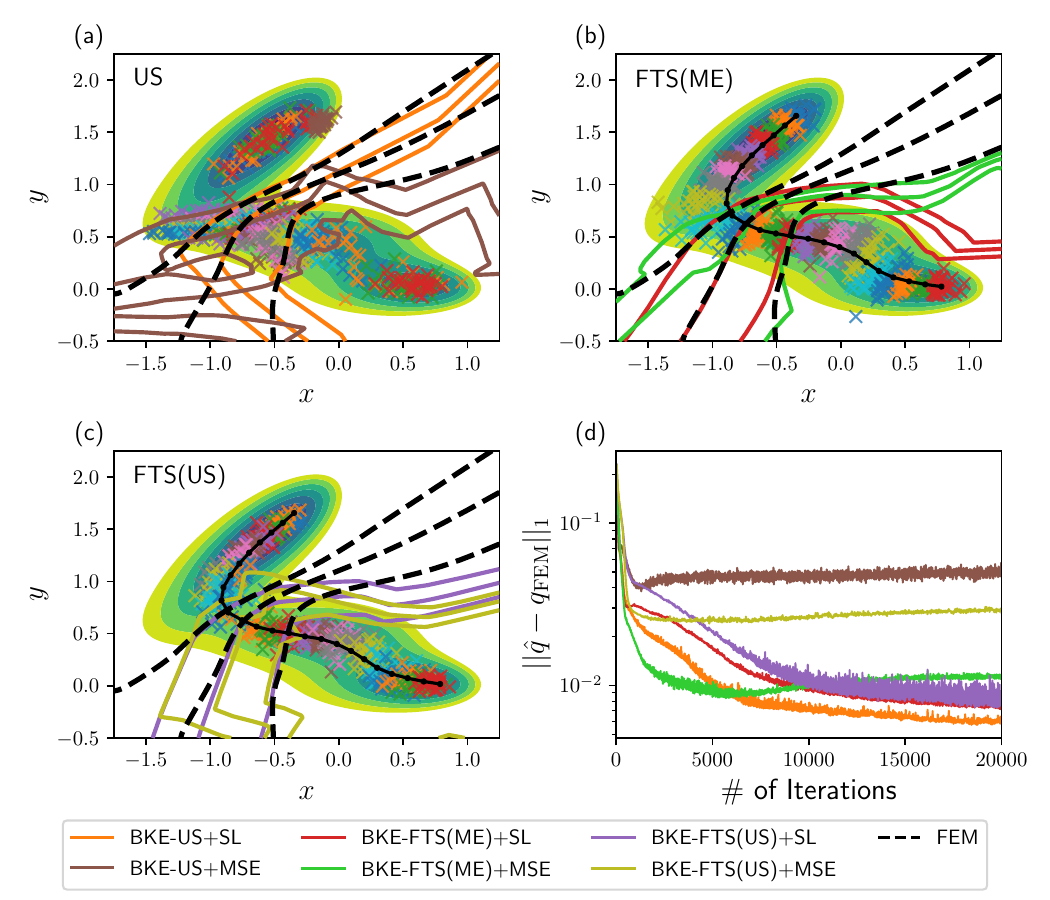}
    \caption{Isocommittor lines for $ q = 0.1 $, $ 0.5 $, and $ 0.9 $ from (a) the BKE--US+SL and BKE--US+MSE method, (b) the BKE--FTS(ME)+SL and BKE--FTS(ME)+SL method, (c) the BKE--FTS(US) and BKE--FTS(US)+MSE method. $\times$ markers denote representative samples obtained from algorithms the SL methods. (d) The $L_1$-norm error as a function of iterations.}
    \label{fig:2dresults_sl}
\end{figure}

\begin{figure}[t]
    \centering
    \includegraphics[width=0.95\linewidth]{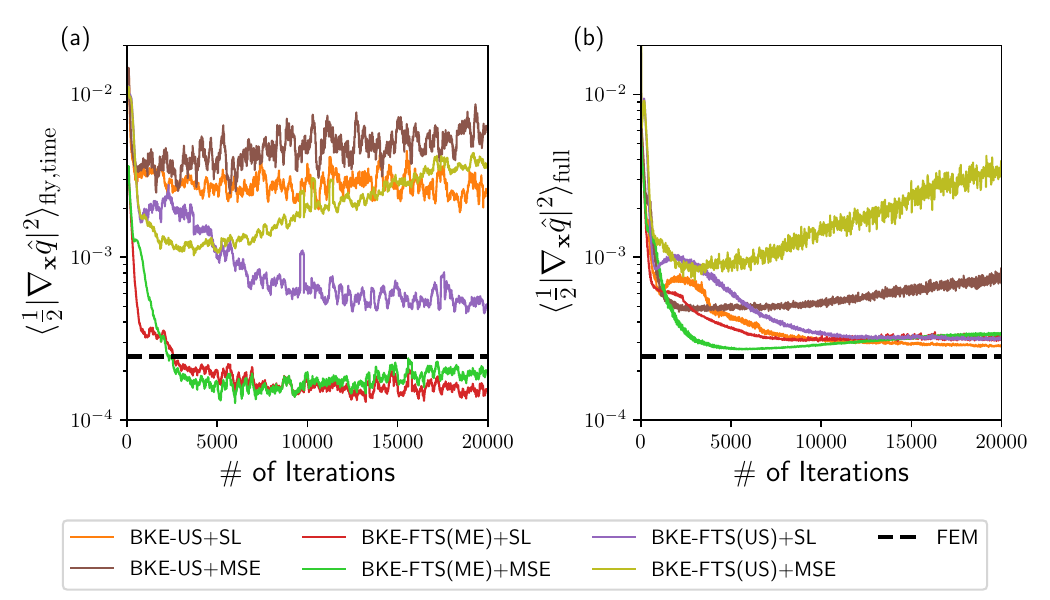}
    \caption{(a) The filtered on-the-fly estimate of the BKE loss obtained at every iteration, with the filtering window set to 200 iterations. (b) The ensemble-averaged loss per \cref{eq:fullloss} obtained at every iteration.}
    \label{fig:mb_metric_sl}
\end{figure}

The results demonstrate that the use of the MSE loss function yields worse accuracy, as shown in the isocommittor lines and $L_1$-norm error in \cref{fig:2dresults_sl}. In fact, the $L_1$-norm error of all methods employing the MSE loss function increases at later iterations. Furthermore, with the exception of the BKE--FTS(ME)+MSE method, both the on-the-fly estimates (\cref{fig:mb_metric_sl}(a)) and ensemble-averaged BKE loss function (\cref{fig:mb_metric_sl}(b)) increase at higher iterations. This suggests that the MSE loss function is prone to overfitting \cite{hardtrecht}, and we provide a sketch for why this occurs. 
To this end, the gradients of the losses with respect to the neural network parameters are evaluated.
For the MSE loss function in \cref{eq:loss_mse} this is
\begin{align}
    \nabla_{\thetab} \hat{L}_{\mathrm{MSE}}(\thetab; \{ \mathcal{C}^\alpha_k \}) &=  \frac{\lambda_{\mathrm{MSE}}}{M} \sum_{\alpha=1}^M \frac{1}{|\mathcal{C}^\alpha_k|} \sum_{(q_{\mathrm{emp}}, \* x) \in \mathcal{C}^\alpha_k}\nabla_{\thetab}  \ell_{\mathrm{MSE}}(q_\mathrm{emp}, \* x; \thetab) 
    \\
    &=\frac{\lambda_{\mathrm{MSE}}}{M} \sum_{\alpha=1}^M \frac{1}{|\mathcal{C}^\alpha_k|} \sum_{(q_{\mathrm{emp}}, \* x) \in \mathcal{C}^\alpha_k} \left( \hat{q}(\* x; \thetab) - q_\mathrm{emp} \right) \nabla_{\thetab} \hat{q}(\* x ; \thetab) \,. \label{eq:grad_mse}
\end{align}
Note that the error $ \hat{q}(\* x; \thetab) - q_\mathrm{emp} $ for every $\* x$ is correlated point-wise with the model's gradient $ \nabla_{\thetab} \hat{q}(\* x ; \thetab)$, which causes large point-wise errors to have more weight in the gradient descent direction. If the global minimum is reached, this results in fitting every datapoint in $\* x$ perfectly, despite the statistical noise in the data. 
In comparison the gradient of the supervised-learning loss \cref{eq:loss_sl} is
\begin{align}
    \nabla_{\thetab} \hat{L}_\mathrm{SL}(\thetab; \{ \mathcal{C}_k^\alpha \}) &= \frac{\lambda_{\mathrm{SL}}}{M} \sum_{\alpha=1}^M \nabla_{\thetab} \ell_{\mathrm{ME}}(\mathcal{C}_{k}^{\alpha}; \thetab) \\ 
                                                                              &= \frac{\lambda_{\mathrm{SL}}}{M} \sum_{\alpha=1}^M \left[ \frac{1}{|\mathcal{C}_{k}^{\alpha}|} \! \sum_{ ( q_\mathrm{emp}, \* x ) \in \mathcal{C}_{k}^{\alpha}} \!\!\!\! (\hat{q}(\* x; \thetab) - q_\mathrm{emp}) \right] \left[ \frac{1}{|\mathcal{C}_{k}^{\alpha}|} \! \sum_{ ( q_\mathrm{emp}, \* x ) \in \mathcal{C}_{k}^{\alpha}} \!\!\!\! \nabla_{\thetab} \hat{q}(\* x; \thetab) \right] \label{eq:grad_mes}  \,,
\end{align}
where the error $ \hat{q}(\* x; \thetab) - q_\mathrm{emp} $ and model gradient $  \nabla_{\thetab} \hat{q}(\* x ; \thetab) $ are now individually averaged with respect to samples in $ \mathcal{C}_{k}^{\alpha} $. This averaging is crucial as it reduces the statistical noise in the empirical committor function $q_\mathrm{emp}(\* x)$.
To see this, we first write $ q_{\mathrm{emp}}(\* x) $ in terms of the exact committor function $ q (\* x) $ as
\begin{equation}
    q_{\mathrm{emp}}(\* x) = q (\* x) + \epsilon (\* x) \,,
\end{equation}
where $ \epsilon (\* x)  $ is some noise.
It is expected that $ \epsilon (\* x) $ has zero mean and some unknown variance related to the number of trajectories used in the estimate.
In the limit of large batch sizes, we can approximate the average over samples with an ensemble average. We then have for a single replica $ \alpha $ 
\begin{align}
    \ell_{\mathrm{ME}}(\mathcal{C}^{\alpha}; \thetab) &= \frac{1}{2}\Bigg[\frac{1}{|\mathcal{C}^{\alpha}|} \sum_{ ( q_\mathrm{emp}, \* x ) \in \mathcal{C}^{\alpha}}( \hat{q}(\* x; \thetab) - q_\mathrm{emp} ) \Bigg]^2 \\
    &\approx \frac{1}{2} \left(\langle \hat{q}(\* x; \thetab) - q_\mathrm{emp}(\*x ) \rangle_\alpha\right)^2 \\
    &\approx \frac{1}{2} \left( \langle \hat{q}(\* x; \thetab) - q (\*x ) \rangle_\alpha - \langle \epsilon (\* x) \rangle_\alpha \right)^2
    \\ 
    &\approx \frac{1}{2} \left( \langle \hat{q}(\* x; \thetab) - q (\*x )\rangle_\alpha \right)^2 \,,
\end{align}
where $ \langle ... \rangle_{\alpha} $ is the ensemble average with respect to replica $ \alpha $.
Note that the noise has been approximately canceled due to the effective matching of negative and positive error terms.
In practice, the locality of the replicas in both umbrella sampling and the FTS method likely ensures that $ \epsilon (\* x) $ is slowly varying.
This leads to the annihilation of noise at the level of summing over batches from every replica without the need for higher quality $ q_{\mathrm{emp}}( \* x ) $.

Returning to the gradient of the supervised learning loss function given in \cref{eq:grad_mes}, we have for large mini-batch sizes
\begin{equation} \label{eq:grad_mes_2}
    \nabla_{\thetab} \hat{L}_\mathrm{SL}(\thetab; \{ \mathcal{C}_k^\alpha \}) \approx \frac{\lambda_{\mathrm{SL}}}{M} \sum_{\alpha=1}^M \langle \hat{q}(\* x; \thetab) - q (\*x ) \rangle_\alpha \langle \nabla_{\thetab} \hat{q}(\* x; \thetab) \rangle_\alpha \,,
\end{equation}
in which the replica average of the gradient is coupled to a noise-reduced measure of the error.
A global minimum is achieved when
\begin{equation}
    \langle \hat{q}(\* x; \thetab) \rangle_\alpha = \langle q (\* x) \rangle_\alpha \quad \forall \alpha \,.
\end{equation}
While this condition can be satisfied for $ \hat{q}(\* x; \thetab) \neq q (\* x) $ in the region sampled by replica $ \alpha $, the additional loss terms in \cref{eq:variationalbke} and continuity between replicas seem to prevent trivial solutions in practice.

In summary, compared to the standard mean-squared loss, the chosen supervised-learning loss function avoids overfitting. 
This is likely due to the polling of empirical committor estimates, which leads to a reduction in the effect of noise on the optimization. 

\section{Examining the Sampling Error in Reweighting Factors} \label{app:ftsme}

In this section, we examine how sampling error in the reweighting factors $z_\alpha$ estimated from all algorithms is reduced in the limit of large batch sizes. For a given batch size, $z_\alpha$ is computed over many iterations of each algorithm while keeping the neural network fixed. Afterwards, the mean of $z_\alpha$ computed from all iterations is compared to the $z_\alpha$ computed from numerical integration of \cref{eq:zalphadef} for umbrella sampling, and \cref{eq:zalphafts} for the master-equation approach. 

\begin{figure}[t]
    \centering
    \includegraphics[width=\linewidth]{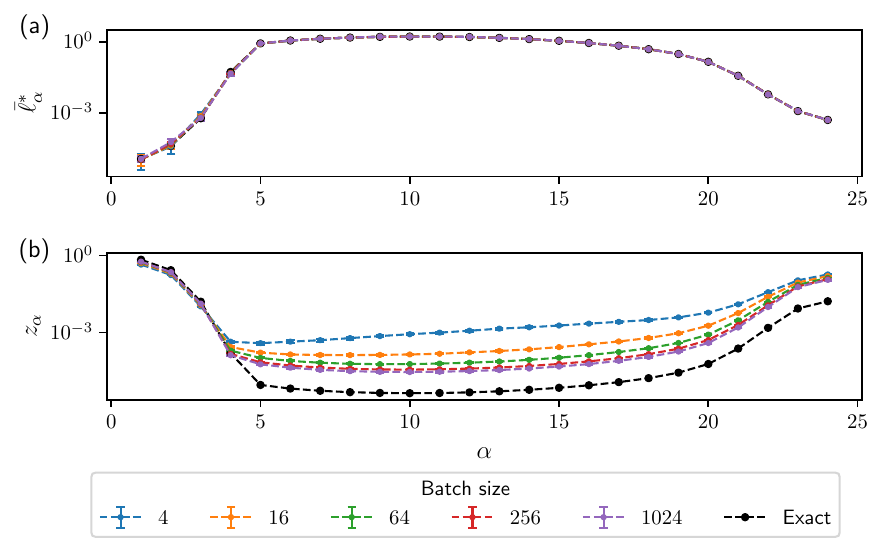}
    \caption{(a) The sample mean of the BKE loss from each replica $\bar{\ell}_\alpha^{*}=\frac{1}{|\mathcal{M}^\alpha_k|} \sum_{\* x \in \mathcal{M}_k^\alpha} \frac{\ell(\* x; \thetab_k)}{c(\* x; \thetab_k)}$, and (b) the estimated reweighting factor $z_\alpha$ from each replica $\alpha$, in comparison to the values obtained by numerical integration (`Exact') for committor-based umbrella sampling. This is done using a fixed neural network for all batch sizes that is obtained from the BKE--US+SL method.}    
    \label{fig:us_z_l_norm}
\end{figure}
\begin{figure}[p]
    \centering
    \includegraphics[width=\linewidth]{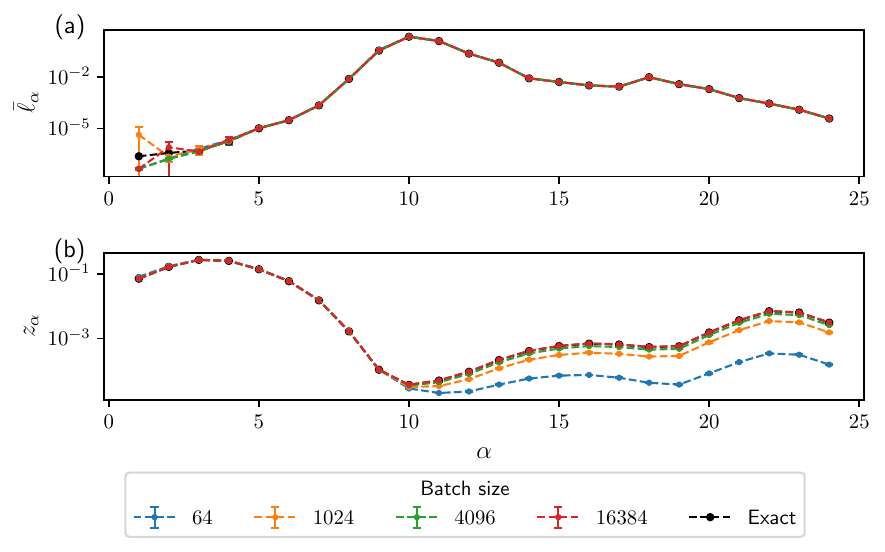}
    \caption{(a) The sample mean of the BKE loss from each replica $\bar{\ell}_\alpha = \frac{1}{|\mathcal{R}^\alpha_k|} \sum_{\* x \in \mathcal{R}_k^\alpha} \ell(\* x; \thetab_k)$, and (b) the estimated reweighting factor $z_\alpha$ from each replica $\alpha$, in comparison to the values obtained by numerical integration (`Exact') for the FTS method with master equation. This is done using a fixed neural network for all batch sizes that is obtained from the BKE--FTS(ME)+SL method.}    
    \label{fig:ftsme_z_l_norm}
    \vspace{12pt}
    \centering
    \includegraphics[width=\linewidth]{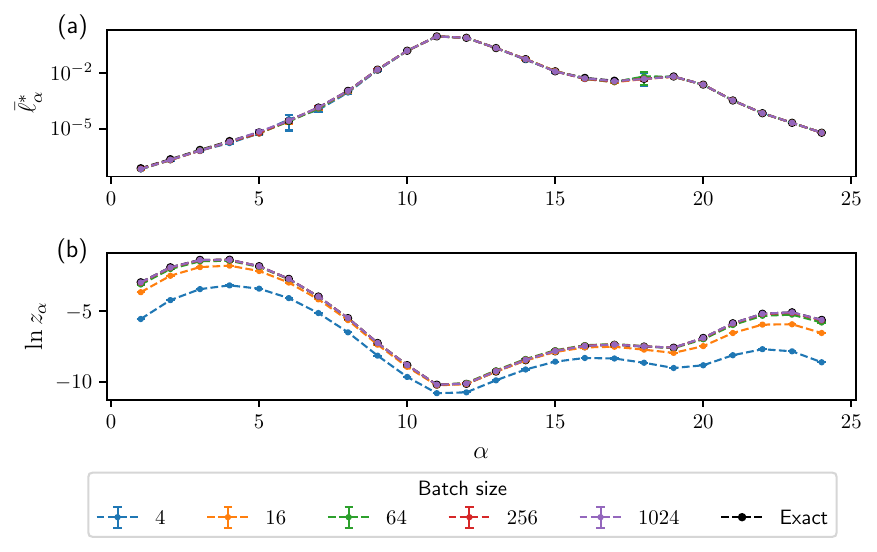}
    \caption{(a) The sample mean of the BKE loss from each replica $\bar{\ell}_\alpha^{*}=\frac{1}{|\mathcal{M}^\alpha_k|} \sum_{\* x \in \mathcal{M}_k^\alpha} \frac{\ell(\* x; \thetab_k)}{c(\* x; \thetab_k)}$, and (b) the estimated reweighting factor $\ln z_\alpha$ from each replica $\alpha$, in comparison to the values obtained by numerical integration (`Exact') for the path-based umbrella sampling. This is done using a fixed neural network for all batch sizes that is obtained from the BKE--FTS(US)+SL method.}    
    \label{fig:ftsus_z_l_norm}
\end{figure}

Figure~\ref{fig:us_z_l_norm}(b) shows $z_\alpha$ for committor-based umbrella sampling, where inaccurate estimates are obtained for $\alpha \in [5,24]$. This result arises due to a lack of overlap in samples obtained from adjacent replicas since $\alpha=5$ coincides with the beginning of non-overlap between samples from the reactant state (1-4) and the transition state, which begins at $\alpha=5$. The inaccuracy in $z_\alpha$ can be contrasted with the sample-mean quantity $\bar{\ell}_\alpha^{*}=\frac{1}{|\mathcal{M}^\alpha_k|} \sum_{\* x \in \mathcal{M}_k^\alpha} \frac{\ell(\* x; \thetab_k)}{c(\* x; \thetab_k)}$ (\cref{fig:us_z_l_norm}(a)), which shows uniform convergence beginning with the smallest batch size. From these results, we may conclude that the large sampling error of the on-the-fly estimates from the BKE--US and BKE--US(SL) method arises from inaccurate reweighting factors due to the lack of overlap in samples between neighboring replicas, and the accuracy may only be improved with prohibitively large batch sizes for training.

Figure~\ref{fig:ftsme_z_l_norm} shows both $z_\alpha$ and the sample mean of the BKE loss from each replica $\bar{\ell}_\alpha=\frac{1}{|\mathcal{R}^\alpha_k|} \sum_{\* x \in \mathcal{R}_k^\alpha} \ell(\* x; \thetab_k)$, as obtained from the FTS method with master equation. We see that the quantity $\bar{\ell}_\alpha$ converges quickly and uniformly, but the error in the reweighting factor $z_\alpha$, which is the largest for $\alpha\in [11,24]$, only diminishes at batch sizes that are too large and impractical to use for neural network training, i.e., at $O(4 \cdot 10^3)$. Meanwhile, $z_\alpha$ computed from path-based umbrella sampling (\cref{fig:ftsus_z_l_norm}(b)) achieves convergence at relatively smaller batch sizes, i.e., at $O(10^2)$, with similar quick convergence for the corresponding $\bar{\ell}_\alpha^{*}$ (\cref{fig:ftsus_z_l_norm}(a)). This demonstrates the advantage of using path-based umbrella sampling for computing accurate reweighting factors, and thus the utility of the BKE--FTS(US) and BKE--FTS(US)+SL method in obtaining accurate on-the-fly estimates of reaction rates at a wide range of batch sizes. 

\section{Additional Figures for Examining Log-Normal Behavior} \label{app:lognorm_figures}

\begin{figure}[t]
    \centering
    \includegraphics[width=\linewidth]{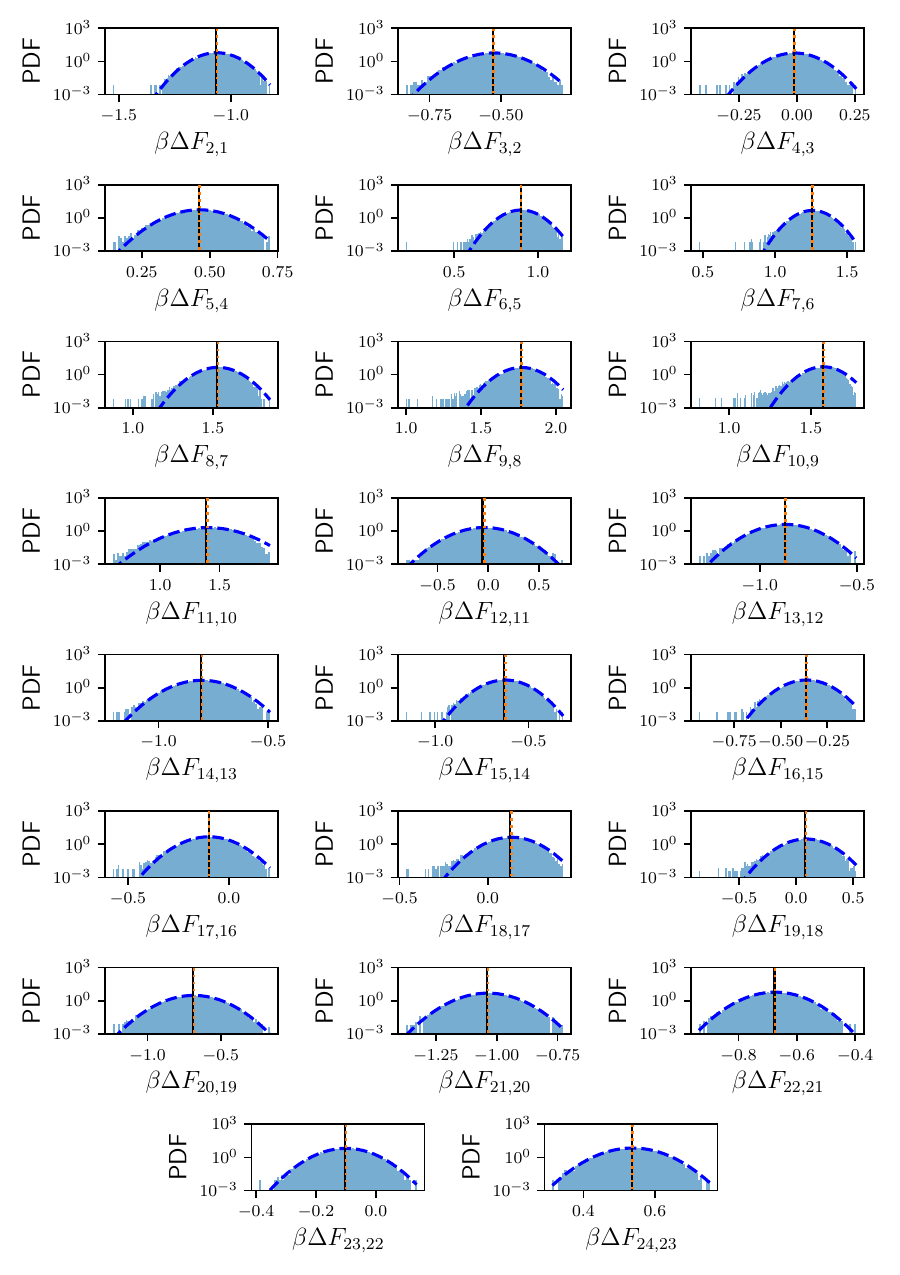}
    \caption{Probability density functions of the forward free-energy differences $\beta \Delta F_{(\alpha+1),\alpha}$.}
    \label{fig:summary_fwd}
\end{figure}

\begin{figure}[t]
    \centering
    \includegraphics[width=\linewidth]{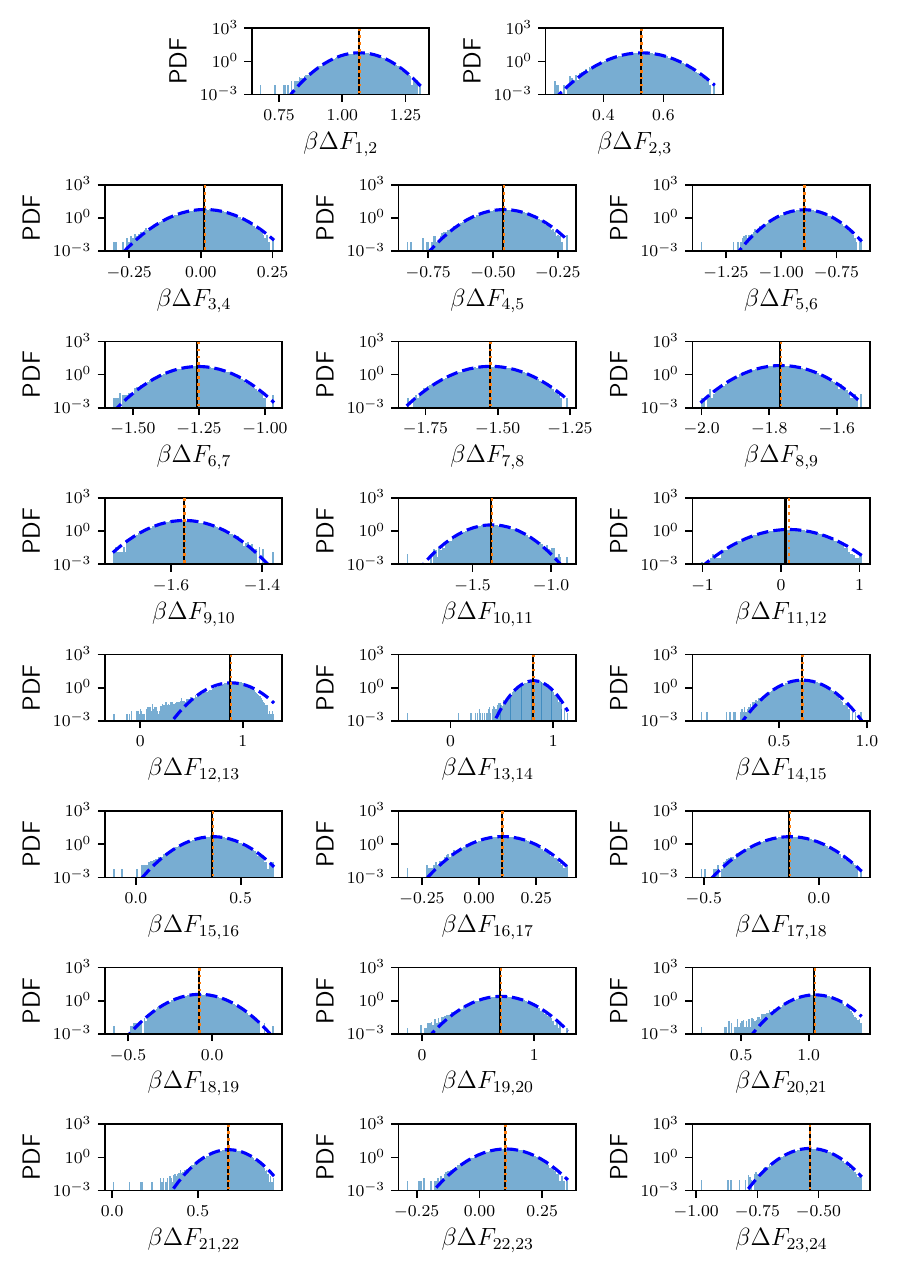}
    \caption{Probability density functions of the backward free-energy differences $\beta \Delta F_{(\alpha-1),\alpha}$.}
    \label{fig:summary_bwd}
\end{figure}

\begin{figure}[t]
    \centering
    \includegraphics[width=\linewidth]{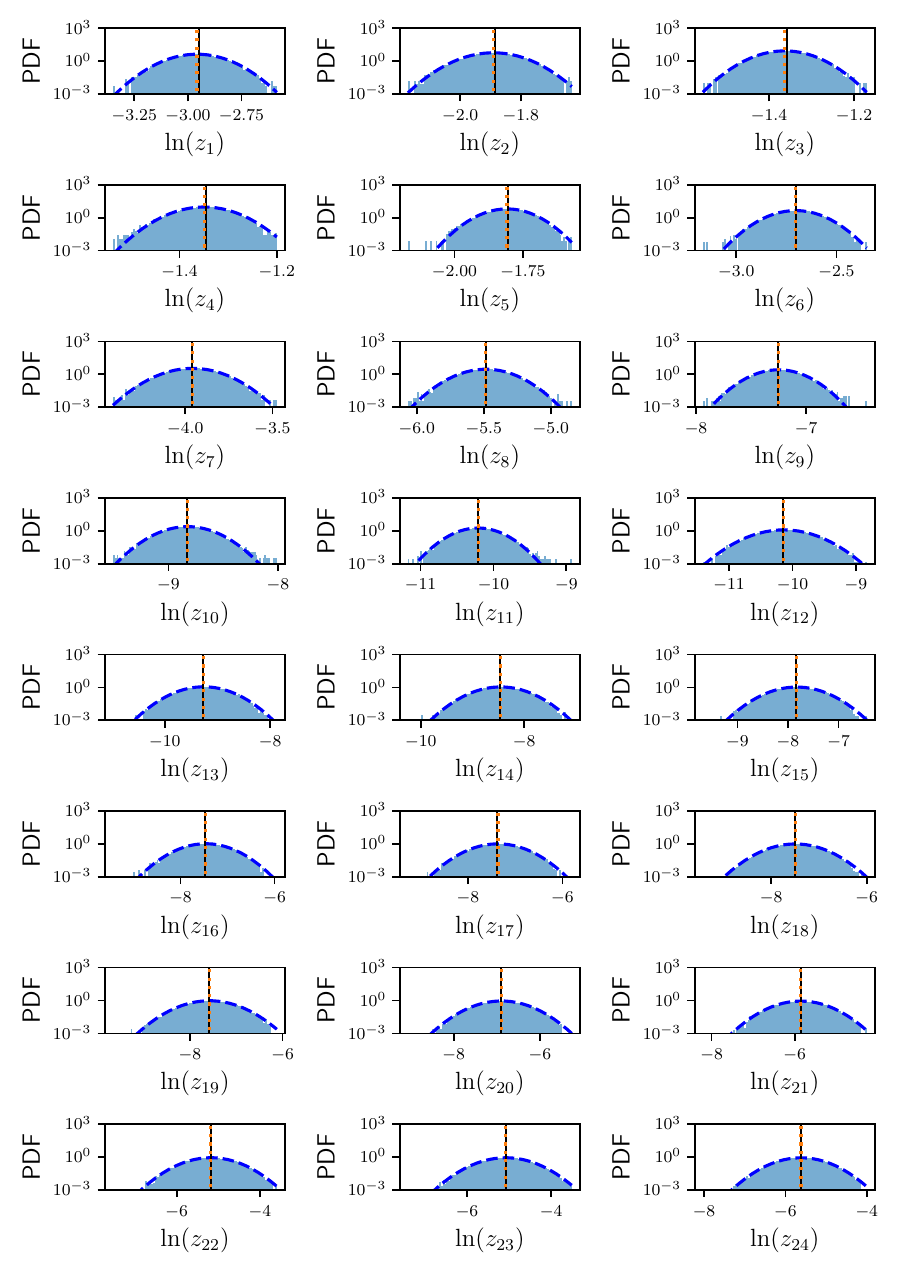}
    \caption{Probability density functions of the log of reweighting factors $\ln z_\alpha$.}
    \label{fig:summary_z}
\end{figure}

\begin{figure}[t]
    \centering
    \includegraphics[width=\linewidth]{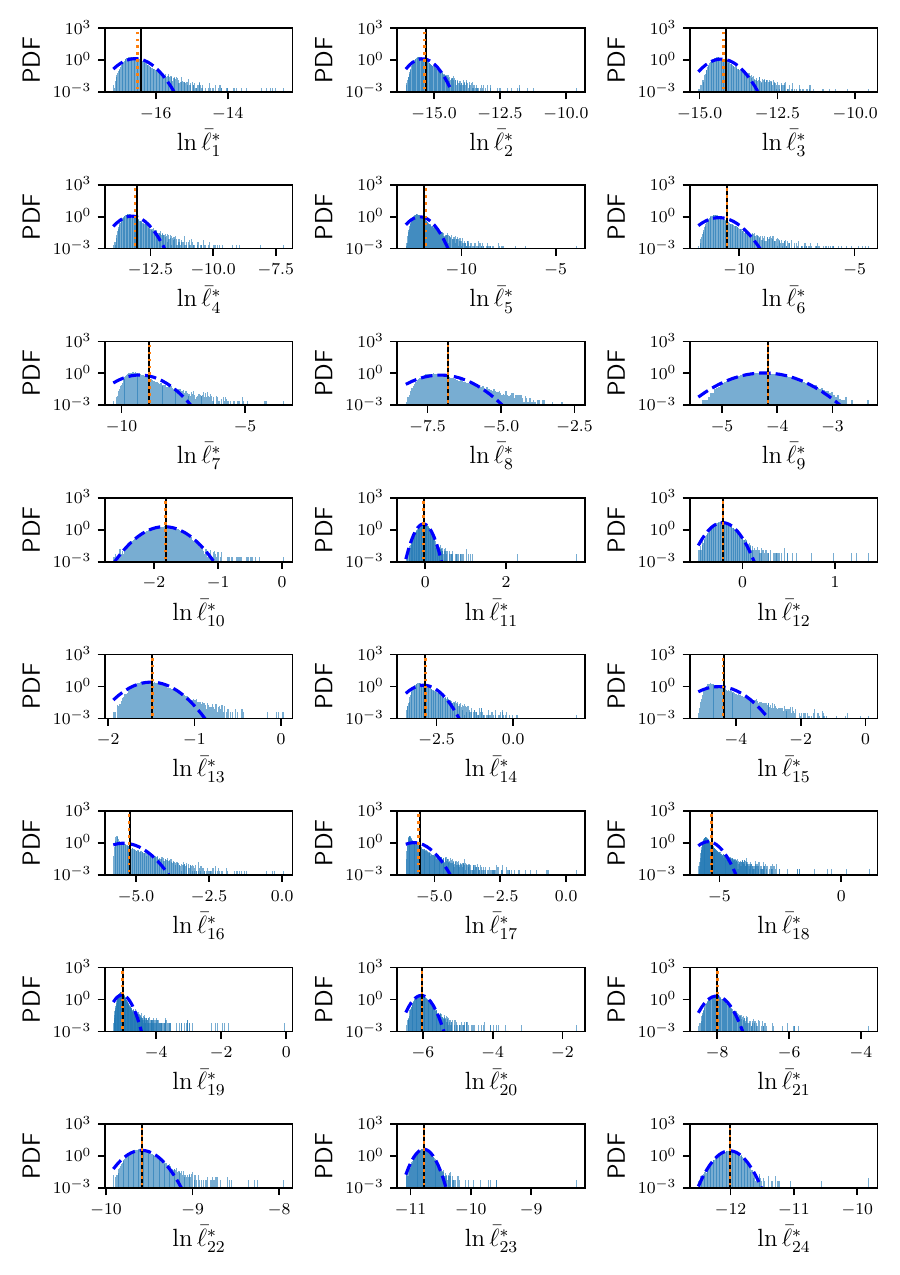}
    \caption{Probability density functions of $ \ln \bar{\ell}_{\alpha}^{*} $.}
    \label{fig:summary_grad}
\end{figure}

\begin{figure}[t]
    \centering
    \includegraphics[width=\linewidth]{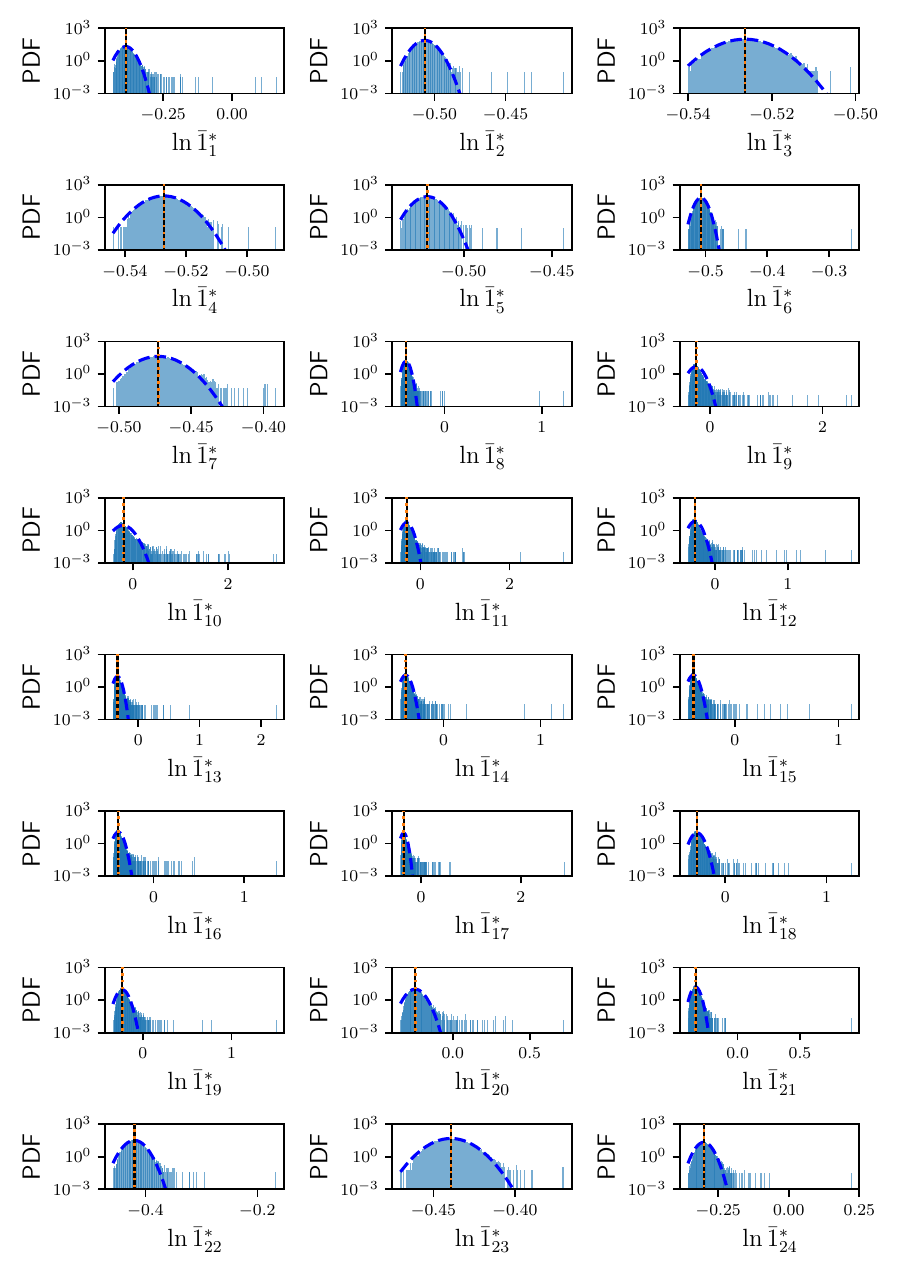}
    \caption{Probability density functions of $ \ln \bar{1}_{\alpha}^{*} $.}
    \label{fig:summary_norm}
\end{figure}

\begin{figure}[t]
    \centering
    \includegraphics[width=\linewidth]{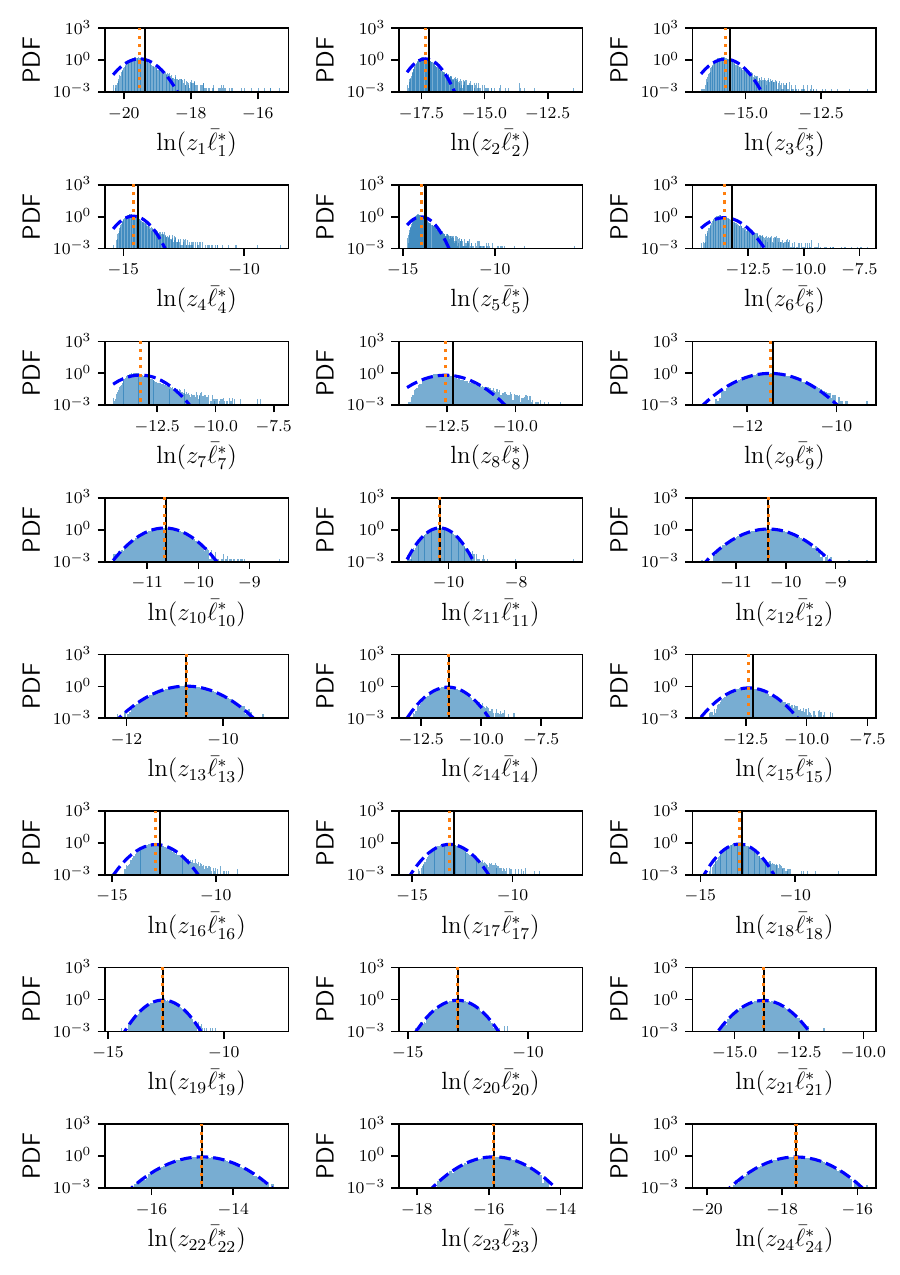}
    \caption{Probability density functions of $ \ln z_{\alpha} \bar{\ell}_{\alpha}^{*} $.}
    \label{fig:summary_grad_z}
\end{figure}

\begin{figure}[t]
    \centering
    \includegraphics[width=\linewidth]{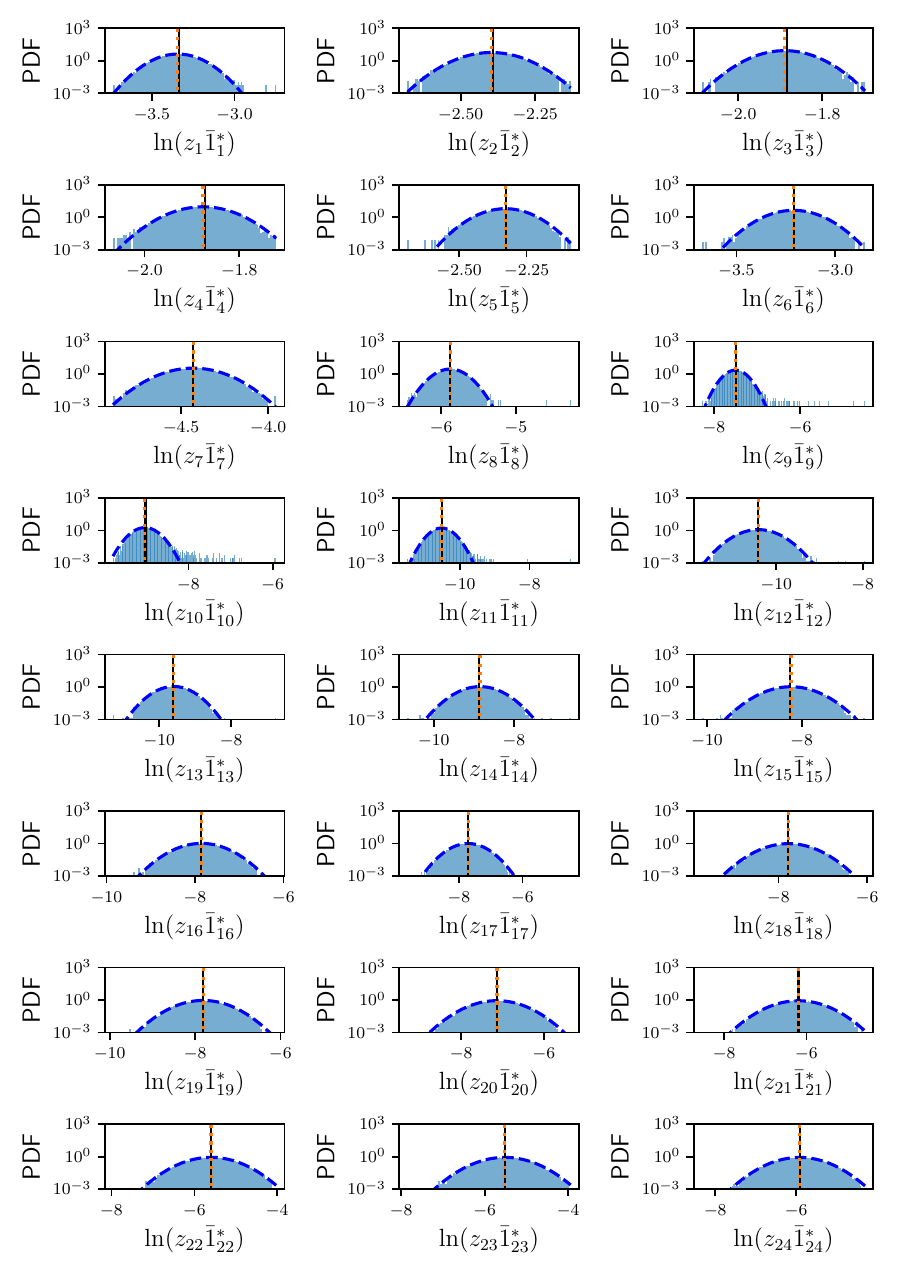}
    \caption{Probability density functions of $ \ln z_{\alpha} \bar{1}_{\alpha}^{*} $.}
    \label{fig:summary_norm_z}
\end{figure}

This section contains additional figures for the probability density functions (PDFs) of all quantities of interest in \cref{sec:lognorm} for all replicas.
The histograms for the forward free-energy differences are given in \cref{fig:summary_fwd}.
The histograms for the backward free-energy differences are given in \cref{fig:summary_bwd}.
The histograms for the reweighting factors are given in \cref{fig:summary_z}.
The histograms for $ \ln \bar{\ell}_{\alpha}^{*} $ and  $ \ln \bar{1}_{\alpha}^{*}$ are given in \cref{fig:summary_grad,fig:summary_norm}, respectively.
The histograms for $ \ln z_{\alpha} \bar{\ell}_{\alpha}^{*} $ and $ \ln z_{\alpha} \bar{1}_{\alpha}^{*} $ are given in \cref{fig:summary_grad_z,fig:summary_norm_z}, respectively. For all histograms, data is obtained by sampling a fixed neural network obtained from the BKE--FTS(US)+SL method at a batch size of $ 1024 $. Dashed blue lines correspond to log-normal distributions fitted using the method of moments \cite{pearson1936method}, while the vertical dotted orange and solid black lines correspond to the mean of the histograms and the corresponding ensemble average computed by numerical integration, respectively.

\begin{figure}[t]
    \centering
    \includegraphics[width=\linewidth]{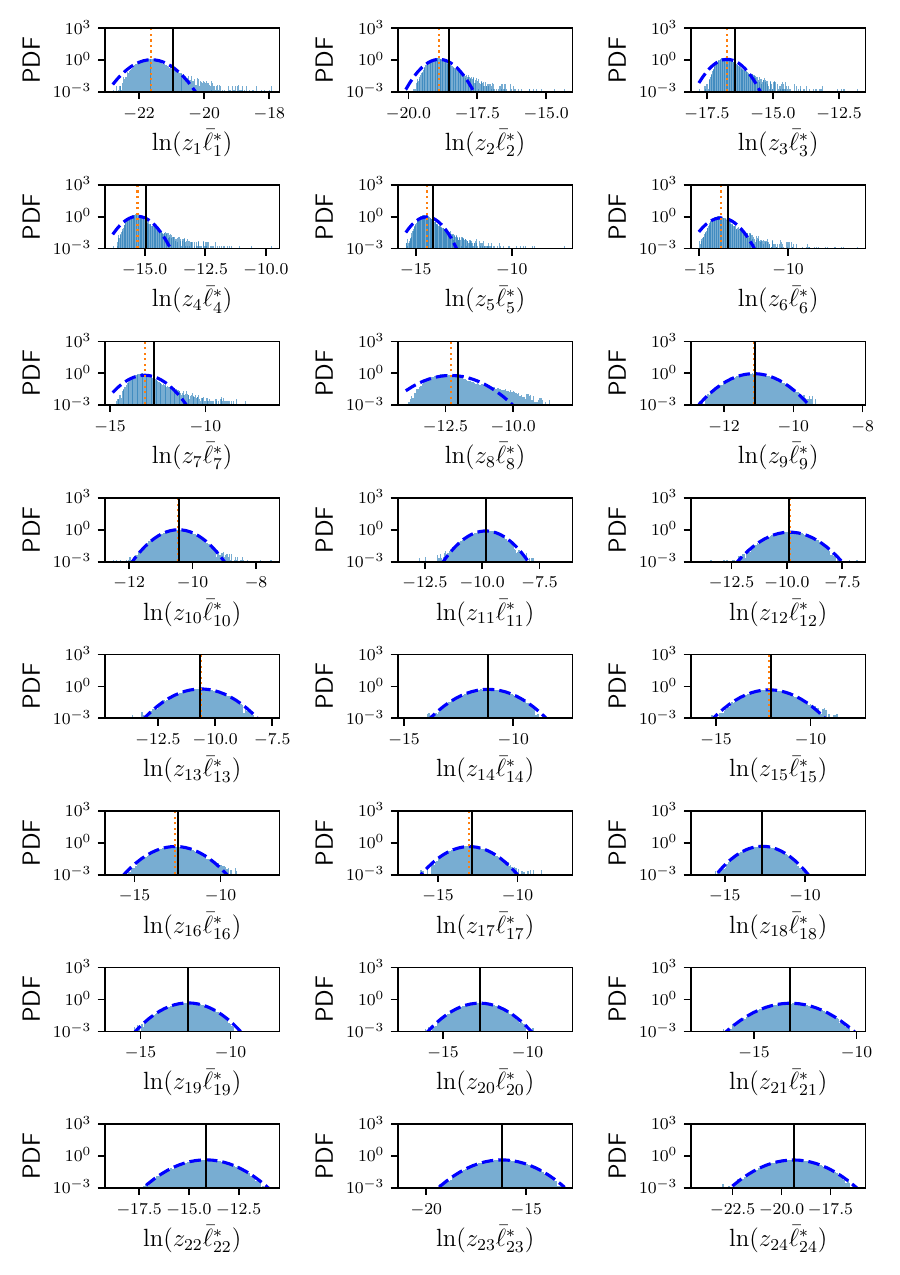}
    \caption{Probability density functions of $ \ln z_{\alpha} \bar{\ell}_{\alpha}^{*} $, where data is obtained from umbrella sampling with bias strengths $\kappa_\alpha^\parallel =2200$ and $\kappa_\alpha^\bot=300$.}
    \label{fig:summary_grad_z_other}
\end{figure}

\begin{figure}[t]
    \centering
    \includegraphics[width=\linewidth]{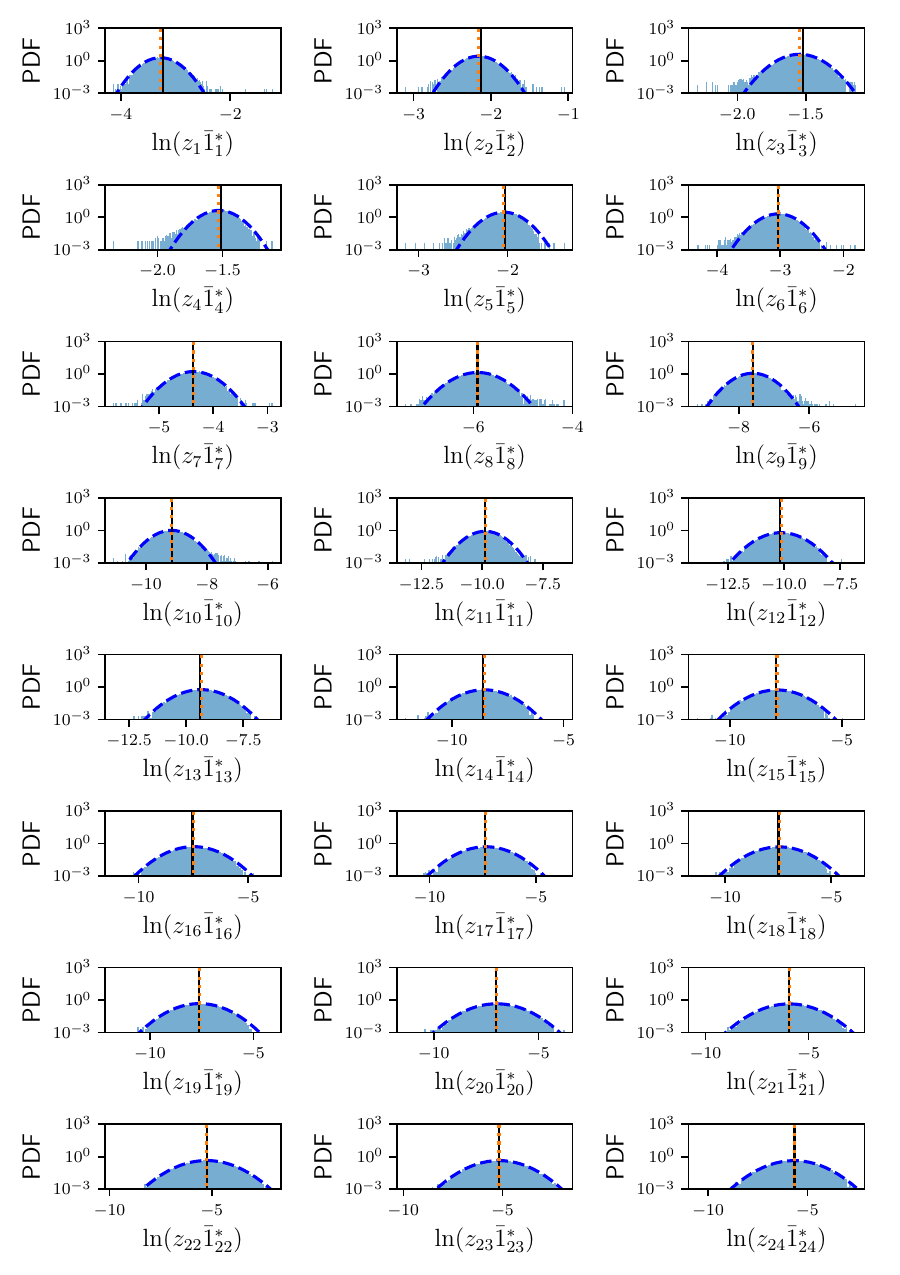}
    \caption{Probability density functions of $ \ln z_{\alpha} \bar{1}_{\alpha}^{*} $, where data is obtained from umbrella sampling with bias strengths $\kappa_\alpha^\parallel =2200$ and $\kappa_\alpha^\bot=300$.}
    \label{fig:summary_norm_z_other}
\end{figure}

The existence of tails in these PDFs is dependent upon the choice of bias potential parameters that are needed for the path-based umbrella sampling.
For instance, \cref{fig:summary_grad_z_other,fig:summary_norm_z_other} show the histograms for $ \ln z_{\alpha} \bar{\ell}_{\alpha}^{*} $ and $ \ln z_{\alpha} \bar{1}_{\alpha}^{*} $ when the bias potential parameters are changed from the ones in \cref{app:optimizer_mb} to $\kappa_\alpha^\parallel =2200$ and $\kappa_\alpha^\bot=300$, where we see that PDFs that originally possess tails, e.g., $\alpha \in [14,18]$ for $ \ln z_{\alpha} \bar{\ell}_{\alpha}^{*} $, are log-normal. It is also expected that any tails in the distributions are suppressed as the batch size is further increased.  

\end{document}